\documentclass[useAMS,usenatbib]{mn2e}

\usepackage{graphicx,color}
\usepackage{amssymb,amsmath}
\usepackage{amsmath} 
\usepackage{multicol}
\usepackage{latexsym}
\usepackage{amsfonts}
\usepackage{amssymb}
\usepackage{wasysym}
\usepackage{textcomp}
\usepackage{longtable,lscape}
\usepackage{rotating}
\usepackage{mathtools}
\usepackage{aas_macros}
\usepackage[dvipsnames]{xcolor}

\usepackage{natbib}
\usepackage{amssymb,amsmath}

\title[Investigating binary clusters with \textit{Gaia} EDR3]{Investigating Galactic binary cluster candidates with \textit{Gaia} EDR3}
\author[M. S. Angelo et al.]{M. S. Angelo$^{1}$\thanks{E-mail:
mateusangelo@cefetmg.br}, J. F. C. Santos Jr.$^{2}$, F. F. S. Maia$^{3}$ and W. J. B. Corradi$^{2,4}$    \\ 
\noindent
$^1$Centro Federal de Educa\c{c}\~ao Tecnol\'ogica de Minas Gerais, Av. Monsenhor Luiz de Gonzaga, 103, 37250-000 Nepomuceno, MG, Brazil\\
$^2$Departamento de F\'isica, ICEx, Universidade Federal de Minas Gerais, Av. Ant\^onio Carlos 6627, 31270-901 Belo Horizonte, MG, Brazil\\
$^3$Universidade Federal do Rio de Janeiro, Instituto de F\'isica, 21941-972, Brazil\\
$^4$Laborat\'orio Nacional de Astrof\'isica, R. Estados Unidos 154, 37530-000 Itajub\'a, MG, Brazil}

\begin{document}

\date{Accepted 2021 December 23. Received 2021 December 20; in original form 2021 October 26}

\pagerange{\pageref{firstpage}--\pageref{lastpage}} \pubyear{XXXX}

\maketitle

\label{firstpage}

\begin{abstract}


A number of stellar open cluster (OC) pairs in the Milky Way occupy similar positions in the phase space (coordinates, parallax and proper motions) and therefore may constitute physically interacting systems. The characterization of such objects based on observational data is a fundamental step towards a proper understanding of their physical status and to investigate cluster pair formation in the Galaxy. In this work, we employed \textit{Gaia} EDR3 data to investigate a set of 16 OCs distributed as seven stellar aggregates. We determined structural parameters and applied a decontamination technique that allowed to obtain unambiguous lists of member stars. The studied OCs span Galactocentric distances and ages in the ranges $7\lesssim\,R_G(\textrm{kpc})\lesssim11$ and $7.3\leq\textrm{log}\,t\leq9.2$. Eight OCs were found to constitute 4 gravitationally bound pairs (NGC\,5617$-$Trumpler\,22, Collinder\,394$-$NGC\,6716, Ruprecht\,100$-$Ruprecht\,101, NGC\,659$-$NGC\,663, the latter being a dynamically unevolved binary) and other 4 clusters constitute 2 interacting, but gravitationally unbound, pairs (King\,16$-$Berkeley\,4, NGC\,2383$-$NGC\,2384, the latter being a dissolving OC). Other 4 OCs (Dias\,1, Pismis\,19, Czernik\,20, NGC\,1857) seem not associated to any stellar aggregates. Apparently, clusters within bound and dynamically evolved pairs tend to present ratios of half-light to tidal radius larger than single clusters located at similar $R_G$, suggesting that mutual tidal interactions may possibly affect their structural parameters. Unbound or dynamically unevolved systems seems to present less noticeable signature of tidal forces on their structure. Moreover, the core radius seems more importantly correlated with the clusters' internal dynamical relaxation process.

\end{abstract}


\section{Introduction}

It is well-known that open clusters (OCs) are essential tools for studying the properties and evolution of stars as well as the Milky Way itself. In this context, previous investigations have suggested that the fraction of OCs located in gravitationally interacting pairs\footnote[1]{Throughout this paper, the terms ``pairs" and ``aggregates"\,are employed in a general context, either referring to gravitationally bound or unbound groups of interacting OCs, or even chance alignments in the sky. In turn, the term ``binary cluster"\,is reserved specifically to gravitationally bound groups of 2 OCs.} is not negligible. The formation, physical properties and evolution of such systems have been a debated topic.

Observationally, a number of works have employed catalogued data to identify star cluster pairs, either based on the small angular separation between the two components or employing known distances. \cite{Subramaniam:1995} employed the catalogues of \cite{Lynga:1987a} and \cite{Mermilliod:1995} to identify a list of 18 binary cluster candidates with separations smaller than 20\,pc. In contrast with earlier findings, which established $h+\chi$ Persei (NGC\,869/NGC\,884) as the only confirmed binary cluster in the Galaxy, \cite{Subramaniam:1995} concluded that $\sim8\%$ of the OCs may constitute actual binaries.   

\citeauthor{de-La-Fuente-Marcos:2009}\,\,(\citeyear{de-La-Fuente-Marcos:2009}, hereafter MM09) employed data from WEBDA \citep{Netopil:2012} and \cite{Dias:2002} catalogues for a volume-limited sample of OCs, located at the solar circle, in order to identify binary candidates. They employed as basic selection criteria the physical (i.e., not projected) distance between pairs of OCs, assuming that 2 objects constitute an interacting system when their separation is smaller than 3 times the average value of the tidal radius for clusters in the Milky Way disk ($\sim$10 pc, \citeauthor{Binney:2008}\,\,\citeyear{Binney:2008}). From the outcomes of their procedure, they concluded that at least $\sim10\%$ of all OCs appear to be experiencing some type of interaction with another cluster, comparable to what is suggested for the Magellanic Clouds (e.g., \citeauthor{Bhatia:1988}\,\,\citeyear{Bhatia:1988}; \citeauthor{Hatzidimitriou:1990}\,\,\citeyear{Hatzidimitriou:1990}; \citeauthor{Pietrzynski:2000}\,\,\citeyear{Pietrzynski:2000}; \citeauthor{Dieball:2002}\,\,\citeyear{Dieball:2002}). 

\citeauthor{Piecka:2021}\,\,(\citeyear{Piecka:2021}, hereafter PP21) employed the OCs data set from \citeauthor{Cantat-Gaudin:2020a}\,\,(\citeyear{Cantat-Gaudin:2020a}, hereafter CG20) catalogue to study aggregates of clusters located at relatively narrow volumes of the phase space (coordinates, parallax and proper motion components), which could indicate some kind of physical interaction. For all selected aggregates, the difference between the median values of both proper motion components is less than half the sum of the corresponding standard deviations. The same criteria was applied for the positions, while the full sum of the standard deviations was used for parallaxes. They identified 60 aggregates, which also share several of the assigned member stars, demonstrating that disentangling each cluster population may not be a trivial task. Moreover, they found that the presence of coincidental members results in parallax distributions that do not coincide with the catalogued values and thus reanalysed these OCs.

From the theoretical point of view, different mechanisms have been proposed to explain the origin of interacting pairs. Such physical processes include: ($i$) simultaneous formation, in which both clusters originated from the same progenitor molecular cloud (e.g., \citeauthor{Bekki:2004}\,\,\citeyear{Bekki:2004}); ($ii$) sequential formation, in which stellar winds or supernova shocks generated within a cluster can induce the collapse of a nearby cloud, triggering the formation of a companion cluster (e.g., \citeauthor{Goodwin:1997}\,\,\citeyear{Goodwin:1997}); ($iii$) resonant trapping followed by tidal capture, that can result in the formation of pairs with a common kinematics, but different ages and chemical composition (e.g., \citeauthor{van-den-Bergh:1996}\,\,\citeyear{van-den-Bergh:1996}; \citeauthor{Dehnen:1998}\,\,\citeyear{Dehnen:1998}).

In this sense, \citeauthor{de-la-Fuente-Marcos:2010}\,\,(\citeyear{de-la-Fuente-Marcos:2010}, hereafter MM10) carried out $N-$body simulations to investigate the evolution of primordial binary open star clusters under the influence of the Milky Way tidal field. They found that the possible evolutionary paths (merging, extreme tidal distortions or dissociation) followed by such systems and their characteristic timescales depend on the initial orbital elements (semi-major axis, eccentricity) and cluster pair mass ratio. In general, binaries with small separations generally result in mergers, while wider pairs are separated under the influence of the external gravitational potential. Their calculations show that a long-term stability of binary star clusters is rare, since few pairs are expected to be observed in close proximity for more than 100\,Myr. Additionally, \cite{Priyatikanto:2016} showed that even binary clusters with the same initial orbital parameters may experience completely different dynamical fates, depending on their initial Galactic orbit orientation and phase. The orbital evolution of binary clusters may involve orbital reversal, spiral-in and vertical oscillation about Galactic plane, prior to obtaining their final configurations (merger or separation).

More recently, $N-$body simulations for a time span of 50\,Myr have been performed by \citeauthor{Darma:2021}\,\,(\citeyear{Darma:2021}; hereafter DAK21) in order to investigate the formation of binary star clusters in the Milky Way under a set of star cluster formation conditions, i.e., different degrees of initial substructures in the spatial distribution of the stellar population and different virial ratios $\alpha_{\textrm{vir}}$. They found that stellar groups with clumpier initial structures are more likely to form binary star clusters at early times, immediately after a phase of violent relaxation (\citeauthor{Takahashi:2000}\,\,\citeyear{Takahashi:2000}; \citeauthor{Baumgardt:2003}\,\,\citeyear{Baumgardt:2003}; \citeauthor{Gieles:2011a}\,\,\citeyear{Gieles:2011a}; \citeauthor{Heggie:2003}\,\,\citeyear{Heggie:2003}). On the other hand, higher stellar velocity dispersion (super-virial) are more likely to form gravitationally unbound systems.  

Also recently, \cite{Dalessandro:2021} employed astrometric and photometric data from the \textit{Gaia}\,EDR3 catalogue \citep{Gaia-Collaboration:2021}, combined with high-resolution optical and near-infrared spectra, to perform a detailed analysis of the dynamics and spectro-photometric properties of stars within a few degrees from the $h+\chi$ Persei pair. These stars were found to be part of a common, substructured stellar complex, which they called LISCA\,I. A set of $N-$body simulations were also performed, generating, as initial conditions, different fractal dimension values to simulate inhomogeneous mass distributions, different stellar velocity regimes and, additionally, the effects of the presence of some primordial angular momentum were explored. They showed that the observational properties of the $h+\chi$ Persei complex fit well within early stages of the modeled hierarchical cluster assembly process, which can eventually evolve in a relatively massive stellar system.

Some other numerical simulations worth mentioning have: ($i$) focused on the impact of the rotation and the presence of a mass spectrum on the clusters’ dynamics during and after the initial violent relaxation stage \citep{Livernois:2021}; ($ii$) investigated the built up process of stellar clusters via mergers of smaller clumps (\citeauthor{Fujii:2021}\,\,\citeyear{Fujii:2021}, who devised an accurate integration of individual stellar orbits without gravitational softening); ($iii$) analysed the dynamical evolution of modeled star-forming regions by detailing their degree of substructrure, mass segregation and local density distribution (\citeauthor{Parker:2014}\,\,\citeyear{Parker:2014}, who introduced the $\mathcal{Q}-\Sigma_{LDR}$ diagram, which allows to differentiate between substructured associations and clusters for a set of fractal dimensions and virial ratios at different ages). It is noticeable that a great effort has been spent in the search of possible unifying principles that govern the formation of different stellar systems \citep{Dalessandro:2021}.

In this framework, the investigation of cluster pairs can provide crucial information about the mechanisms of cluster formation and evolution \citep{Dalessandro:2018}. Since binary clusters seem not an uncommon occurrence in the Galaxy, they are potentially interesting objects that could contribute for a proper comprehension on the dynamics of groups of stellar systems \citep{Subramaniam:1995}. Additionally, merger of cluster pairs may explain the presence of multiple stellar populations, as in the case of iron-complex globular clusters (e.g., $\omega$\,Cen, NGC\,1851) in the Milky Way, leaving a signature in their rotation curves \citep{Gavagnin:2016}, or in the Magellanic Clouds, as in the case of the intermediate-age clusters NGC\,411 and NGC\,1806 \citep{Hong:2017}. This way, it is clear that observational studies aimed at establishing memberships of stars within cluster pairs and estimation of their physical properties are really essential.



Ideally, observational investigations should employ updated data and uniform analysis methods, in order to allow unbiased comparisons among, e.g., dynamical parameters of binary and single stellar clusters. The present paper is a contribution in this sense. Here we investigated a set of 7 cluster aggregates, previously classified as candidates or confirmed physical binaries: NGC\,5617 $-$ Trumpler\,22, Collinder\,394 $-$ NGC\,6716, NGC\,2383 $-$ NGC\,2384, Ruprecht\,100 $-$ Ruprecht\,101, King\,16 $-$ Dias\,1, NGC\,659 $-$ NGC\,663, Czernik\,20 $-$ NGC\,1857. The OC Pismis\,19 is projected in the same field of NGC\,5617$-$Trumpler\,22 and thus has been incorporated into our analysis. The same for the OC Berkeley\,4, which was proved to be a closer companion of King\,16 than Dias\,1 (further details in Section~\ref{sec:results}). In what follows, we provide brief informations about previous investigations related to these systems.

\cite{De-Silva:2015a} performed photometric and spectroscopic analysis of the OCs NGC\,5617 and Trumpler\,22 and concluded that they constitute a genuine binary cluster. The same conclusion was then drawn by \cite{Bisht:2021}, who identified member stars for both OCs from \textit{Gaia}\,EDR3 data and estimated their orbits around the Galactic centre. Their close values of orbital parameters indicate that they are physically connected. By means of \textit{Gaia}\,DR2 \citep{Gaia-Collaboration:2018} data, \cite{Naufal:2020} investigated the morphology and identified member stars of the pair Collinder\,394$-$NGC\,6716; based on their close physical proximity and similar kinematics, they concluded that both OCs might be interacting with each other. 

In turn, the pair NGC\,2383$-$NGC\,2384 is part of the sample analysed by \cite{Vazquez:2010}. They concluded that both are physically close to each other, but discarded the hypothesis of a common origin, given their different ages and metallicities, in agreement with the previous study of \cite{Subramaniam:1999}. Particularly, NGC\,2384 was considered a sparse young remnant group of a recent star formation episode. The other 4 pairs (Ruprecht\,100 $-$ Ruprecht\,101, King\,16 $-$ Dias\,1, NGC\,659 $-$ NGC\,663, Czernik\,20 $-$ NGC\,1857) have been taken from the list of PP21. We have selected OCs from PP21's  list based on the following criteria: the OCs are free from their parental molecular clouds and show notable contrast in relation to the Galactic field population in their direction. This allowed us to derive structural parameters and to build decontaminated colour-magnitude diagrams (CMD) with recognizable evolutionary sequences.    
 
By employing our membership assignment technique and by deriving astrophysical and structural properties for this sample in an homogeneous way, we avoided systematic biases regarding the relative dynamical stage of each pair. Our objective is twofold: firstly, to establish unambiguous lists of member stars for the analysed OCs and to determine their fundamental astrophysical parameters (distance, age, reddening), thus allowing to conclude about the physical nature of each pair (physically interacting or just chance alignments along the line of sight); then, for the physical pairs, we aim at establishing their dynamical state based on relations between structural parameters (core ($r_c$), half-light ($r_h$) and tidal ($r_t$) radii) and their possible connection with time-related parameters (half-light relaxation times ($t_{rh}$) and age/$t_{rh}$ ratios). Comparisons with theoretical results are also outlined.        

This paper is structured as follows: in Section~\ref{sec:sample}, we describe the collected data and present the sample of studied OCs; in Section~\ref{sec:method}, we present our method; the results are shown in Section~\ref{sec:results} and discussed in Section~\ref{sec:discussion}; our conclusions are outlined in Section~\ref{sec:conclusions}.

\section{Sample and data description}
\label{sec:sample}

\begin{table*}
\centering
\tiny
\rotcaption{ Central Coordinates, Galactocentric distances, structural and fundamental parameters, mean proper motion components and half-light relaxation times ($t_{rh}$) for the studied sample. }
\begin{sideways}
\begin{minipage}{240mm}

\begin{tabular}{lllccccccccccccl}

 Cluster             & $RA$                            & $DEC$                           & $\ell$             & $b$               & R$_G$                  & $r_c$                & $r_{h}$             & $r_t$                  & $(m-M)_0$              & $E(B-V)$             & log\,$t$              & $[Fe/H]$                & $\langle\mu_{\alpha}\,\textrm{cos}\,\delta\rangle^{\dag\dag}$   & $\langle\mu_{\delta}\rangle^{\dag\dag}$     & $t_{rh}$                   \\    
                     &($h$:$m$:$s$)                    & ($^{\circ}$:$'$:$''$)           & $(^\circ)$         & $(^\circ)$        &  (kpc)                 &  (pc)                &     (pc)            &  (pc)                  & (mag)                  & (mag)                & (dex)                 & (dex)                   & (mas yr$^{-1}$)                                                 & (mas yr$^{-1}$)                             & (Myr)                      \\     
                                                                                                                                                                                                                                                                                                                                                                                                                                                                                
\hline                                                                                                                                                                                                                                                                                                                                                                                                                                                                          
NGC\,5617            & 14:29:44                        & -60:42:42                        &  314.67           & -0.10             & 6.87\,$\pm$\,0.52      & 3.58\,$\pm$\,0.41    & 2.82\,$\pm$\,0.21   & 5.51\,$\pm$\,0.51      & 11.24\,$\pm$\,0.30     & 0.62\,$\pm$\,0.10    & 8.30\,$\pm$\,0.15     & -0.13\,$\pm$\,0.23      & -5.68\,$\pm$\,0.10                                              & -3.18\,$\pm$\,0.08                          & 143\,$\pm$\,18             \\ 
Trumpler\,22         & 14:31:13                        & -61:08:50                        &  314.64           & -0.58             & 6.87\,$\pm$\,0.52      & 2.57\,$\pm$\,0.51    & 2.50\,$\pm$\,0.34   & 5.66\,$\pm$\,1.03      & 11.24\,$\pm$\,0.30     & 0.67\,$\pm$\,0.05    & 8.20\,$\pm$\,0.10     & -0.32\,$\pm$\,0.24      & -5.15\,$\pm$\,0.05                                              & -2.69\,$\pm$\,0.05                          & 92 \,$\pm$\,20             \\
Pismis\,19$^{*}$     & 14:30:42                        & -60:53:27                        &  314.70           & -0.30             & 6.79\,$\pm$\,0.52      & 0.62\,$\pm$\,0.10    & 0.98\,$\pm$\,0.12   & 3.30\,$\pm$\,0.56      & 11.42\,$\pm$\,0.30     & 1.40\,$\pm$\,0.10    & 9.00\,$\pm$\,0.10     & -0.06\,$\pm$\,0.20      & -5.48\,$\pm$\,0.10                                              & -3.40\,$\pm$\,0.07                          & 44 \,$\pm$\,9              \\
\hline                                                                                                                                                                                                                                                                                                                                                                                                                                                                          
NGC\,659             & 01:44:18                        &  60:39:22                        &  129.38           & -1.54             & 9.88\,$\pm$\,0.58      & 1.45\,$\pm$\,0.23    & 2.20\,$\pm$\,0.37   & 7.04\,$\pm$\,1.91      & 12.10\,$\pm$\,0.40     & 0.79\,$\pm$\,0.10    & 7.55\,$\pm$\,0.30     &  0.00\,$\pm$\,0.23      & -0.84\,$\pm$\,0.07                                              & -0.34\,$\pm$\,0.08                          & 78 \,$\pm$\,21             \\
NGC\,663             & 01:46:22                        &  61:13:51                        &  129.47           & -0.94             & 9.67\,$\pm$\,0.56      & 3.22\,$\pm$\,0.41    & 3.93\,$\pm$\,0.38   & 10.62\,$\pm$\,1.37     & 11.86\,$\pm$\,0.40     & 0.90\,$\pm$\,0.10    & 7.35\,$\pm$\,0.15     & -0.13\,$\pm$\,0.31      & -1.13\,$\pm$\,0.09                                              & -0.33\,$\pm$\,0.09                          & 317\,$\pm$\,47             \\
\hline                                                                                                                                                                                                                                                                                                                                                                                                                                                                          
Collinder\,394       & 18:52:08                        & -20:17:43                        &  14.88            & -9.24             & 7.35\,$\pm$\,0.51      & 1.24\,$\pm$\,0.20    & 1.24\,$\pm$\,0.17   & 2.86\,$\pm$\,0.60      & 9.19\,$\pm$\,0.30      & 0.35\,$\pm$\,0.10    & 7.85\,$\pm$\,0.20     &  0.05\,$\pm$\,0.20      & -1.47\,$\pm$\,0.09                                              & -5.90\,$\pm$\,0.11                          & 21\,$\pm$\,5               \\   
NGC\,6716            & 18:54:30                        & -19:54:59                        &  15.40            & -9.59             & 7.36\,$\pm$\,0.51      & 0.99\,$\pm$\,0.20    & 1.07\,$\pm$\,0.17   & 2.63\,$\pm$\,0.59      & 9.16\,$\pm$\,0.30      & 0.23\,$\pm$\,0.10    & 7.90\,$\pm$\,0.30     &  0.05\,$\pm$\,0.20      & -1.50\,$\pm$\,0.08                                              & -6.04\,$\pm$\,0.09                          & 14\,$\pm$\,4               \\    
\hline                                                                                                                                                                                                                                                                                                                                                                                                                                                                          
NGC\,2383            & 07:24:41                        & -20:56:09                        &  235.27           & -2.46             & 9.79\,$\pm$\,0.51      & 1.88\,$\pm$\,0.39    & 2.12\,$\pm$\,0.24   & 5.40\,$\pm$\,0.63      & 12.15\,$\pm$\,0.20     & 0.35\,$\pm$\,0.05    & 8.50\,$\pm$\,0.10     & -0.22\,$\pm$\,0.19      & -1.61\,$\pm$\,0.04                                              &  1.91\,$\pm$\,0.03                          & 77\,$\pm$\,14              \\   
NGC\,2384$^{\dag}$   & 07:25:08                        & -21:01:04                        &  235.39           & -2.39             & 9.95\,$\pm$\,0.58      & 5.61\,$\pm$\,0.77    & 17.0\,$\pm$\,0.9    & 31\,$\pm$\,6           & 12.33\,$\pm$\,0.40     & 0.34\,$\pm$\,0.10    & 7.30\,$\pm$\,0.15     &  0.00\,$\pm$\,0.23      & -2.29\,$\pm$\,0.07                                              &  3.10\,$\pm$\,0.07                          & 1883\,$\pm$\,177           \\    
\hline
King\,16             & 00:43:45                        &  64:10:57                        &  122.09           & 1.33              & 9.61\,$\pm$\,0.50      & 1.41\,$\pm$\,0.22    & 1.73\,$\pm$\,0.20   & 4.69\,$\pm$\,0.74      & 12.04\,$\pm$\,0.15     & 0.85\,$\pm$\,0.10    & 8.20\,$\pm$\,0.20     &  0.14\,$\pm$\,0.16      & -3.04\,$\pm$\,0.05                                              & -0.54\,$\pm$\,0.07                          & 42\,$\pm$\,9               \\   
Berkeley\,4$^{**}$   & 00:45:11                        &  64:23:46                        &  122.25           & 1.53              & 9.50\,$\pm$\,0.50      & 0.98\,$\pm$\,0.21    & 1.13\,$\pm$\,0.14   & 2.93\,$\pm$\,0.42      & 11.90\,$\pm$\,0.15     & 0.80\,$\pm$\,0.10    & 7.40\,$\pm$\,0.20     & -0.06\,$\pm$\,0.26      & -2.41\,$\pm$\,0.09                                              & -0.32\,$\pm$\,0.08                          & 22\,$\pm$\,5               \\   
Dias\,1              & 00:42:31                        &  64:03:16                        &  121.97           & 1.20              & 9.28\,$\pm$\,0.50      & 0.73\,$\pm$\,0.12    & 0.90\,$\pm$\,0.10   & 2.43\,$\pm$\,0.36      & 11.60\,$\pm$\,0.15     & 1.20\,$\pm$\,0.10    & 7.45\,$\pm$\,0.20     &  0.00\,$\pm$\,0.23      & -2.73\,$\pm$\,0.06                                              & -0.60\,$\pm$\,0.04                          & 14\,$\pm$\,3               \\   
\hline                                                                                                                                                                                                                                                                                                                                                                                                                                                                          
Czernik\,20          & 05:20:31                        &  39:32:57                        &  168.28           & 1.43              & 11.00\,$\pm$\,0.57     & 1.42\,$\pm$\,0.27    & 2.08\,$\pm$\,0.37   & 6.47\,$\pm$\,1.77      & 12.42\,$\pm$\,0.20     & 0.59\,$\pm$\,0.05    & 9.15\,$\pm$\,0.10     & -0.13\,$\pm$\,0.16      &  0.55\,$\pm$\,0.12                                              & -1.59\,$\pm$\,0.09                          & 69 \,$\pm$\,20             \\
NGC\,1857            & 05:20:05                        &  39:20:28                        &  168.44           & 1.22              & 10.20\,$\pm$\,0.54     & 2.87\,$\pm$\,0.39    & 3.35\,$\pm$\,0.36   & 8.73\,$\pm$\,1.30      & 11.75\,$\pm$\,0.20     & 0.60\,$\pm$\,0.10    & 8.45\,$\pm$\,0.10     & -0.47\,$\pm$\,0.33      &  0.46\,$\pm$\,0.11                                              & -1.38\,$\pm$\,0.05                          & 157\,$\pm$\,27             \\
\hline                                                                                                                                                                                                                                                                                                                                                                                                                                                                          
Ruprecht\,100        & 12:06:04                        & -62:38:04                        &  297.71           & -0.21             & 7.14\,$\pm$\,0.51      & 1.13\,$\pm$\,0.20    & 2.24\,$\pm$\,0.38   & 8.98\,$\pm$\,2.43      & 12.22\,$\pm$\,0.30     & 0.49\,$\pm$\,0.10    & 8.75\,$\pm$\,0.15     &  0.05\,$\pm$\,0.15      & -8.26\,$\pm$\,0.04                                              &  0.95\,$\pm$\,0.02                          & 72 \,$\pm$\,20             \\
Ruprecht\,101        & 12:09:34                        & -62:59:10                        &  298.22           & -0.50             & 7.12\,$\pm$\,0.51      & 2.96\,$\pm$\,0.48    & 3.33\,$\pm$\,0.39   & 8.41\,$\pm$\,1.36      & 12.20\,$\pm$\,0.30     & 0.62\,$\pm$\,0.05    & 9.00\,$\pm$\,0.15     &  0.05\,$\pm$\,0.15      & -9.52\,$\pm$\,0.07                                              &  0.48\,$\pm$\,0.09                          & 207\,$\pm$\,40             \\
\hline

\end{tabular}                                                                                                                                                                                                                                                                                                                                                                                                                                                                     
                                                                                                                                                                                                                                                                                                                                                                                                                                                                                  
$^{*}$  Projected in the field of NGC\,5617 and Trumpler\,22 (Fig.~\ref{Fig:skymap_members_binaries_part1}). \\                                                                                                                                                                                                                                                                                                                                                                                                                  
$^{**}$ Not present in the list of \cite{Piecka:2021}, but it is projected in the field of King\,16 and Dias\,1.\\
$^{\dag}$ $r_{h}$ for NGC\,2384 was derived directly from the list of members, by taking their individual masses and distances to the cluster centre, under the assumption that light traces mass. Its overall size ($\sim36'$) was taken from the literature (CG20; see Section~\ref{sec:particular_procedures_N2383_N2384} for details). Its limiting radius ($R_{\textrm{lim}}=6.60\pm0.90\,$arcmin) was assumed as a rough upper limit for its \textit{core} radius, since $R_{\textrm{lim}}$ delimits the region with higher stellar density contrast compared to the background, comprising the central part of its extended structure (see Section~\ref{sec:particular_procedures_N2383_N2384}). \\         
$^{\dag\dag}$ The numbers after the ``$\pm$"\,signal correspond to the intrinsic (i.e., corrected for measurement uncertainties) dispersions derived from the member stars data. \\

\end{minipage}
\end{sideways}
\label{tab:investig_sample}
\end{table*}

\begin{table*}
\centering
\rotcaption{ Physical parameters for each binary system: separation between centres, relative and escape velocities, total masses and estimated Roche radii. }
\begin{sideways}
\begin{minipage}{240mm}

\begin{tabular}{cllccccrrr}
 
 Pair\,\#   & Cluster 1       & Cluster 2       & $\Delta\,r$        &  $v_{rel}^{\dag}$   & $v_{esc}^{\dag\dag}$                      & $M_{clu1}$                     & $M_{clu2}$                    & $R_{roc}^{clu1}$    & $R_{roc}^{clu2}$  \\   
            &                 &                 &     (pc)           &  (km.s$^{-1}$)      & (km.s$^{-1}$)                             & ($\times10^3\,M_{\odot}$)      & ($\times10^3\,M_{\odot}$)     &      (pc)           & (pc)              \\                                   
\hline                                                                                                                                                                                                                                                          
                                                                                                                                                                                                           
 1$^{*}$    & NGC\,5617       & Trumpler\,22    &  14.6\,$\pm$\,1.4  &  6.1\,$\pm$\,5.2    & 1.5\,$\pm$\,0.1                           & 2.42\,$\pm$\,0.08              & 1.18\,$\pm$\,0.05             & 8.6\,$\pm$\,0.8     &  6.4\,$\pm$\,0.6  \\
 2          & NGC\,659        & NGC\,663        &  27.3\,$\pm$\,3.6  &  3.5\,$\pm$\,2.6    & 1.7\,$\pm$\,0.1                           & 1.71\,$\pm$\,0.09              & 7.7 \,$\pm$\,0.2              & 9.9\,$\pm$\,1.2     & 18.3\,$\pm$\,2.2  \\
 3          & Collinder\,394  & NGC\,6716       &  8.0\,$\pm$\,0.8   &  0.5\,$\pm$\,2.7    & 0.8\,$\pm$\,0.1                           & 0.40\,$\pm$\,0.04              & 0.18\,$\pm$\,0.03             & 4.8\,$\pm$\,0.5     &  3.5\,$\pm$\,0.3  \\
 4          & NGC\,2383       & NGC\,2384       &  6.5\,$\pm$\,0.7   &  18.2\,$\pm$\,5.0   & 2.2\,$\pm$\,0.1                           & 1.10\,$\pm$\,0.06              & 2.5 \,$\pm$\,0.1              & 2.9\,$\pm$\,0.3     &  4.0\,$\pm$\,0.4  \\
 5$^{**}$   & King\,16        & Berkeley\,4     & 11.4\,$\pm$\,0.6   &  7.9 \,$\pm$\,2.4   & 1.02\,$\pm$\,0.04                         & 0.62\,$\pm$\,0.05              & 0.75\,$\pm$\,0.06             & 5.8\,$\pm$\,0.3     &  6.2\,$\pm$\,0.3  \\
 6          & Czernik\,20     & NGC\,1857       &                    &                     &                  \multicolumn{3}{c}{Not a physical pair (Section~\ref{sec:results})}                                                   &                     &                   \\
 7          & Ruprecht\,100   & Ruprecht\,101   & 25.7\,$\pm$\,2.5   &  17.5\,$\pm$\,15.2  & 1.0\,$\pm$\,0.1                           & 0.79\,$\pm$\,0.05              & 2.5\,$\pm$\,0.1               & 8.9\,$\pm$\,0.6     &  14.2\,$\pm$\,0.8 \\

\hline

\multicolumn{10}{l}{ \textit{Note 1}: The Roche radius $R_{clu}$ for each cluster was estimated from eq.~\ref{eq:numerical_solution_R_roche_maintext}. }\\ 

\multicolumn{10}{l}{ \textit{Note 2}: The separation $\Delta r$ between the clusters' centre has been determined from their angular separation and taking the mean} \\

\multicolumn{10}{l}{of the two cluster distances.} \\

\multicolumn{10}{l}{$^{\dag}$ The uncertainties have been obtained from propagating the cluster distance errors and also the dispersion of proper motion } \\
\multicolumn{10}{l}{ components (Table~\ref{tab:investig_sample}).} \\

\multicolumn{10}{l}{$^{\dag\dag}$ Section~\ref{sec:escape_velocity}. } \\

\multicolumn{10}{l}{$^{*}$ Pismis\,19\, seems located in the background and therefore not a member of this system (Section~\ref{sec:results}).} \\

\multicolumn{10}{l}{$^{**}$ Dias\,1 seems located in the foreground and therefore not a member of this system (Section~\ref{sec:results}). } \\

\end{tabular}

\end{minipage}
\end{sideways}
\label{tab:params_binarios}
\end{table*}

Photometric ($G$, $G_{BP}$, $G_{RP}-$bands) and astrometric ($\varpi, \mu_{\alpha}\,\textrm{cos}\,\delta$, $\mu_{\delta}$) data were extracted from the \textit{Gaia} EDR3 catalogue by means of the Gaia@AIP\footnote[2]{https://gaia.aip.de/query/} service. For each investigated cluster, we adopted an extraction radius of $\sim2\degr$ centered on the cluster's coordinates, as given in the SIMBAD database \citep{Wenger:2000}.

Following the recipes specified in the \textit{Gaia} EDR3 release papers\footnote[3]{https://www.cosmos.esa.int/web/gaia/edr3-papers}, we employed the available source codes\footnote[4]{https://www.cosmos.esa.int/web/gaia/edr3-code} in order to correct: ($i$) our clusters' data for parallax zero-point biases \citep{Lindegren:2021}, ($ii$) the $G-$band magnitudes for sources with 6-parameter astrometric solutions \citep{Riello:2021} and ($iii$) the $G_{BP}/G_{RP}$ flux excess factors (\citeauthor{Riello:2021}\,\,\citeyear{Riello:2021}). After this step, to ensure the best quality of our data, we removed sources with spurious astrometric and/or photometric solutions by keeping those for which ($i$) the Renormalised Unit Weight Error (RUWE) astrometric parameter is smaller than 1.4 \citep{Lindegren:2021a} and ($ii$) $|C^{*}|<5\sigma_{C^*}$, where $\sigma_{C^*}$ is given by eq. 18 of \cite{Riello:2021} and $C^*$ is the corrected flux excess factor. We also restricted our analysis to sources with $G\leq19\,$mag, in order to avoid incompleteness for magnitudes fainter than this limit \citep{Fabricius:2021}.

\begin{figure*}
\begin{center}

\parbox[c]{1.00\textwidth}
  {
   \begin{center}
        \includegraphics[width=0.352\textwidth]{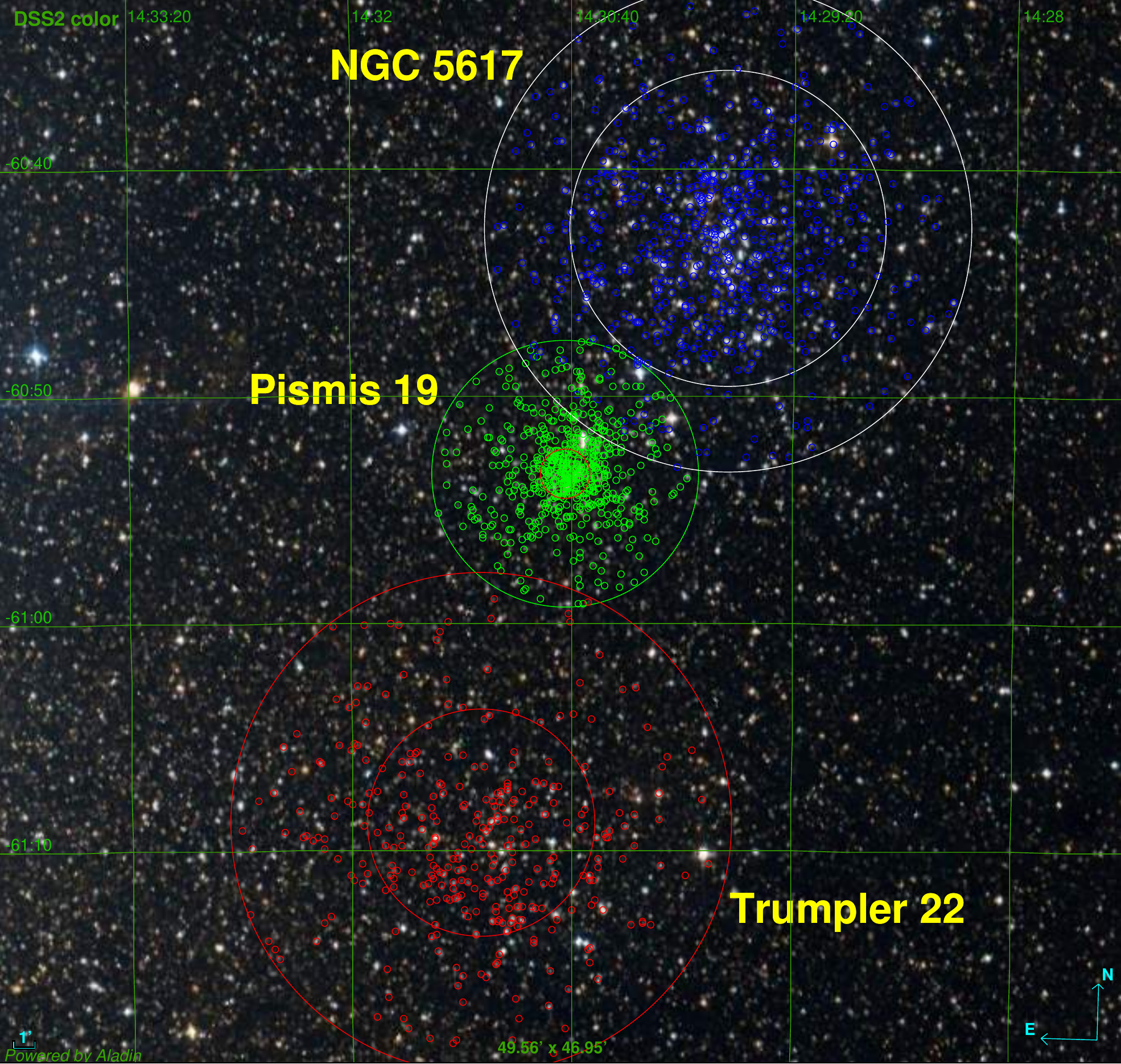}  
        \includegraphics[width=0.330\textwidth]{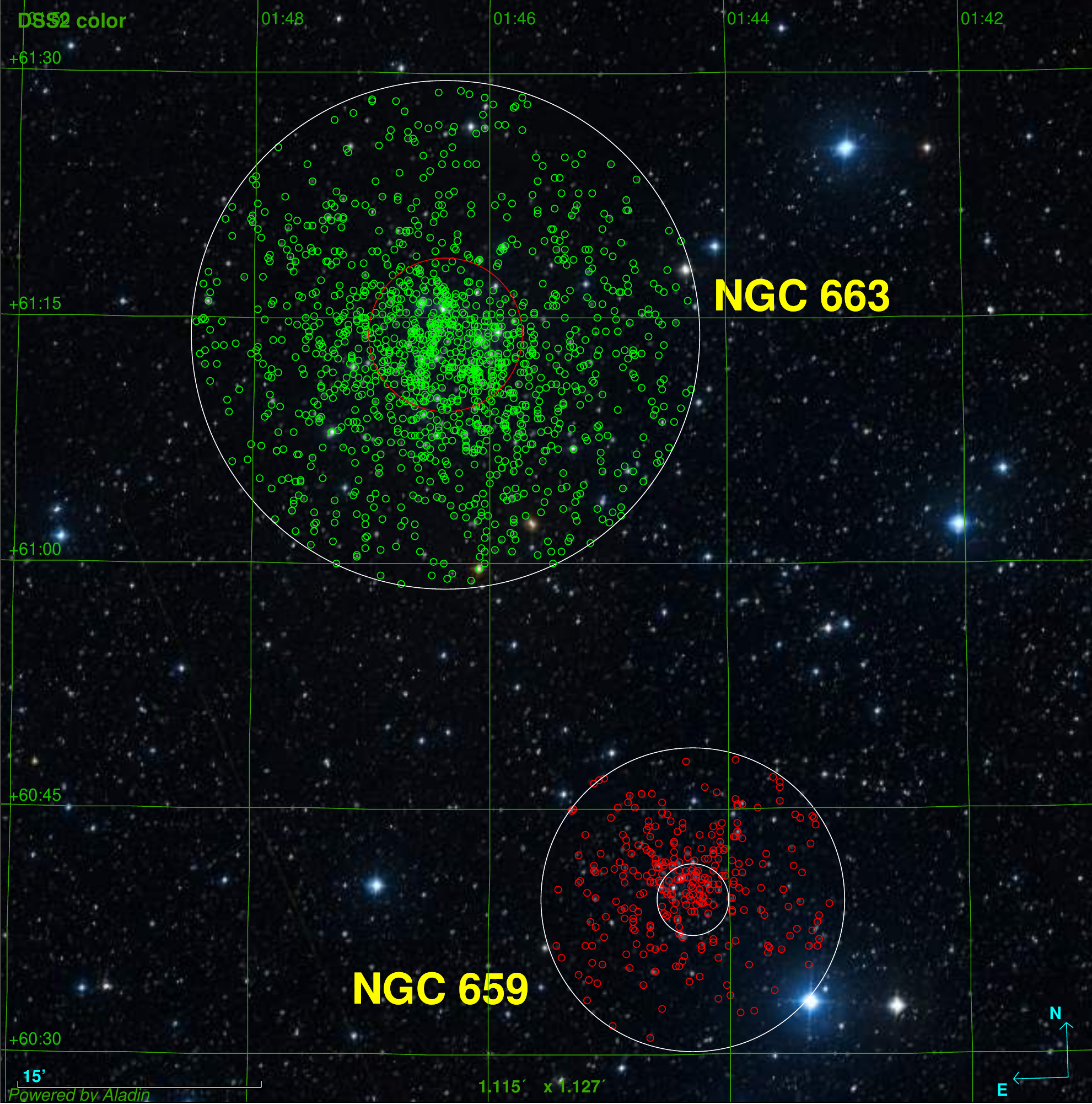}     
        \includegraphics[width=0.342\textwidth]{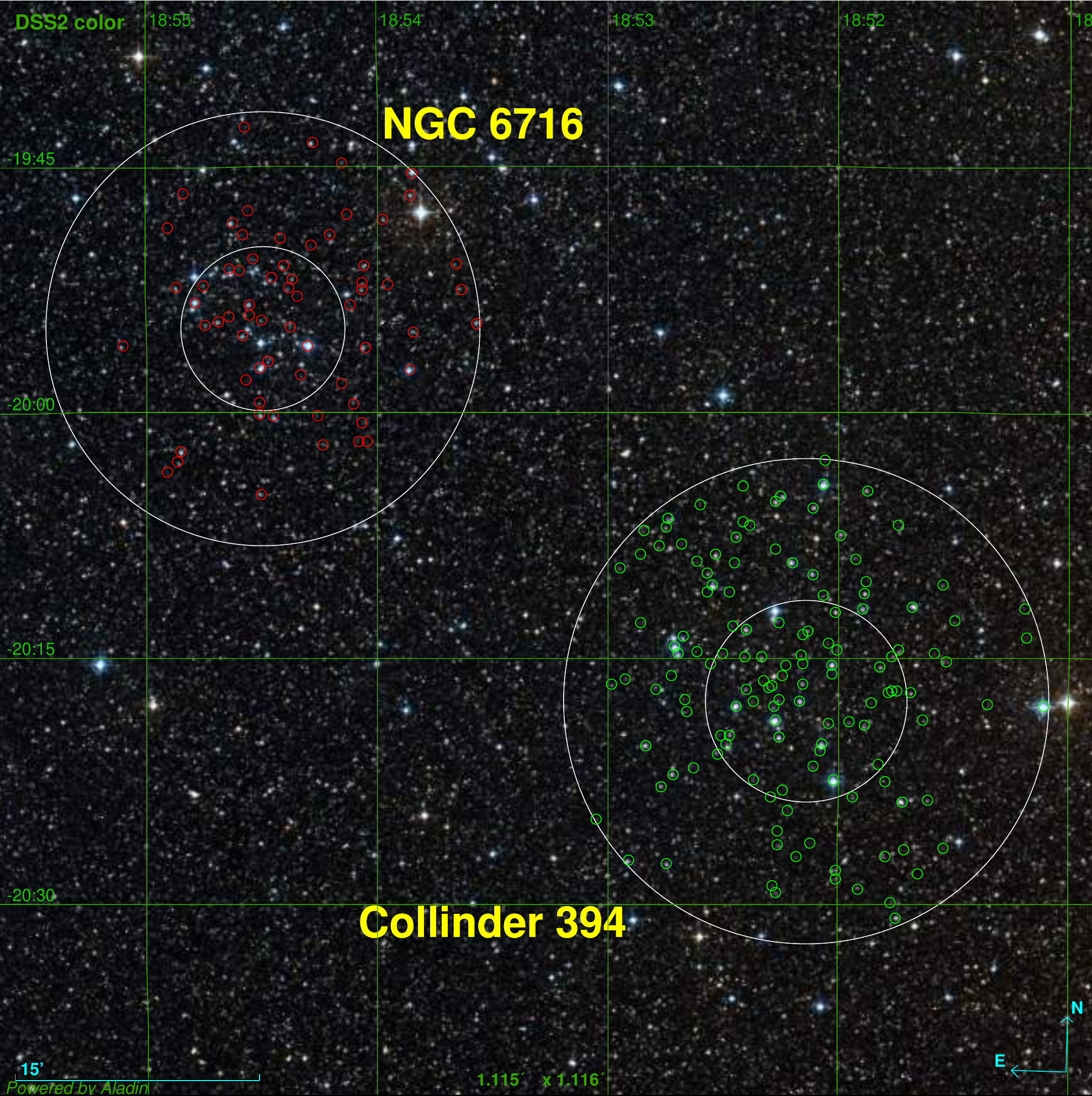} 
        \includegraphics[width=0.336\textwidth]{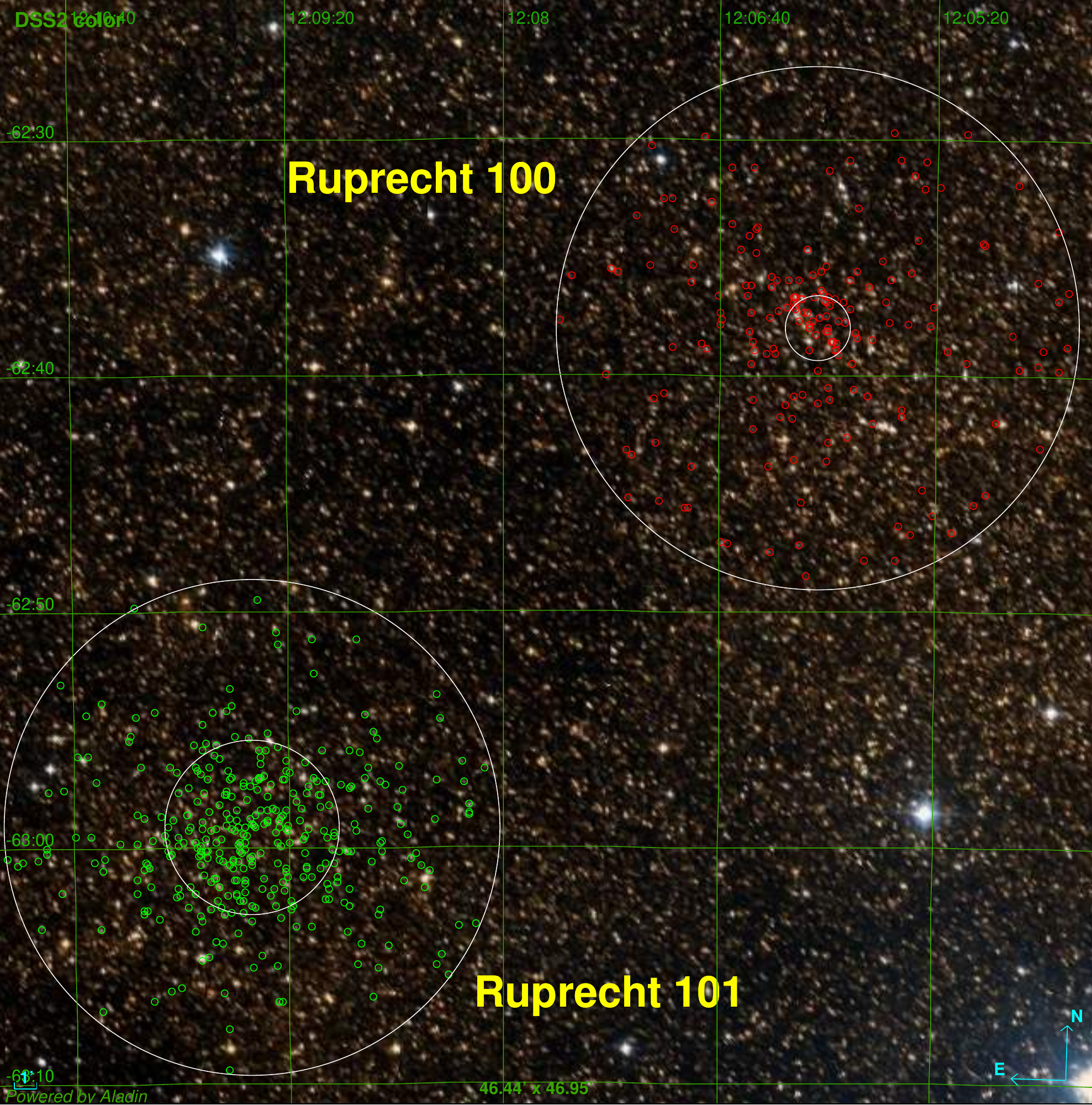} 
    \end{center}    
  }
\caption{ DSS2 images for 4 of the studied cluster pairs with member stars encircled. From top left to bottom right panel:  NGC\,5617 (blue)$-$Trumpler\,22 (red; image size: 50$\arcmin\times47\arcmin$); the green circles identify Pismis\,19's member stars, projected in the same area, but not in close proximity with the other two; see Section~\ref{sec:results}); NGC\,659 (red)$-$NGC\,663 (green; image size: 1$\degr\times1\degr$), Collinder\,394 (green)$-$NGC\,6716 (red; image size: 1$\degr\times1\degr$), Ruprecht\,100 (red)$-$Ruprecht\,101 (green; image size: 46$\arcmin\times47\arcmin$). In all panels, the larger circles represent the clusters' core and tidal radii (see Fig.~\ref{fig:RDPs_part1} and Table~\ref{tab:investig_sample}). North is up and east to the left. }

\label{Fig:skymap_members_binaries_part1}
\end{center}
\end{figure*}

Fig.~\ref{Fig:skymap_members_binaries_part1} shows 9 of the 16 investigated OCs, distributed as 4 stellar aggregates. Member stars (Section~\ref{sec:memberships}) are highlighted in each case. Other 7 OCs are shown in Appendix~\ref{sec:supplementary_figures}. Henceforth, the same strategy will be employed for other figures. Table~\ref{tab:investig_sample} shows the astrophysical parameters of each OC and the physical parameters of each binary system are shown in Table~\ref{tab:params_binarios} (see Sections~\ref{sec:struct_params}, \ref{sec:fundamental_params_determ} and \ref{sec:dyn_properties}). Complementarly, we searched for radial velocity data ($V_r$) for the member stars in the \textit{Gaia} EDR3 catalogue and other sources, as shown in Table~\ref{tab:Vrads}.

\section{Method}
\label{sec:method}
\subsection{Structural analysis}
\label{sec:struct_params}

After applying the photometric and astrometric quality filters to our data, we performed the structural analysis by  (I) applying a proper motions filter to the cluster data, in order to remove the excess of field contamination, (II) determining the central coordinates and structural parameters by means of \cite{King:1962} model fitting of the cluster's radial density profile (RDP). These procedures are described below.

\subsubsection*{I: Proper motions filtering}
\label{sec:prop_motions_filtering}

\begin{figure*}
\begin{center}

\parbox[c]{1.00\textwidth}
  {
   \begin{center}
       \includegraphics[width=0.280\textwidth]{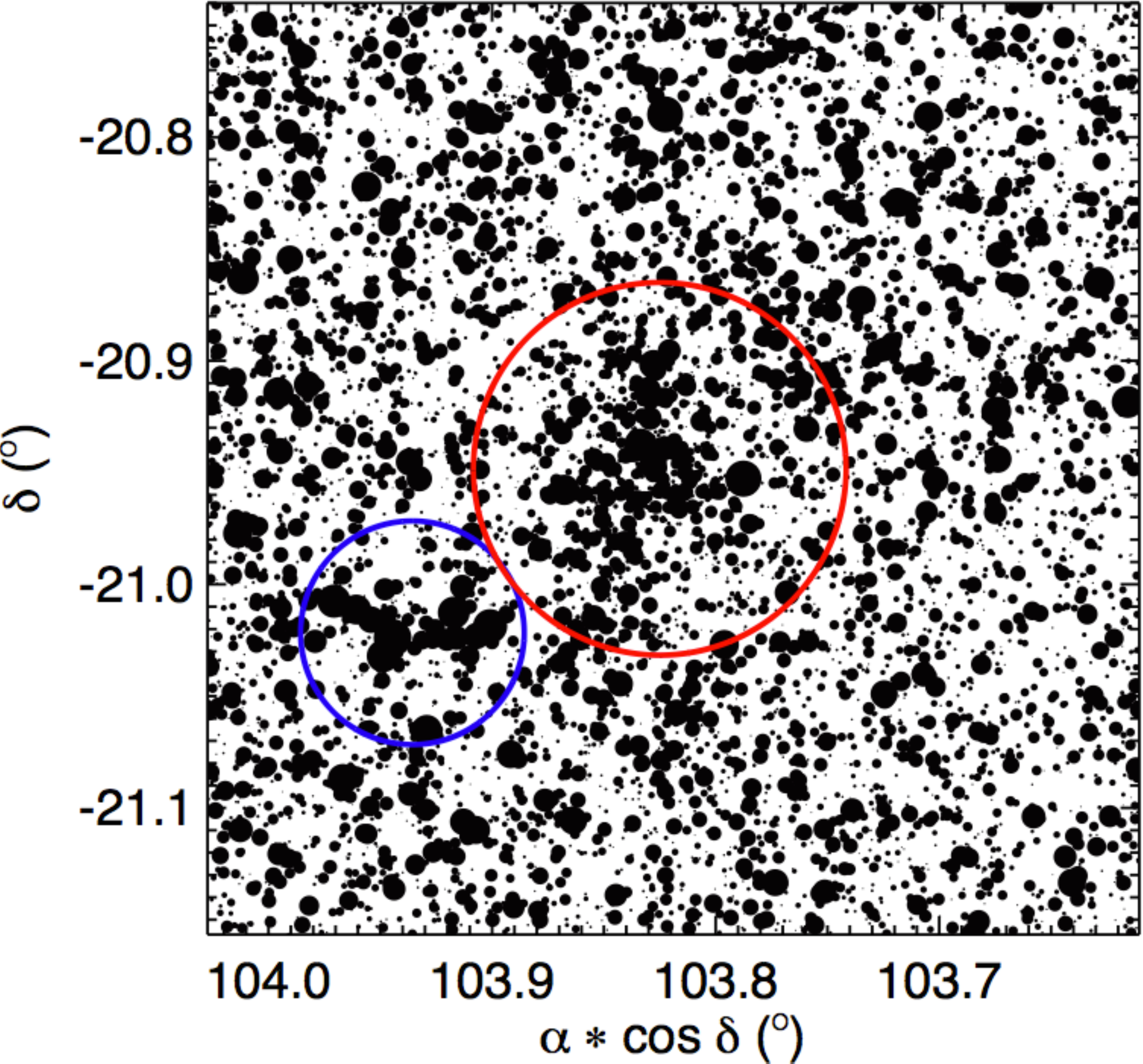}   
       \includegraphics[width=0.262\textwidth]{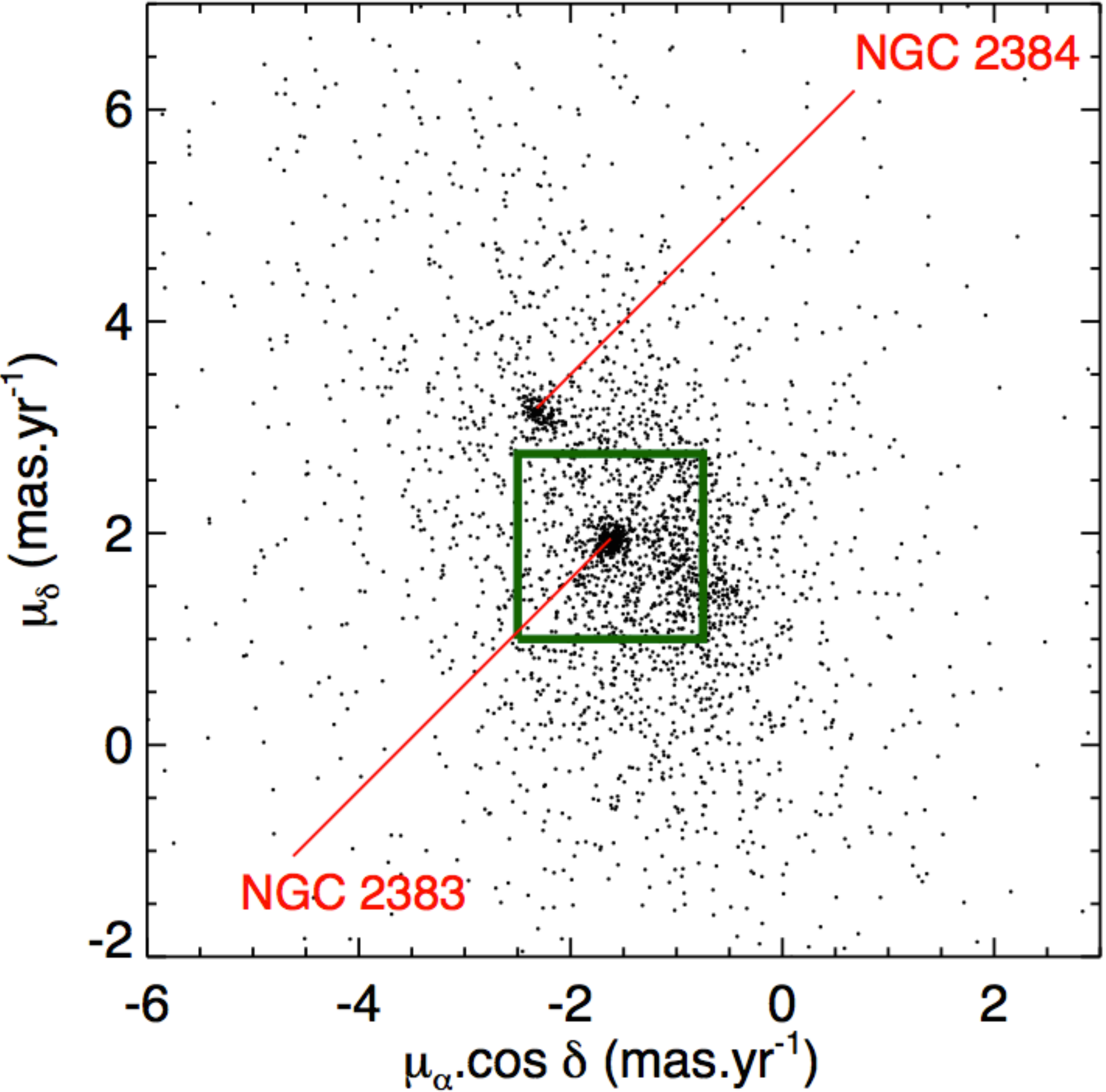}  
       \includegraphics[width=0.280\textwidth]{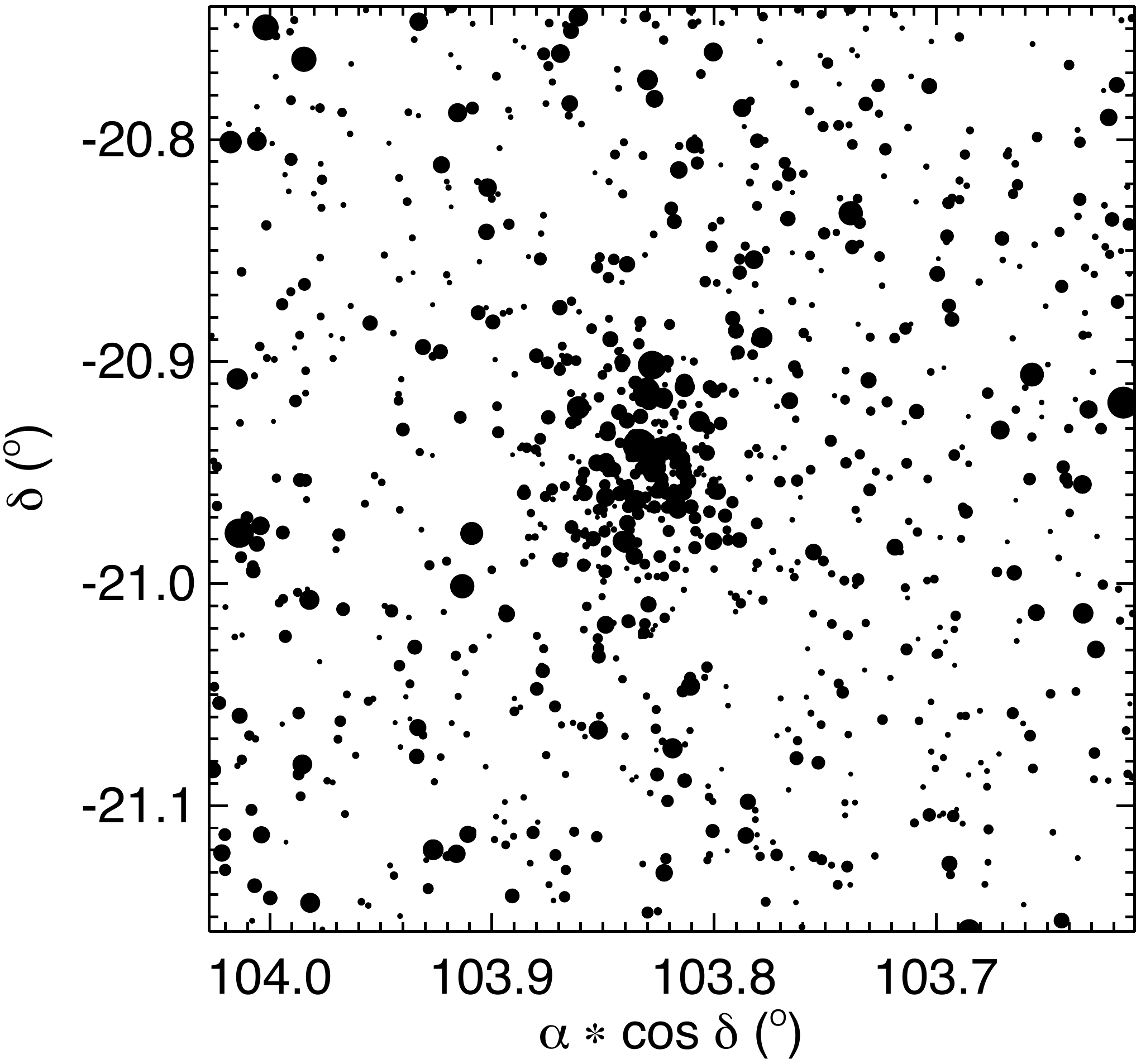}   
    \end{center}    
  }
\caption{ Left panel: skymap ($25\arcmin\times25\arcmin$) for stars centered on NGC\,2383's coordinates. The red and blue circles (radius $5\arcmin$ and $3\arcmin$, respectively) delimit the inner central structures of NGC\,2383 and NGC\,2384, respectively. Middle panel: VPD for this sample of stars. The two identified concentrations are defined mainly by member stars of each OC. The green square is the proper motions filter. Right panel: same as left panel, but after applying the proper motions filter. }

\label{fig:ilustra_preanalise_NGC2383}
\end{center}
\end{figure*}

We firstly looked for the signature of each cluster in its vector-point diagram (VPD). This is illustrated in Fig.~\ref{fig:ilustra_preanalise_NGC2383} for the OC NGC\,2383. The left panel shows a skymap ($25\arcmin\times25\arcmin$; symbol sizes are proportional to the stars' brightnesses in the $G-$band) centered on NGC\,2383's coordinates. The red circle marks its visual radius. The blue circle marks the location of the companion cluster NGC\,2384. The central panel in this figure shows the VPD for this sample of stars, where it is noticeable the presence of 2 proeminent agglomerations, defined mainly by member stars of both OCs. We then applied a proper motion filter, keeping those stars located within the green square (with side $\sim2$mas.yr$^{-1}$), in order to eliminate most of the contamination by field stars and also most of NGC\,2384's members.

For all investigated OCs, after setting memberships (Section~\ref{sec:memberships}), we note that the size of the proper motions filter is at least 1 order of magnitude larger than the dispersions in $\mu_{\alpha}\,\textrm{cos}\,\delta$ and $\mu_{\delta}$ (Table~\ref{tab:investig_sample}) for the member stars. The right panel in Fig.~\ref{fig:ilustra_preanalise_NGC2383} shows the skymap of NGC\,2383 after applying the proper motions filter. Analogous procedure has been employed for our complete sample. 

\subsubsection*{II: Determining central coordinates and structural parameters}

For each OC, we built a uniform grid of coordinates encompassing the central part of the filtered skymap (right panel of Fig.~\ref{fig:ilustra_preanalise_NGC2383}). Typically $\sim$200 ($\alpha, \delta$) pairs were employed in this task, with an even spacing of $\sim0.5\arcmin$. For each ($\alpha, \delta$) pair, a RDP is built by counting the number ($N_*$) of stars within concentric rings and dividing this number by the ring area, that is, $\sigma(\bar{r})=N_*/[\pi(r_{k+1}^2-r_{k}^2)]$, where $\bar{r}=(r_{k+1}+r_{k})/2$. 

At this stage, different ring widths were employed and the corresponding densities were overplotted in the same RDP. The background density ($\sigma_{\textrm{bg}}$) was determined from the average of densities located beyond the limiting radius ($R_{\textrm{lim}}$), defined as the distance from the cluster centre beyond which the stellar densities are almost constant. The loci of points in the background-subtracted RDP was then fitted via $\chi^2$ minimization using the \cite{King:1962} model defined by

\begin{equation}
   \sigma(r)\,\propto\,\left(\frac{1}{\sqrt{1+(r/r_c)^2}} - \frac{1}{\sqrt{1+(r_t/r_c)^2}}\right)^2
   \label{eq:King_profile}
\end{equation}

\noindent
where $r_c$ is the \textit{core radius}, which is a length scale of the cluster's  central structure, and $r_t$ is the \textit{tidal radius}, which is the truncation radius of the King profile, thus providing an empirical length scale of the cluster's overall size. 

\begin{figure*}
\begin{center}

\parbox[c]{1.0\textwidth}
  {
   \begin{center}
    \includegraphics[width=0.60\textwidth]{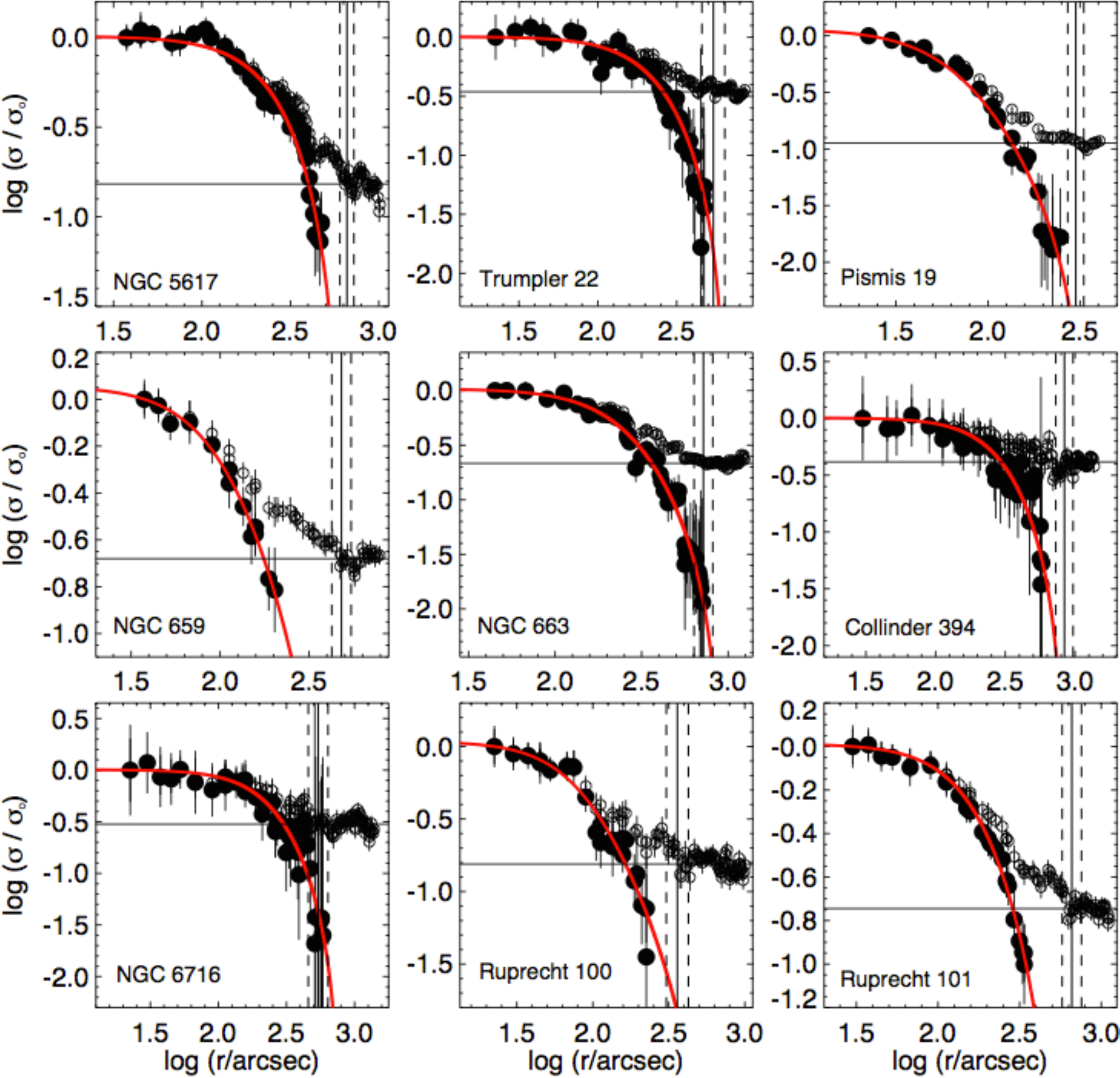} 
    \end{center}
    
  }
\caption{ The filled (open) circles represent the background-subtracted (non-background-subtracted) RDPs for 9 investigated OCs. In all cases, the profiles have been normalized to unity at the innermost radial bins. The red lines represent the fitted King\,\,(\citeyear{King:1962}) profile. The horizontal continuous line represents $\sigma_{\textrm{bg}}$ and the vertical one indicates $R_{\textrm{lim}}$ (uncertainties in $R_{\textrm{lim}}$ are indicated by dashed lines). Poissonian error bars are overplotted. For some OCs (e.g., NGC\,5617, NGC\,659, Ruprecht\,100 and Ruprecht\,101), the King profile fit was truncated at inner radial bins (see text for details). }

\label{fig:RDPs_part1}
\end{center}
\end{figure*}

The adopted central coordinates (Table~\ref{tab:investig_sample}) correspond to the ($\alpha, \delta$) pair that resulted, at the same time, in the largest central density with the minimal residuals during the King profile fit. Part of the investigated OCs (e.g., NGC 5617, NGC 659, Ruprecht 100 and Ruprecht 101) present significant fluctuations in the background-subtracted RDPs (filled symbols in Fig.~\ref{fig:RDPs_part1}) in their outermost parts (i.e., at $r\sim R_{\textrm{lim}}$), therefore the King profile fit was truncated at inner radial bins, for better convergence.

The results of this procedure (which has also been employed in Angelo et al.\,\citeyear{Angelo:2020}, \citeyear{Angelo:2021}) are shown in Fig.~\ref{fig:RDPs_part1}, where the RDPs have been normalized to unity at the innermost radial bins and the fitted King profile is represented by red lines. The error bars come from Poisson statistics. We estimated the projected half-light radius ($r_{hp}$) of each OC from the fitted $r_c$ and $r_t$ using eq.\,9 of \cite{Santos:2020}. Then we obtained the three-dimensional half-light value from the relation $r_h=1.33\,r_{hp}$ \citep{Baumgardt:2010}. The structural parameters (converted to the physical scale) are shown in Table~\ref{tab:investig_sample}.

In order to check the robustness of the derived structural parameters, we rederived them this time employing relaxed quality filters (RUWE $<$ 4.2 and $|C^{*}|$ $<$ 10\,$\sigma_{C^*}$; see above Section~\ref{sec:sample}) on our data. After reconstructing the OCs' RDPs and performing again the King profile fits, we found differences in $r_t$ and $r_c$ that are well within the uncertainties informed in Table~\ref{tab:investig_sample}.

\subsection{Membership assignment}
\label{sec:memberships}

At this stage of the analysis, our sample was restricted to those stars selected by the proper motions filter and located within the cluster tidal radius. For each OC, we also selected photometric and astrometric data for stars in a concentric annular control field, whose area is $\sim3\times$ the cluster area and its internal radius is large enough to not intersect any parts of the companion cluster. The same filtering process was applied to the field stars. 

Then we executed a decontamination algorithm (detailed in \citeauthor{Angelo:2019a}\,\,\citeyear{Angelo:2019a}) that sweeps the 3-dimensional parameters space ($\varpi, \mu_{\alpha}\,\textrm{cos}\,\delta, \mu_{\delta}$) looking for statistically significant overdensities of stars in the cluster direction that are more concentrated than the field stars' astrometric data. The main assumption of the method is that member stars share common parallaxes and proper motions, being more tightly distributed than field stars. 

Briefly, the method consists in 4 steps: (i) firstly, the 3D astrometric space is divided in a grid of cells with widths proportional to the mean sample (cluster+field) uncertainties in proper motions and parallax; (ii) then, for each star present in the cluster sample within a given cell, a membership likelihood $l_{\textrm{star}}$ (eq. 1 of \citeauthor{Angelo:2019a}\,\,\citeyear{Angelo:2019a}) is computed, taking into account the astrometric parameters uncertainties, their intrinsic dispersion and the correlation coefficients; the same procedure is applied for field stars and both sets of $l_{\textrm{star}}$ values are statistically compared by means of \textit{entropy-like functions} ($S_{\textrm{cluster}}$ and $S_{\textrm{field}}$; eq. 3 of \citeauthor{Angelo:2019a}\,\,\citeyear{Angelo:2019a}); (iii) for those cluster stars within cells where $S_{\textrm{cluster}}<S_{\textrm{field}}$, an exponential factor is evaluated (eq. 4 of \citeauthor{Angelo:2019a}\,\,\citeyear{Angelo:2019a}) considering the local overdensity of stars compared to the whole set of cells; (iv) finally, cell sizes are then increased and decreased by one-third of their mean sizes and the procedure is repeated. The final membership likelihoods correspond to the median of values for the whole set of grid configurations. 

This procedure ensures that appreciable membership likelihoods are assigned only to groups of cluster stars that define significant overdensities in the astrometric space and statistically distinguishable from the field sample. Only decontaminated samples are employed in the subsequent parts of our analysis.

\subsection{Fundamental parameters determination}
\label{sec:fundamental_params_determ}

\begin{figure*}
\begin{center}

\parbox[c]{0.65\textwidth}
  {
   \begin{center}
     \includegraphics[width=0.65\textwidth]{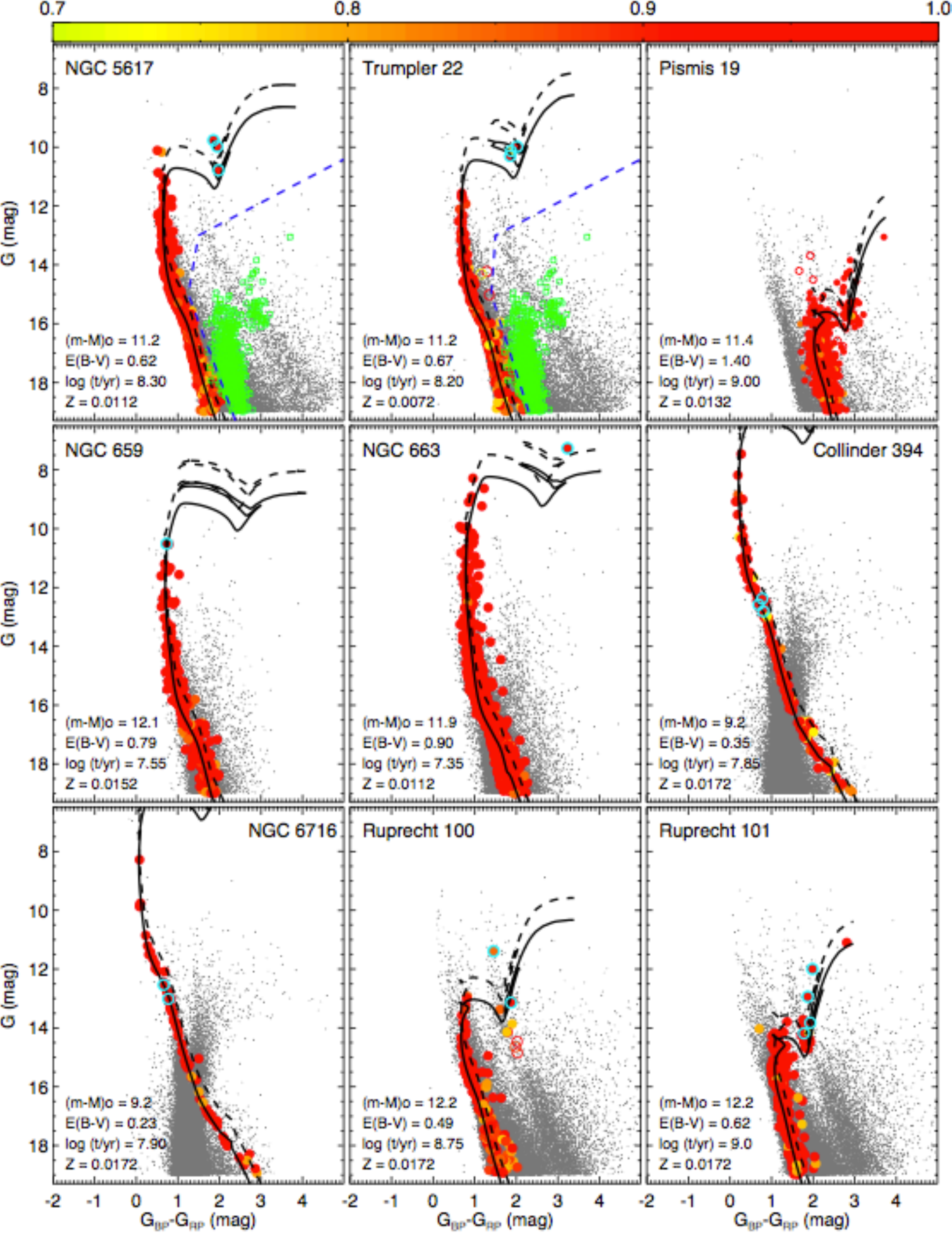}    
    \end{center}
    
  }
\caption{ Decontaminated CMDs for 9 investigated OCs. The continous lines represent the best-fitted PARSEC isochrones and the black dashed ones are the locus of unresolved binaries with equal-mass components. In all panels, member (non-member) stars are represented by filled (open) circles, with colours attributed according to the membership likelihood scale, as indicated in the colourbar. Small gray dots are stars in a control field. The fundamental astrophysical parameters are indicated. Stars marked with turquoise circles present radial velocity available in \textit{Gaia} EDR3 (Table~\ref{tab:Vrads}). The blue dashed lines in the CMDs of NGC\,5617 and Trumpler\,22 are colour filters employed to remove the contamination by Pismis\,19's member stars (overplotted with small open squares, for better visualization). Pismis\,19's CMD is also shown, for completeness. }

\label{fig:decontam_CMDs_part1}
\end{center}
\end{figure*}

The outcomes of the method outlined above have been employed to build decontaminated colour-magnitude diagrams (Fig.~\ref{fig:decontam_CMDs_part1}). Stars with high membership likelihoods ($\gtrsim0.7$) define recognizable evolutionary sequences, from which we determined the fundamental astrophysical parameters (log\,($t.$yr$^{-1}$), $(m-M)_0$, $E(B-V)$ and overall metallicity $Z$) via isochrone fitting.


In each case, we superimposed PARSEC isochrones \citep{Bressan:2012} in the \textit{Gaia} EDR3 photometric system by carefully inspecting the matching of some key evolutionary regions in the CMD, like the main sequence morphology, the turnoff and the red giant branch. Initial guesses for the astrophysical parameters were obtained from \cite{Dias:2002}, \citeauthor{Dias:2021a}\,\,(\citeyear{Dias:2021a}, hereafter DMML21), and CG20. Keeping $E(B-V)$ fixed, we vertically shifted the isochrone to match the cluster main sequence, to derive $(m-M)_0$. The extinction relations from \cite{Cardelli:1989} and \cite{ODonnell:1994} have been employed, with $R_V=3.1$.

For the cluster age determination, the turnoff was a fundamental constraint. Further refinements were then obtained by varying the reddening and also the metallicity $Z$ (converted to $[Fe/H]$ via the relation $[Fe/H]=$log$(Z/Z_{\odot})$). In this last step, the relative distance between the cluster turnoff and the red clump (if present) is a useful constraint. The parameters uncertainties were evaluated by shifting the best-matched isochrone to get a maximum deviation from the central solution that still encompasses the data.

\subsection{Particular procedures for some pairs}
\subsubsection*{ NGC\,5617$-$Trumpler\,22}

The OC Pismis\,19 is projected in the same area of the pair NGC\,5617$-$Trumpler\,22 (Fig.~\ref{Fig:skymap_members_binaries_part1}) and presents similar proper motions (Table~\ref{tab:investig_sample}). Therefore a specific data filtering procedure was necessary in this case. We took advantage from the fact that Pismis\,19 presents considerably redder sequences in its CMD (Fig.~\ref{fig:decontam_CMDs_part1}), which allowed us to employ colour filters to remove the contamination by its member stars prior to the structural analysis of NGC\,5617 and Trumpler\,22. The detailed procedure is presented in Appendix~\ref{sec:N5617_Tr22_Pis19}.

\subsubsection*{Czernik\,20$-$NGC\,1857}
Besides presenting small angular separation in the sky, the OCs Czernik\,20 and NGC\,1857 are also close to each other in the astrometric space (see Section~\ref{sec:results}). Therefore, the use of proper motion filters (Section~\ref{sec:prop_motions_filtering}) was not enough to eliminate the contamination by the companion and special care had to be taken for this pair during the RDPs construction procedure,  since contamination by a nearby cluster can affect star counts within each radial bin. We employed an iterative process following the steps described below:

\begin{itemize}
  
  \item I: We applied our decontamination algorithm (Section~\ref{sec:memberships}) to stars in Czernik\,20 area within a visual radius of 5$\arcmin$, centered on the literature coordinates ($\rmn{RA}$=05:20:31, $\rmn{DEC}$=+39:32:42; CG20). This inner limit defines the more contrasting part of the cluster against the field population, therefore most of Czernik\,20's member stars are expected to be found within this circular area. We then built a preliminary decontaminated CMD, from which we determined a first guess to the fundamental parameters and identified member stars, after isochrone fitting (Sections~\ref{sec:memberships} and \ref{sec:fundamental_params_determ});
  
  \item II: We then proceeded to the analysis of NGC\,1857. We took all stars in this OC area (after applying the quality and proper motions filter, as outlined in Sections~\ref{sec:sample} and ~\ref{sec:prop_motions_filtering}) and filtered out the members of Czernik\,20, as identified in step I above. The removal of most ot the contamination caused by Czernik\,20' stars resulted in a smoother RDP, from which we determined the first guess for the central coordinates and structural parameters of NGC\,1857. From its decontaminated CMD, we also obtained initial values for its fundamental parameters and a first list of member stars;
  
  \item III: We removed the set of member stars of NGC\,1857, as identified in the previous step, from Czernik\,20' skymap. This filtered skymap was used for the complete structural analysis of Czernik\,20 (central coordinates and structural parameters determination) and a new set of member stars was obtained;
  
  \item IV: We then reanalysed (structural part, decontamination and isochrone fit) the OC NGC\,1857, after the removal of Czernik\,20's member stars identified in the step above.  
        
\end{itemize}

After that, steps III and IV are repeated until no further changes are verified in the structural parameters and member lists. The convergence of this procedure was obtained after 3 iterations for each OC. Obviously, this procedure is not completely free of some possible residuals (i.e., a star flagged as a member of a cluster might actually be a member of the companion OC), but in a statistical sense, both populations could be disentangled. This procedure could be improved with the use of radial velocities and metallicities data.

\subsubsection*{ NGC\,2383$-$NGC\,2384}
\label{sec:particular_procedures_N2383_N2384}
NGC\,2383 and NGC\,2384 are present in the catalogue of DMML21, but the latter is absent in CG20. Instead of NGC\,2384, they list the OC UBC\,224 (recently reported by \citeauthor{Castro-Ginard:2020}\,\,\citeyear{Castro-Ginard:2020} as a new Galactic OC) as the closest one to NGC\,2383. A preliminary analysis based on the list of members of CG20 revealed that UBC\,224 presents an extended structure, which was then confirmed from the outcomes of our analysis (Section~\ref{sec:indiv_comments_N2383_N2384}). UBC\,224 central coordinates are located $\sim10\arcmin$ Northeastwards of NGC\,2384's centre, as listed in DMML21. Furthermore, as NGC\,2384 bright stars were considered members of UBC\,224 by \cite{Castro-Ginard:2020} and CG20, we checked the possibility that UBC\,224 and NGC\,2384 are the same object.


We performed the structural analysis for this object using the coordinates listed in CG20 as an initial guess; no significant overdensity with respect to the field population was verified. By stepping through the grid of tentative central coordinates (Section~\ref{sec:struct_params}), a larger contrast was found by adopting the coordinates of Table~\ref{tab:investig_sample}, which are close to those informed by DMML21 for NGC\,2384. The object's RDP shows large fluctuations in stellar densities, precluding an adequate fit to the data. Therefore, we took the astrometric data for stars within the object limiting radius ($R_{\textrm{lim}}\sim6.6'$) and ran our decontamination algorithm in order to build a preliminary list of member stars.

A direct comparison between this preliminary list (centered on NGC\,2384) with that of CG20 (for UBC\,224)  revealed almost identical mean values for the astrometric parameters (within $\sim0.01\,$mas for the parallax and $\sim0.01\,$mas.yr$^{-1}$ for the proper motion components; see Section~\ref{sec:compara_results_lit}) and the same evolutionary sequences on its  decontaminated CMD. Additionally, we noted that most of the literature members lie outside the object's $R_{\textrm{lim}}$. In fact, the compact group of bright stars close to NGC\,2384 centre (Fig.~\ref{fig:ilustra_preanalise_NGC2383}) defines visually the spatial extension of its central part, presenting larger contrast with the field, but its extended overall structure demanded a larger decontamination radius for a proper a member stars determination. 

We identified member stars within a $\sim5\times R_{\textrm{lim}}$ (or $\sim36\arcmin$) search radius, which is compatible with the largest angular distance of CG20 member stars with respect to our redetermined centre for NGC\,2384 (Table~\ref{tab:investig_sample}). The use of decontamination radii larger than this limit resulted in an increasing amount of outliers (that is, stars with large membership likelihoods, but with photometric data incompatible with the clusters' evolutionary sequences on its CMD), due to the decreasing  cluster$-$field contrast in the astrometric space. At the end of this procedure, we confirmed that our member stars for NGC\,2384 and those of CG20 for UBC\,224 define almost the same loci of data (for $G\lesssim17\,$mag) in its CMD (Fig.~\ref{fig:decontam_CMDs_compCGaudin_part1} in Appendix~\ref{sec:supplementary_figures}), the same kinematics and compatible parallaxes (as detailed in Section~\ref{sec:compara_results_lit}), consistent with the hypothesis that we are in fact dealing with the same object. In fact, as already pointed out by \cite{Monteiro:2020}, some of the newly discovered `UBC'\,clusters by \cite{Castro-Ginard:2020} were already known OCs.

\section{Results}
\label{sec:results}

\begin{figure}
\begin{center}

\parbox[c]{0.48\textwidth}
  {

   \begin{center}
     \includegraphics[width=0.48\textwidth]{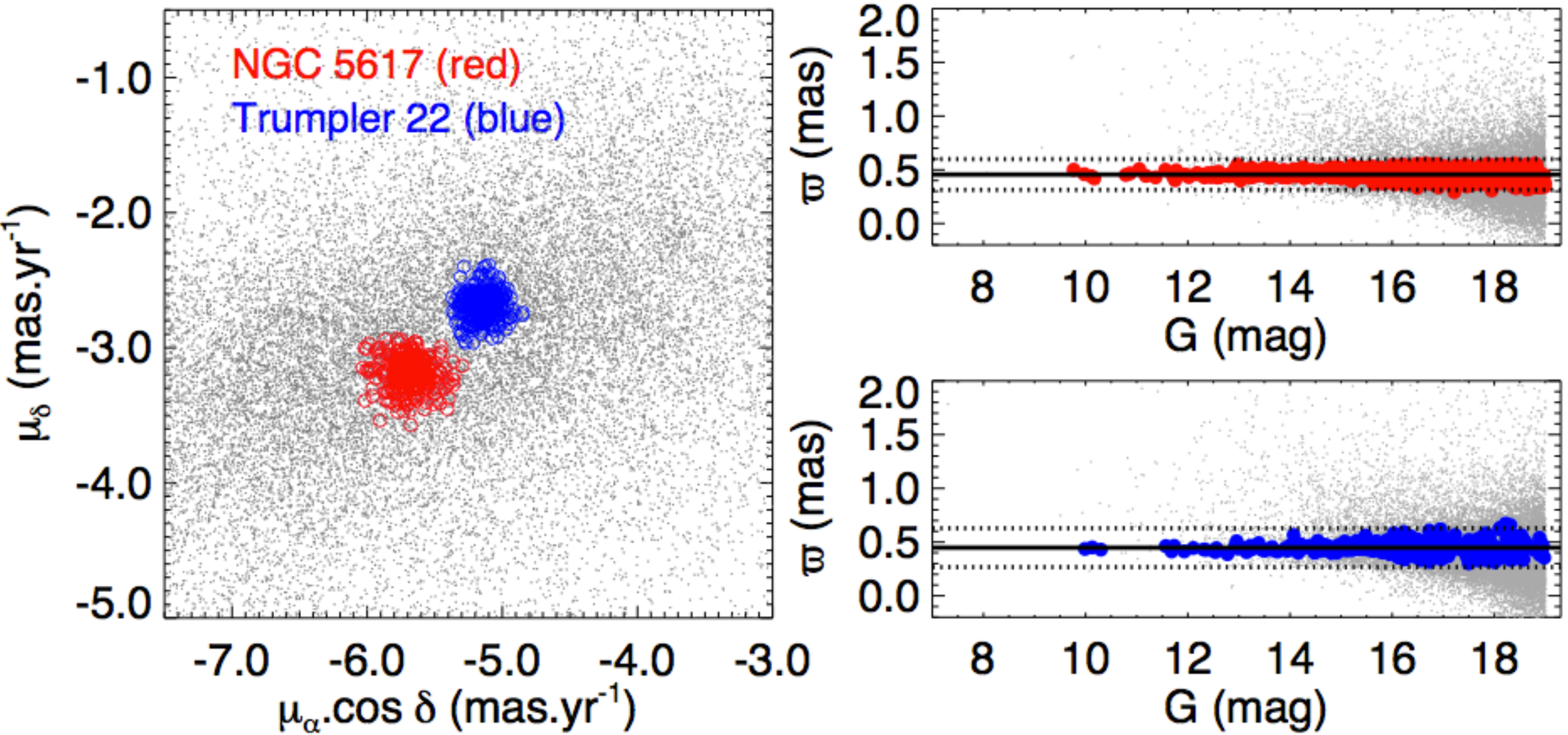}    
     
   \begin{center}
     \includegraphics[width=0.48\textwidth]{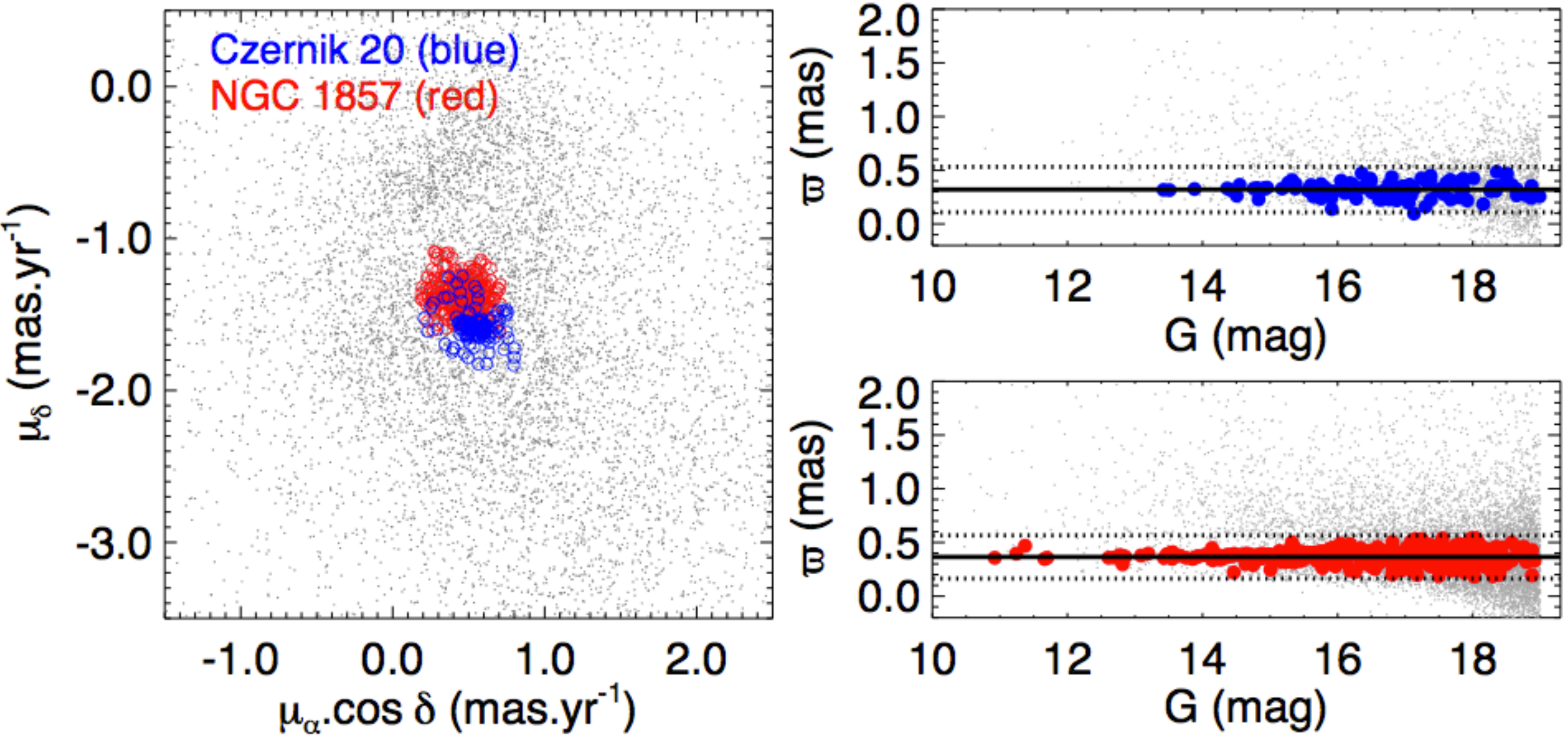}    
    \end{center}     
    \end{center}
    
  }
\caption{ VPDs (left column) and $\varpi\times G$ (right column) for member stars of the binaries NGC\,5617$-$Trumpler\,22 and Czernik\,20$-$NGC\,1857. Symbol colours identify each OC (see the legend). The horizontal continuous line in the rightmost plots represents the mean $\varpi$ value for member stars and the dotted ones represent the 3$\sigma$ limits. In all panels, small grey dots represent control field stars. }

\label{fig:VPD_plx_Gmag_examples}
\end{center}
\end{figure}

The derived parameters for the investigated OCs are presented in Table~\ref{tab:investig_sample}. Fig.~\ref{fig:VPD_plx_Gmag_examples} exhibits the VPDs and $\varpi\times G$ plots after applying our decontamination procedure for the pairs NGC\,5617$-$Trumpler\,22 and Czernik\,20$-$NGC\,1857, taken here as illustrative examples. A more detailed version of these figures (including membership likelihoods and also other investigated OCs) can be found in Appendices~\ref{sec:N5617_Tr22_Pis19} and \ref{sec:supplementary_figures}. Its noticeable that, besides establishing evolutionary sequences in the decontaminated CMDs (Fig.~\ref{fig:decontam_CMDs_part1}), the high membership stars ($\gtrsim70\%$) define conspicuous concentrations in the VPDs and also present similar parallaxes, as expected.

Considering the results in Table~\ref{tab:investig_sample}, the pairs of OCs that present compatible distance modulus and reddening (within uncertainties) are: NGC\,5617 $-$ Trumpler\,22, NGC\,659 $-$ NGC\,663, Collinder\,394 $-$ NGC\,6716, NGC\,2383 $-$ NGC\,2384, King\,16 $-$ Berkeley\,4 and Ruprecht\,100 $-$ Ruprecht\,101. In each case, the separation between both OCs is $<\,30\,$pc (Table~\ref{tab:params_binarios}), consistent with the maximum separation found in binary cluster candidates according to previous studies (e.g., MM09; \citeauthor{Subramaniam:1995}\,\,\citeyear{Subramaniam:1995}; see also Section~\ref{sec:comp_with_other_pairs_lit}). Therefore, these 6 listed pairs could constitute physically interacting systems.

Czernik\,20 and NGC\,1857 present very different values for $(m-M)_0$ (discrepancy of $\sim$0.70\,mag), which is consistent with the hypothesis of a chance alignment along the line of sight. \cite{Piecka:2021} listed King\,16$-$Dias\,1 among their catalogued stellar aggregates, but from the outcomes of our analysis, Berkeley\,4 (instead of Dias\,1) was found to form a physical pair with King\,16. Dias\,1 presents $(m-M)_0$ incompatible with King\,16 (discrepancy of $\sim$0.4\,mag) and only marginally compatible with Berkeley\,4; besides, Dias\,1 is submitted to a considerably larger interstellar reddening, while the other 2 two OCs present almost the same $E(B-V)$. That is an analogous situation for the case of Pismis\,19, which is a probable background cluster subject to a larger interstellar reddening in comparison with NGC\,5617$-$Trumpler\,22.  

In what follows (Section~\ref{sec:dyn_properties}), we determined dynamical parameters for the OCs sample and, for the probable interacting pairs, we analysed their physical status (Section~\ref{sec:discussion}).

\subsection{Dynamical properties}
\label{sec:dyn_properties}
\subsubsection{Total mass from mass functions}
\label{sec:mass_functions}

\begin{figure}
\begin{center}

\parbox[c]{0.45\textwidth}
  {
   \begin{center}
     \includegraphics[width=0.45\textwidth]{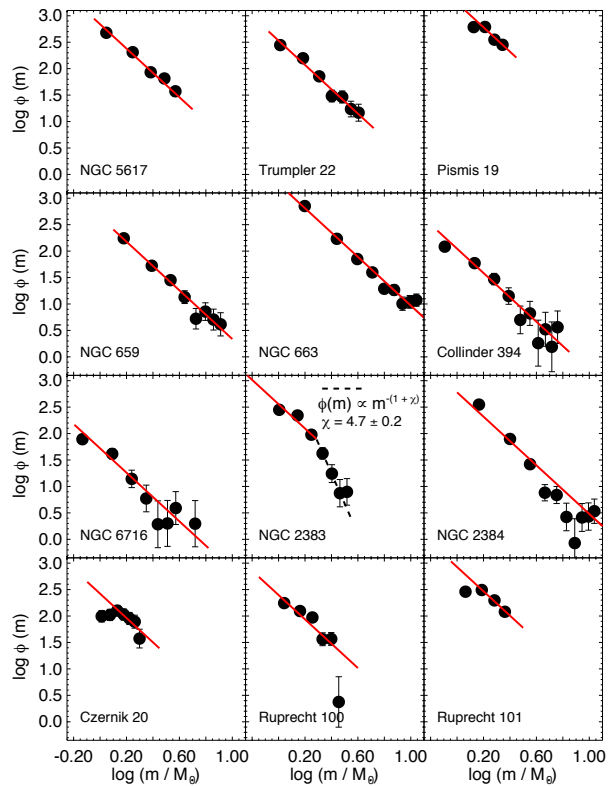}    
    \end{center}
    
  }
\caption{ Mass functions (black dots) for 12 investigated OCs. The scaled IMF of \citeauthor{Kroupa:2001}\,\,(\citeyear{Kroupa:2001}) is overplotted with red lines. In the case of NGC\,2383, the dashed line represents a linear fit to the higher mass bins. Poissonian error bars are shown. }

\label{Fig:MF_binaries_part1}
\end{center}
\end{figure}

For each OC, we took the sample of member stars (filled symbols in Fig.~\ref{fig:decontam_CMDs_part1}) and estimated individual masses from interpolation in $G$ magnitude along the best-fit isochrone. Then the number of stars within bins of mass was obtained by summing up their membership likelihoods, in order to build each cluster mass function $\phi(m)=dN/dm$, which was then converted to the log scale (Fig.~\ref{Fig:MF_binaries_part1}).

Moreover, we overplotted the initial mass function (IMF) of \citeauthor{Kroupa:2001}\,\,(\citeyear{Kroupa:2001}; red lines in Fig.~\ref{Fig:MF_binaries_part1}), scaled according to each OC observed mass. It is indistinguishable, within uncertainties, from Salpeter's\,\,(\citeyear{Salpeter:1955}) IMF for the observed mass range. Considering the higher mass bins, we can see that, with the exception of NGC\,2383, the observed mass functions are compatible with Kroupa and Salpeter IMFs. Some OCs (Pismis\,19, Ruprecht\,101 and Czernik\,20) present more notable depletion of lower mass stars, which may be consequence of preferential evaporation during their dynamical evolution (e.g., \citeauthor{Portegies-Zwart:2010}\,\,\citeyear{Portegies-Zwart:2010}).   

In order to take into account possible member stars below the detection limit, we used Kroupa IMF along with the zero-points in Fig.~\ref{Fig:MF_binaries_part1} to estimate the cluster total (M$_{clu}$; Table~\ref{tab:params_binarios}) and mean mass ($\langle m\rangle$; see eq.~\ref{define_trh}) down to a stellar mass of $\sim0.1\,M_{\odot}$. Uncertainties come from error propagation.

\subsubsection{Relaxation time}

The cluster's half-light relaxation time ($t_{rh}$, Table~\ref{tab:investig_sample}) was obtained from the following expression \citep{Spitzer:1971}:

\begin{equation}
   t_{rh}=(8.9\times10^5\,\textrm{yr})\frac{M_{clu}^{1/2}\,r_{h}^{3/2}}{\langle m\rangle\,\textrm{log}_{10}(0.4\,M_{clu}/\langle m\rangle)}
   \label{define_trh}
,\end{equation}

\noindent
where $\langle m\rangle$ is the mean stellar mass and $M_{clu}$ is the cluster's total mass, both expressed in $M_{\odot}$. The half-light radius $r_{h}$ is expressed in pc. From a theoretical point of view, $t_{rh}$ is the time-scale on which the system tends to thermal equilibrium, with its stars establishing a Maxwellian velocity distribution, continuously repopulating its high-velocity tail and thus losing stars by evaporation \citep{Portegies-Zwart:2010}.

\subsubsection{Escape velocity}
\label{sec:escape_velocity}

After performing the total mass estimates, we checked if each of the potentially interacting pairs are gravitationally bound (i.e, genuine binary clusters) by comparing their relative velocity ($v_{\textrm{rel}}$) and the system's escape velocity ($v_{\textrm{esc}}$). An order of magnitude estimate of $v_{\textrm{esc}}$ can be obtained from the expression (e.g., \citeauthor{Naufal:2020}\,\,\citeyear{Naufal:2020})

\begin{equation}
   v_{esc} = \left[\frac{2G(M_1+M_2)}{\Delta\,r}\right]^{1/2} = (0.093\,\,\textrm{km.s}^{-1})\,\left[\frac{(M_1+M_2)}{\Delta\,r}\right]^{1/2},
   \label{eq:v_esc}
\end{equation}

\noindent
where $M_1$ and $M_2$ are the clusters' total masses (expressed in $M_{\odot}$) and $\Delta r$ is the separation (in pc) between their central coordinates (Table~\ref{tab:params_binarios}). Obviously, the above formula is an oversimplification, since a cluster is not a point mass and a detailed modeling of the gravitational potential would be required for a better estimation.

For each of the interacting pairs in Table~\ref{tab:params_binarios}, the mean projected movement of \textit{cluster\,2} relatively to \textit{cluster\,1} was derived from the simple expression 

\begin{equation}
    \vec{V}_{\textrm{rel}}= \langle\vec{\mu}_{2}\rangle - \langle\vec{\mu}_1\rangle, 
    \label{eq:veloc_relativa}
\end{equation}

\noindent
where $\langle\vec{\mu}\rangle$ is the cluster systemic motion, as determined from the set of member stars  after excluding from the calculation those stars that deviate by more than 3$\sigma$ from the median of $\mu_{\alpha}\,$cos\,$\delta$ and $\mu_{\delta}$. The relative velocities were then converted to the linear scale using the distance moduli in Table~\ref{tab:params_binarios}; the uncertainties were obtained from  propagating the clusters distance errors and also the dispersion of proper motion components. Due to the scarcity of radial velocity data for the investigated sample (Table~\ref{tab:Vrads}), the determination of $V_{\textrm{rel}}$ only incorporates the proper motion components.

\subsubsection{Roche limit for clusters in a physical pair}
\label{sec:estima_roche_limit}

The Roche limit ($R_{roc}$) is assumed here as a characteristic length scale that delimits the Roche volume, that is, the region (centered on the cluster coordinates) within which a star is more proeminently affected by the cluster's tidal pull, taking into account the gravitational influence of another cluster (physically close to the first one) and also the external tidal field.  A simplified approach is presented in Appendix~\ref{sec:simple_approach_Roche_limit}, where it is shown that $R_{roc}$ for a cluster \textit{clu1} can be obtained by solving the equation below:

\begin{equation}
  \left[\frac{M_2}{(s-r)^2}-\frac{M_1}{r^2}\right]_{r=R_{roc,clu1}}\,=\,\left[\frac{M_2}{s^2}-\frac{2M_G\,r}{D^3}\right]_{r=R_{roc,clu1}},
   \label{eq:numerical_solution_R_roche_maintext}
\end{equation}   

\noindent
where $M_1$ and $M_2$ are, respectively, the masses of \textit{clu1} and of its companion cluster (\textit{clu2}), $s$ is their separation, $M_G$ is the Galaxy mass (treated here as a point mass $M_G\sim1.0\times10^{11}\,M_{\odot}$; \citeauthor{Carraro:1994}\,\,\citeyear{Carraro:1994}; \citeauthor{Bonatto:2005}\,\,\citeyear{Bonatto:2005}; \citeauthor{Taylor:2016}\,\,\citeyear{Taylor:2016}) and $D$ is the Galactocentric distance of the pair. Interchanging the values of $M_{1}$ and $M_{2}$ in eq.~\ref{eq:numerical_solution_R_roche_maintext} allows to estimate the Roche limit for \textit{clu2}. The comparison between the empirical $r_t$ value, as determined from King model fits (Fig.~\ref{fig:RDPs_part1}), and $R_{roc}$ is an indicative of the degree of tidal filling of a cluster.

\subsection{Comparison with recent catalogues}
\label{sec:compara_results_lit}

\begin{figure*}
\begin{center}

\parbox[c]{0.70\textwidth}
  {
   \begin{center}
     \includegraphics[width=0.70\textwidth]{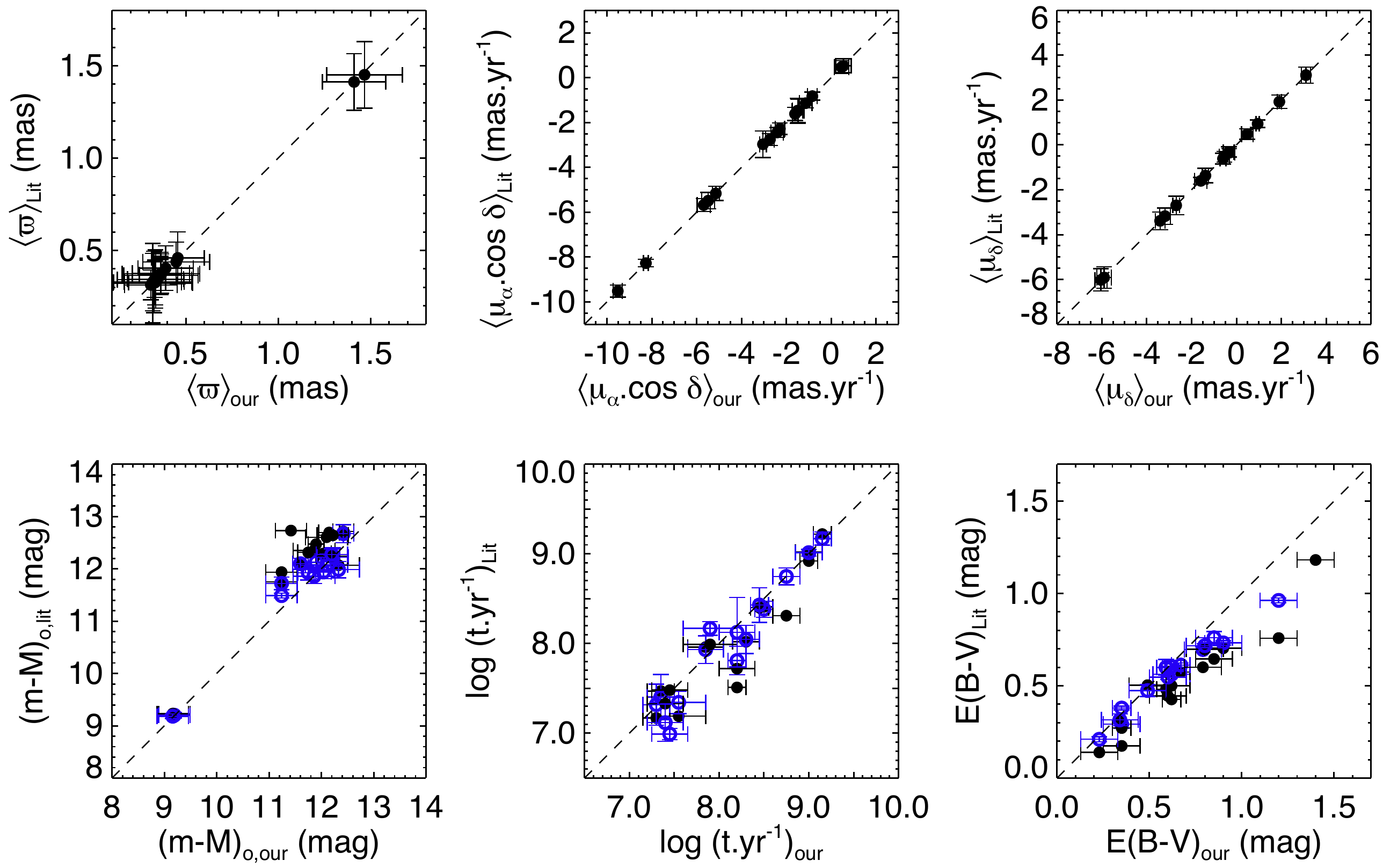}    
    \end{center}
    
  }
\caption{ Top row: comparisons between the mean astrometric parameters ($\langle\varpi\rangle$, $\langle\mu_{\alpha}\,\textrm{cos}\,\delta\rangle$, $\langle\mu_{\delta}\rangle$) obtained in this work (x-axis) and in CG20 (y-axis). Bottom row: comparisons between the fundamental parameters ($(m-M)_0$, log\,$t$ and $E(B-V)$) derived in the present paper and in the literature (black symbols: CG20 (no uncertainties informed); blue symbols: DMML21). In all panels, the dashed line is the identity locus. }

\label{fig:compara_our_results_lit}
\end{center}
\end{figure*}

In Fig.~\ref{fig:compara_our_results_lit}, the set or parameters derived for the investigated OCs is compared with the results obtained from the catalogues of CG20 (2017 OCs) and DMML21 (1743 OCs), both studies based on \textit{Gaia}\,DR2 data. CG20 provide the list of member stars for all OCs in their sample. This allowed us to look for their correspondence in \textit{Gaia} EDR3 (the cross-match table \textit{gaiaedr3.dr2\_neighbourhood}, available in the \textit{Gaia} archive, was used for this task), correct the data for parallax zero-point biases and $G-$band fluxes, apply the quality filters as specified in Section~\ref{sec:sample} and determine the literature values for the mean astrometric parameters ($\langle\varpi\rangle_{\textrm{lit}}$, $\langle\mu_{\alpha}\,\textrm{cos}\,\delta\rangle_{\textrm{lit}}$, $\langle\mu_{\delta}\rangle_{\textrm{lit}}$). The top row of Fig.~\ref{fig:compara_our_results_lit} shows the comparison with our results, where we can note almost identical results in both studies. The error bars in both axes correspond to the 3 times the dispersion, obtained from the member star lists, for each astrometric parameter (in the case of the parallaxes, we have summed in quadrature an uncertainty of $\sim$0.01 mas systematically affecting the astrometric solution in the \textit{Gaia} EDR3 catalogue; \citeauthor{Maiz-Apellaniz:2021}\,\,\citeyear{Maiz-Apellaniz:2021}). 

The bottom row in Fig.~\ref{fig:compara_our_results_lit} compares the fundamental astrophysical parameters ($(m-M)_0$, log\,$t$ and $E(B-V)$) derived here with those obtained from the literature. Regarding CG20's results, in the leftmost panel we note severe discrepancies in distance modulus (which can reach $\sim1.3\,$mag, in the case of the OC Pismis\,19), the literature values being systematically larger than ours. In the case of $E(B-V)$ (rightmost panel), their values are sistematically smaller than ours and a slightly better agreement is found for the log\,$t$ (middle panel), although with some strong discrepancies. 

In order to investigate the origin of such differences, we compared directly the original lists of member stars of CG20 (that is, as obtained by them from the \textit{Gaia} DR2, in order to avoid any biases related to the cross-match with the EDR3 catalogue and/or related to the data quality filters) with ours in the CMDs of Figs.~\ref{fig:decontam_CMDs_compCGaudin_part1} and \ref{fig:decontam_CMDs_compCGaudin_part2} (Appendix~\ref{sec:supplementary_figures}). We have also overplotted \cite{Bressan:2012} isochrones of solar metallicity (magnitudes converted to the \textit{Gaia} DR2 photometric system) properly shifted according to CG20's parameters, derived from the same set of isochrones. We can see a mismatch between the evolutionary sequences defined in the decontaminated CMDs and the isochrones (our solution providing a better fit), which can supposedly be attributed to different isochrone fitting methods, degeneracy in the parameters solution and, at least partially, different extinction laws employed in each study (see CG20 and references therein). It is particularly interesting the case of Trumpler\,22 (Fig.~\ref{fig:decontam_CMDs_compCGaudin_part1}), which presents a giant member star ($G\sim10.3\,$mag) among the literature members list, but it has apparently been negleted during their isochrone fitting procedure.    

The comparisons with DMML21's results show an overall better agreement, the more critical discrepancy is noted for OCs with higher insterstellar extinction ($E(B-V)_{\textrm{our}}\gtrsim0.75\,$mag in the case of NGC\,659, NGC\,663, King\,16, Berkeley\,4 and Dias\,1; Pismis\,19 is absent in their catalogue), for which our values are sistematically larger than the literature ones. The largest discrepancies (reaching $\sim$0.2\,mag) in this parameter are verified for NGC\,663 and Dias\,1. In both cases, adopting the literature parameters results in the isochrone being placed exaggeratedly to the left of the decontaminated sequences in the CMDs, therefore do not providing a proper fit to the data.

\section{Discussion}
\label{sec:discussion}
\subsection{Individual comments on some clusters}
\label{sec:indiv_comments}
\subsubsection*{NGC\,5617$-$Trumpler\,22}
\label{sec:NGC5617_Trumpler22}
\cite{De-Silva:2015a} carried out high-resolution spectroscopy for both OCs and confirmed their physical connection based on their common mean metallicities ($[Fe/H]_{N5617}=-0.18\pm0.02$ and $[Fe/H]_{Tr22}=-0.17\pm0.04$) and radial velocities ($V_{r,N5617}=-38.63\pm2.25\,$km.s$^{-1}$ and $V_{r,Tr22}=-38.46\pm2.08\,$km.s$^{-1}$). According to Table~\ref{tab:Vrads}, the mean radial velocity for 3 member stars of NGC\,5617 with available $V_r$ in \textit{Gaia} EDR3 is in agreement with the mean value of \cite{De-Silva:2015a}, while the mean $V_r$ for Trumpler\,22 from \textit{Gaia} data presents a discrepancy of $\sim5\,$km.s$^{-1}$. Despite this, a more meaningful comparison is precluded given the small number of stars with available $V_r$ in \textit{Gaia} EDR3.

Comparatively to \citeauthor{Bisht:2021}\,\,(\citeyear{Bisht:2021})'s results for NGC\,5617, we found agreement only on the $E(B-V)$ estimates. Our distance modulus and log\,$t$ are, respectively, $\sim0.7\,$mag smaller and 0.35 larger. Compared to their results for Trumpler\,22, our derived distance modulus and log\,$t$ are, respectively, $\sim0.85\,$mag smaller and 0.3\,dex larger. The $E(B-V)$ values agree within uncertainties. 


In the case of NGC\,5617's RDP, there is a fluctuation in stellar density in the range $7.5\arcmin\lesssim r\lesssim10.5\arcmin$ (2.65\,$\lesssim\,$\textrm{log}\,($r$/arcsec)$\,\lesssim$\,2.80, Fig.~\ref{fig:RDPs_part1}); Trumpler\,22's RDP also presents some less notable fluctuations close to its $R_{\textrm{lim}}$. It is important to emphasize that these fluctuations are not due to contamination by Pismis\,19' stars, which have been properly filtered out of the skymaps before the structural analysis. NGC\,5617 and Trumpler\,22 present the same heliocentric distance, compatible ages and metallicities (Table~\ref{tab:investig_sample}), which indicates a physically interacting pair with common origin.

Their relative velocity is about the same order as the system escape velocity (eq.~\ref{eq:v_esc}). This result maginally favor the hypothesis of a binary cluster. This conclusion remains unaltered even if we include radial velocities to estimate $v_{esc}$, since both OCs present compatible $V_r$ according to \cite{De-Silva:2015a}. For both OCs, the empirical tidal radii (Table~\ref{tab:investig_sample}) are smaller than the estimated $R_{roc}$. This means that their stellar content are located well within the allowed tidal volume and therefore less susceptible to mass-loss due to tidal effects (\citeauthor{Heggie:2003}\,\,\citeyear{Heggie:2003}; \citeauthor{Ernst:2015}\,\,\citeyear{Ernst:2015}).

\subsubsection*{Collinder\,394$-$NGC\,6716}

\cite{Naufal:2020} employed \textit{Gaia} DR2 data to investigate three binary cluster candidates, including the pair Collinder\,394$-$NGC\,6716. They found that both OCs present very similar proper motions, compatible heliocentric distances and ages, thus entailing an interacting pair. Their fundamental parameters derived for Collinder\,394 were: $(m-M)_0=9.25\,$mag and log\,($t.$yr$^{-1}$)=8.30. For NGC\,6716, they obtained $(m-M)_0=9.21\,$mag and log\,($t.$yr$^{-1}$)=8.30. In both cases, no $E(B-V)$ values and no uncertainties were informed. These values for distance modulus and age are in agreement with ours (Table~\ref{tab:investig_sample}). Both OCs seem stable against tidal disruption, since their estimated Roche limit is greater than the respective $r_t$ (Table~\ref{tab:params_binarios}). 


The escape velocity for the pair Collinder\,394 $-$ NGC\,6716 is comparable to their relative velocity (Table~\ref{tab:params_binarios}), thus suggesting that we are facing a binary cluster. The median $V_r$ for Collinder\,394 is 21.9\,km.s$^{-1}$, which is almost the same value for one of the two NGC\,6716's member stars with available $V_r$ (\textit{Gaia}\_source\_id: 4085290948306525312). This suggests that both OCs present compatible radial movements. Another NGC\,6716's member (\textit{Gaia}\_source\_id: 4086041536793495808) presents incompatible $V_r$, which can be consequence of binarity or even due to the presence of an outlier among this OC members list. Additionally, both OCs present the same age ($t\sim80\,$Myr) and metallic content, which suggests a common origin.

\subsubsection*{NGC\,2383$-$NGC\,2384}
\label{sec:indiv_comments_N2383_N2384}

Four alleged binary clusters, including the pair NGC\,2383$-$NGC\,2384, were investigated by \cite{Vazquez:2010}, who employed deep CCD $UBVRI$ photometric data and searched for member stars within the clusters limits by inspection of photometric diagrams. Membership control was done by requiring that stars have consistent positions in several diagrams and by using published spectral types (see references therein). For both OCs, our derived fundamental parameters (Table~\ref{tab:investig_sample}) are consistent with their results, which are: $(m-M)_{0,N2383}=12.46\pm0.20\,$mag, $E(B-V)_{N2383}=0.30\pm0.05\,$mag and log\,($t.$yr$^{-1}$)$_{N2383}$=$8.45\pm0.05$; $(m-M)_{0,N2384}=12.39\pm0.22\,$mag, $E(B-V)_{N2384}=0.31\pm0.03\,$mag and log\,($t.$yr$^{-1}$)$_{N2384}$=$7.15\pm0.05$.

\cite{Monteiro:2020} characterized a set of 45 Galactic OCs (including NGC\,2384) listed in \cite{Dias:2002} by means of \textit{Gaia} DR2 data and an isochrone fitting code that incorporates ($i$) an improved treatment of interstellar extinction for the \textit{Gaia} photometric band-passes and ($ii$) the Galactic abundance gradient as a prior for metallicity. Their results for NGC\,2384 are: $(m-M)_0=12.00\pm0.16\,$mag, $E(B-V)=0.31\pm0.03\,$mag and log\,($t.$yr$^{-1}$)=$7.32\pm0.23$, which are consistent with ours (Table~\ref{tab:investig_sample}), within uncertainties. 


The projected relative velocity between NGC\,2383 and NGC\,2384 is about one order of magnitude larger than the system's escape velocity (Table~\ref{tab:params_binarios}). Therefore, this pair of OCs seems gravitationally unbound, which is also suggested from the large differences in radial velocities ($\gtrsim10\,$km.s$^{-1}$), as reported in Table~\ref{tab:Vrads}. Both OCs present very different ages and estimated metallicities. This suggests that NGC\,2383 and NGC\,2384 are OCs of distincit origin undergoing a close encounter. Their Roche limit is smaller than their respective $r_t$; apparently, their close proximity causes shrinkage of their respective Roche volumes, making them tidally overfilled and thus subject to episodes of mass loss.

\begin{figure}
\begin{center}

\parbox[c]{0.34\textwidth}
  {
   \begin{center}
     \includegraphics[width=0.34\textwidth]{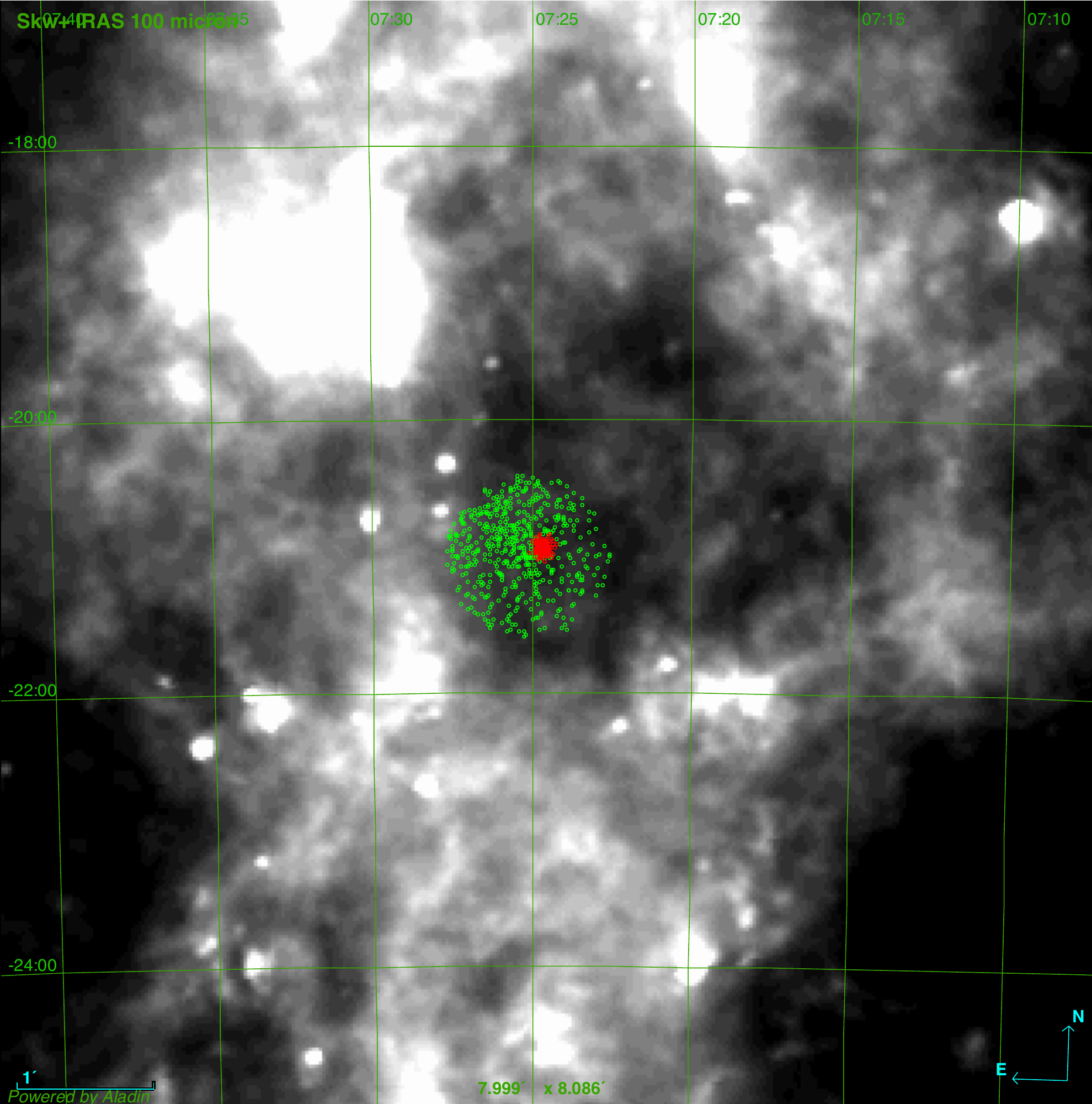}    
    \end{center}
    
  }
\caption{ IRAS 100\,$\mu$m dust continuum emission map ($8^{\circ}\times8^{\circ}$), showing the member stars of NGC\,2383 and NGC\,2384 (red and green dots, respectively). }

\label{Fig:IRAS_NGC2384}
\end{center}
\end{figure}

In the case of the OC NGC\,2384, the concentrated group of bright stars in its centre is located within a much larger structure (Figs.~\ref{fig:ilustra_preanalise_NGC2383} and \ref{Fig:IRAS_NGC2384}) that does not present a well-defined visual contrast with respect to the field, which can also be seem in figure 2 of \cite{Vazquez:2010}. They pointed out that NGC\,2384 is supposedly a young and sparse remnant group of a recent star formation episode. This is consistent with the presence of B stars among its members and also of some cavities in the local dust distribution (Fig.~\ref{Fig:IRAS_NGC2384}) as inferred from IRAS\,100\,$\mu$m emission map \citep{Schlegel:1998}. Besides, the asymmetry in the spatial distribution of NGC\,2384's member stars (Fig.~\ref{Fig:IRAS_NGC2384}), located preferentially Northeastwards of its centre, is supposedly the result of star formation process combined with stellar feedback mechanisms.    


\subsubsection*{Other investigated pairs}

The pairs NGC\,659$-$NGC\,663, King\,16$-$Berkeley\,4  and Ruprecht\,100$-$Ruprecht\,101 (the OCs Czernik\,20 and NGC\,1857 are single ones, as mentioned in Section~\ref{sec:results}) do not present, to the best of our knowledge, previous investigations devoted to explore their physical connection. In what follows, we highlight some brief comments regarding their properties, based on the outcomes of our analsys.

\subsubsection*{NGC\,659$-$NGC\,663}
This seems a young coeval ($t_{N659}=35\,$Myr and $t_{N663}=22\,$Myr; Table~\ref{tab:investig_sample}) and gravitationally bound pair (therefore a genuine binary candidate), as their projected relative velocity is comparable to the system escape velocity (Table~\ref{tab:params_binarios}). Both do not seem dynamicallly evolved clusters, since they are younger than their respective half-light relaxation times. Based on the outcomes from eq.~\ref{eq:numerical_solution_R_roche_maintext}, NGC\,659 and NGC\,663 seem tidally underfilled, since their $r_t$ values are smaller than their respective Roche radii (Tables~\ref{tab:investig_sample} and \ref{tab:params_binarios}). 


\subsubsection*{Ruprecht\,100$-$Ruprecht\,101}

The difference between the mean radial velocity (Table~\ref{tab:Vrads}) for each OC in this pair ($\langle V_r\rangle_{\textrm{Rup100}}$=-11.2\,km.s$^{-1}$ and $\langle V_r\rangle_{\textrm{Rup100}}=$-14.4\,km.s$^{-1}$) is $\sim3\,$km.s$^{-1}$, which is about the same order of the system escape velocity (Table~\ref{tab:params_binarios}). Therefore both clusters may be gravitationally bound to each other, which makes the pair Ruprecht\,100$-$Ruprecht\,101 a possible binary. Nonetheless, this statement is not absolutely conclusive, given the small number of stars with available $V_r$ and the large uncertainty in the value of $V_{rel}$ obtained from proper motions (Table~\ref{tab:params_binarios}).  

From eq.~\ref{eq:numerical_solution_R_roche_maintext}, the pair Ruprecht\,100$-$Ruprecht\,101 is a  semidetached binary candidate, since the former seems a tidally filled cluster, while the latter seems tidally underfilled. Ruprecht\,100 is less massive, more dynamically evolved (larger age/$t_{rh}$ ratio) and presents a more compact internal structure compared to Ruprecht\,101 (see the discussions in Section~\ref{sec:aggregates_in_Gal_context}). Moreover, both OCs present compatible distance, metallicity and comparable ages (Table~\ref{tab:investig_sample}), considering uncertainties.


\subsubsection*{King\,16$-$Berkeley\,4}

Both OCs in this pair present different ages (Table~\ref{tab:investig_sample}) and seem gravitationally unbound, since their projected relative velocity is $\sim8\,\times$ the system escape velocity (Table~\ref{tab:params_binarios}; no $V_r$ available for both clusters). Besides, King\,16 seems more metal-rich than Berkeley\,4, although more precise $[Fe/H]$ data would be needed to state this more firmly.  This pair seems to  consist of OCs with distinct origins experiencing a close encounter and will probably be disassociated to each other in the future (see Section~\ref{sec:comp_with_other_pairs_lit}).


\subsection{Aggregates of clusters in the Galactic context}
\label{sec:aggregates_in_Gal_context}
 
This section is devoted to comparisons between the dynamical properties of the potentially interacting pairs studied here and those of a set of Galactic non-binary OCs previously investigated. We have also included those OCs classified here as single ones (namely: Pismis\,19, Dias\,1, Czernik\,20 and NGC\,1857). To warrant uniformity in our treatment, the astrophysical parameters of the comparison OCs have been taken from the samples of Angelo et al.\,\,(\citeyear{Angelo:2020}, \citeyear{Angelo:2021}; table 1 of both papers), who employed \textit{Gaia}\,DR2 data to characterize a set of 65 OCs with the same analysis methods as those presented in this paper. 

\begin{table}
 \small
\begin{minipage}{85mm}
  \caption{ Symbol convention and colours used in Figs.~\ref{Fig:plots_radius_age_mass1} and \ref{Fig:plots_radius_age_mass2}. }
  \label{tab:symbols_convention}
 \begin{tabular}{lcccc}

\hline

                                                 &  \multicolumn{4}{c}{$R_G$ (kpc) intervals}                                                                                                                     \\ \cline{2-5}
                                                 &                                    &                                      &                                            &                                       \\   
  Sample                                         &   6.0$-$7.0                        & 7.0$-$9.0                            &  9.0$-$11.0                                & 11.0$-$13.0                           \\ 
\hline                                                                                                                                                                                                            
 Angelo et al.\,\,                               &  \color{gray}{\Large{$\bullet$}}   &  \color{gray}{$\blacksquare$}        & \color{gray}{$\blacktriangle$}             &  \color{gray}{$\blacklozenge$}        \\     
(\citeyear{Angelo:2020}, \citeyear{Angelo:2021}) &                                    &                                      &                                            &                                       \\
\hline

\multicolumn{5}{c}{Interacting pair candidates} \\

\hline
N5617\,;\,Tr22                                    &  \color{red}\Large{$\Circle$}      &                                      &                                          &                                       \\
Col394\,;\,N6716                                  &                                    & \color{violet}\Large{$\square$}      &                                          &                                       \\
N2383\,;\,N2384                                   &                                    &                                      & \color{green}{\Large$\blacktriangle$}    &                                       \\
Rup100                                            &                                    & \color{cyan}\Large{$\blacksquare$}   &                                          &                                       \\
Rup101                                            &                                    & \color{cyan}\Large{$\square$}        &                                          &                                       \\
K16\,;\,Berk4                                     &                                    &                                      & \color{black}{\Large$\triangle$}         &                                       \\
N659\,;\,N663                                     &                                    &                                      & \color{blue}{\Large$\triangle$}          &                                       \\
\hline
\multicolumn{5}{c}{Single clusters} \\
\hline
Pis19                                             & \color{red}{$\bigotimes$}          &                                      &                                          &                                       \\
Dias1                                             &                                    &                                      & \color{black}{\Large$\bigtriangledown$}  &                                       \\
Cz20                                              &                                    &                                      &                                          & \color{orange}{\Large$\lozenge$}      \\
N1857                                             &                                    &                                      & \color{orange}{\Large$\triangle$}        &                                       \\

\hline

\multicolumn{5}{l}{\textit{Note}: Filled symbols represent tidally filled clusters; open} \\
\multicolumn{5}{l}{symbols represent tidally underfilled ones (Section~\ref{sec:estima_roche_limit}).} \\
\multicolumn{5}{l}{For the single clusters, we employed the Jacobi radius:} \\
\multicolumn{5}{l}{$R_{J}=R_G\,(M_{clu}/3M_G)^{1/3}$. }

\end{tabular}

\end{minipage}
\end{table}

\begin{figure*}
\begin{center}

\parbox[c]{1.00\textwidth}
  {
   \begin{center}
    \includegraphics[width=0.75\textwidth]{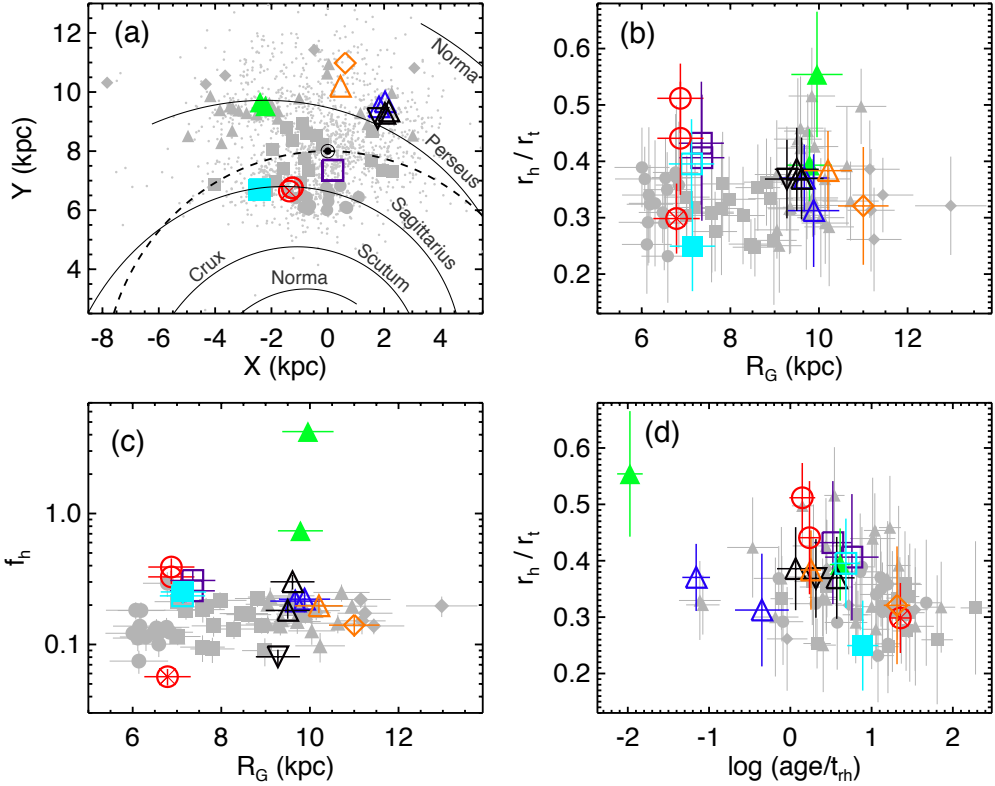}  
    \end{center}    
  }
\caption{ (a) Galactic plane showing the disposal of the clusters investigated in this paper. Symbol convention and colours are shown in Table~\ref{tab:symbols_convention}. Small dots are OCs taken from \protect\cite{Dias:2021a} sample. Schematic position of the spiral arms from \protect\cite{Vallee:2008}. The Sun's position and the solar circle (dashed lines) are indicated. (b) $r_h/r_t$ versus $R_G$ plot. (c) Roche volume filling factor ($f_h$) versus $R_G$ plot. (d) $r_h/r_t$ versus log\,(age/$t_{rh}$) plot. }

\label{Fig:plots_radius_age_mass1}
\end{center}
\end{figure*}

Panel (a) of Fig.~\ref{Fig:plots_radius_age_mass1} shows the disposal of our sample along the Galactic plane. The symbol convention is shown in Table~\ref{tab:symbols_convention}. The small dots are OCs taken from \cite{Dias:2021a} catalogue. The set of OCs investigated  here are located close to the Sagittarius (NGC\,5617$-$Trumpler\,22; Ruprecht\,100$-$Ruprecht\,101) and Perseus arms (NGC\,2383$-$NGC\,2384; King\,16$-$Berkeley\,4 and Dias\,1; NGC\,659$-$NGC\,663) or in the interarms region (Collinder\,394$-$NGC\,6716; Czernik\,20 and NGC\,1857), spanning a range of Galactocentric distances ($R_G$) between $\sim$6.5$-$11.0\,kpc along the four Galactic quadrants (Table~\ref{tab:investig_sample}).       

Panel (b) of Fig.~\ref{Fig:plots_radius_age_mass1} shows the $r_{h}/r_t$ ratio (which can be interpreted as  the OC internal structure length scale relative to its overall size) as function of $R_G$ for the investigated sample (coloured symbols) and for the literature OCs (grey filled symbols). The binaries NGC\,5617$-$Trumpler\,22 and Collinder\,394$-$NGC\,6716 present $r_{h}/r_t$ ratios that are larger compared to their counterparts located at similar $R_G$. The same is verified for Ruprecht\,101 in the Ruprecht\,100$-$Ruprecht\,101 binary. A Kolmogorov-Smirnov (KS) two-sample text relating this 5 OCs and the whole set of single OCs located in the range of $R_G$ between $\sim6.5-7.5\,$kpc resulted in a probability smaller than $\sim0.1\%$ for these two sample being statistically similar. However, this statement should be considered with some care, given the uncertainties and the low number of gravitationally bound pairs present in our analysis. The high $r_{h}/r_t$ ratios for these clusters may possibly be attributed to mutual tidal forces between them and their respective companion OC during their orbital motions, which can affect their internal mass distributions through mass loss. In fact, as stated by \cite{Innanen:1972}, tidal interactions are significant when the separation of two star clusters is less than $\sim$3 times their tidal radii.

Although these 3 pairs (NGC\,5617$-$Trumpler\,22, Collinder\,394$-$NGC\,6716 and Ruprecht\,100$-$Ruprecht\,101) are located at relatively small Galactocentric distances ($R_G<7.5\,$kpc), and therefore subject to stronger external tidal fields, they are tidally underfilled (except Ruprecht\,100), that is, their whole stellar content seems located well within the allowed Roche volume (Section~\ref{sec:estima_roche_limit}). In turn, Ruprecht\,100 is tidally filled, but with a small $r_{h}/r_t$ ratio ($\simeq$\,0.25) compared to other OCs at $R_G\sim 7\,$kpc. Its internal compact structure seemingly prevents this OC from being tidally disrupted. Since Ruprecht\,100 and Ruprecht\,101 present marginally compatible ages, the former being more dynamically evolved, differences in their current physical structures and dynamical stages may be traced back to their formation conditions. Analogous comments can be stated for the NGC\,5617$-$Trumpler\,22 and Collinder\,394$-$NGC\,6716, both pairs presenting coeval OCs. 

The $r_{h}/r_t$ ratio of the binary NGC\,659$-$NGC\,663 is compatible with those of other OCs at the same $R_G$. These two OCs are young ($t_{\textrm{N659}}\sim35\,$Myr, $t_{\textrm{N663}}\sim22\,$Myr) and dynamically unevolved systems (age/$t_{rh}<1$). This means non-relaxed internal structures and therefore their present dynamical state may reflect more closely their initial formation conditions. Their mutual tidal interactions seemingly have not severely biased their structural parameters.

Our results for these OCs can be compared to those of DAK21. Their figure 14 (top panels) shows the orbital properties of separated aggregates, that is, modeled pairs of clusters with increasingly wider separations under the influence of the Galactic tidal field. The current separation between NGC\,659 and NGC\,663 ($\Delta\,r\simeq27\,$pc; Table~\ref{tab:params_binarios}) and mass ratio ($q=M_{N659}/M_{N663}\sim0.2$) are comparable with their simulated realizations under initial conditions ``C8" (supervirialized, $\alpha_{vir}=0.7$; see their table 2), which results in $\Delta\,r$ between 25$-$50\,pc and $q<0.25$ for ages between $\sim20-30\,$Myr. According to DAK21, higher initial $\alpha_{vir}$ results in a higher probability of forming a separated system. Based on this direct comparison, we can suggest that the pair NGC\,659$-$NGC\,663 will be disassociated from each other in the future. The same conclusion about this system is established by \citeauthor{de-la-Fuente-Marcos:2010}\,\,(\citeyear{de-la-Fuente-Marcos:2010}; their section 4.1).

In the case of the gravitationally unbound pair Berkeley\,4$-$King\,16, their mutual tidal interactions apparently have not severely impacted their internal structures in such a way that could differentiate this pair from isolated OCs at similar $R_G$ (panel (b) of Fig.~\ref{Fig:plots_radius_age_mass1}) or comparable dynamical stage (i.e., similar age/$t_{rh}$ ratio; panel (c) of Fig.~\ref{Fig:plots_radius_age_mass1}). In turn, the single clusters Pismis\,19, Dias\,1, Czernik\,20 and NGC\,1857 do not define particular locus of data in panel (b), (c) and (d) of Fig.~\ref{Fig:plots_radius_age_mass1}, being compatible with their analogues from the literature.

\begin{figure*}
\begin{center}

\parbox[c]{1.00\textwidth}
  {
   \begin{center}
    \includegraphics[width=0.75\textwidth]{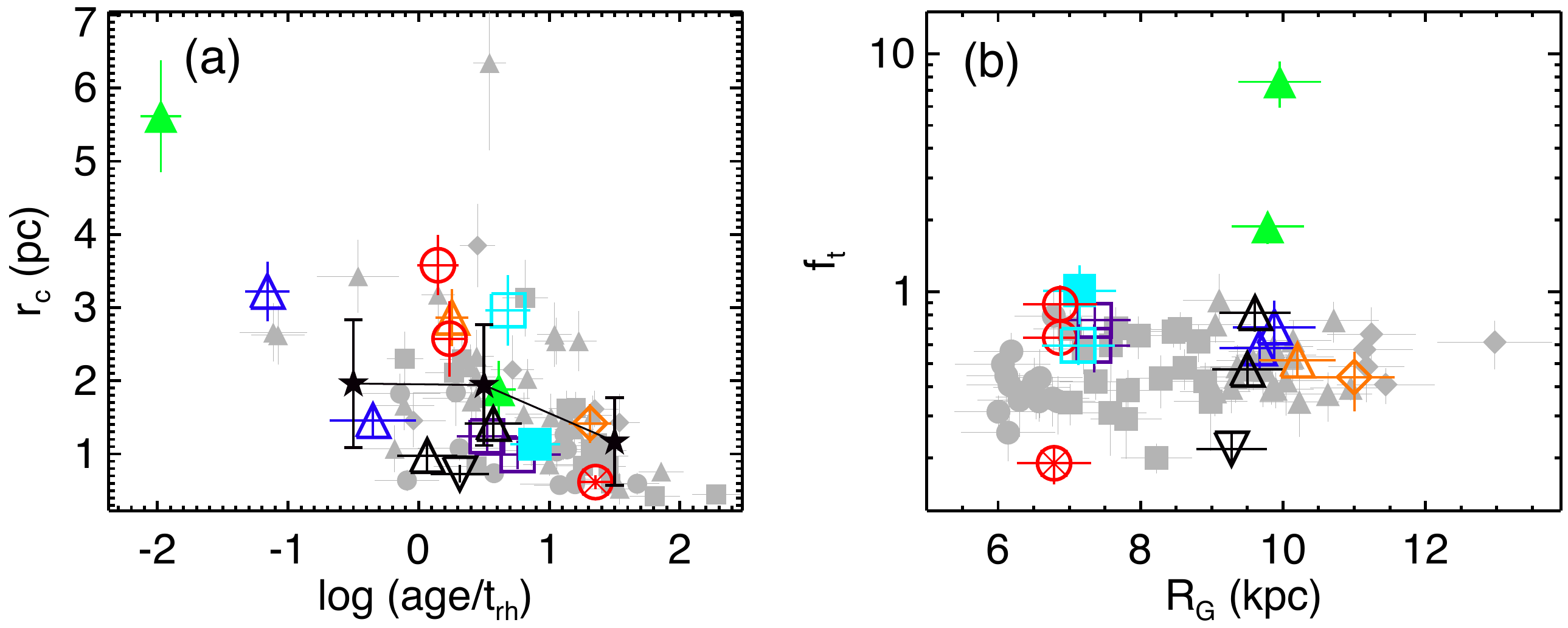}  
    \end{center}    
  }
\caption{ (a) $r_c$ versus log\,(age/$t_{rh}$) plot. In this last panel, the $X-$axis has been divided in three bins: log\,(age/$t_{rh}$) below 0.0, between 0.0$-$1.0 and greater than 1.0. In each case, the mean and dispersion of the $r_c$ values for the \textit{literature} OCs (grey filled symbols) have been determined, as indicated by the filled black stars (in the central bin, the outlier Collinder\,110 ($r_c=6.3\,$pc) has been excluded from the calculation). (b) Tidal volume filling factor ($f_t$) in terms of the ratios $r_t/R_{roc}$ (for those physical pair candidates) and $r_t/R_J$ (for single OCs). }

\label{Fig:plots_radius_age_mass2}
\end{center}
\end{figure*}

NGC\,2384 has a very inflated structure, its $r_{hm}/r_t$ ratio (=0.55) being the highest in relation to other OCs at $R_G\sim10\,$kpc and with most of its stars located beyond the estimated Roche radius (Table~\ref{tab:params_binarios}). As alreadly mentioned in Section~\ref{sec:indiv_comments_N2383_N2384}, its close proximity with NGC\,2383 can favor episodes of mass loss by tidal stripping and the system dynamics may also be affected by stellar feedback mechanisms. In this complex scenario, interactions with the Galactic disc probably contributes to the dispersion of NGC\,2384's stellar content across the cluster area, enhancing the process of mass loss.

Compared to NGC\,2384, the OC NGC\,2383 is older, less massive, considerably more evolved ((age/$t_{rh}$)$_{N2383}\gg$(age/$t_{rh}$)$_{N2384}$) and presents a more compact structure, with its $r_{h}/r_t$ ratio compatible with other OCs at similar $R_G$. It is also tidally filled, which suggests that mass loss process is ongoing due to tidal stripping. Interestingly, its mass function (Fig.~\ref{Fig:MF_binaries_part1}) is steeper than the IMFs of Salpeter and Kroupa in the higher mass bins ($M\gtrsim2\,M_{\odot}$), which indicates an absence of stars in this mass range possibly due to an interplay between stochastic effects (e.g., \citeauthor{Santos:1997}\,\,\citeyear{Santos:1997}; \citeauthor{Lim:2015}\,\,\citeyear{Lim:2015}), shorter evolutionary timescale (e.g., \citeauthor{Valegard:2021}\,\,\citeyear{Valegard:2021}) and dynamical interactions (e.g., \citeauthor{Fujii:2011}\,\,\citeyear{Fujii:2011}).

Panel (c) of Fig.~\ref{Fig:plots_radius_age_mass1} shows the Roche volume filling factor ($f_h$) as function of the Galactocentric distance. In the case of the interacting pair candidates (Table~\ref{tab:params_binarios}), we employed $f_h=r_h/R_{roc}$, where $R_{roc}$ is obtained from eq.~\ref{eq:numerical_solution_R_roche_maintext}; this formulation, although theoretically rather simple, is useful since it predicts the shrinking of the Roche volume due to the presence of a companion OC. For single OCs we adopted $f_h=r_h/R_J$, where $R_J$ is the Jacobi radius ($R_J=R_G(M_{clu}/3M_G)^{1/3}$; \citeauthor{von-Hoerner:1957}\,\,\citeyear{von-Hoerner:1957}). $f_h$ is a useful quantity, since it allows to evaluate the effect of the tidal environment on the clusters' properties (e.g., \citeauthor{Dalessandro:2018}\,\,\citeyear{Dalessandro:2018}). According to \cite{Gieles:2008}, it is related to the fraction of escaping stars at each $t_{rh}$ and therefore to the cluster' star loss rate. 


We especulate that, due to their close proximity, OCs within evolved physical pair candidates tend to be slightly more tidally filled than single OCs at compatible $R_G$ and therefore more subject to tidal stresses (e.g., \citeauthor{Santos:2020}\,\,\citeyear{Santos:2020} and references therein). In the case of the unbound pair King\,16$-$Berkeley\,4, the former seems more tidally influenced (that is, larger $f_h$) than its companion, possibly due to its smaller total mass. Analogously to panel (b), the location of the dynamically unvevolved pair NGC\,659$-$NGC\,663 is compatible with single OCs counterparts, while the high $f_h$ for NGC\,2383$-$NGC\,2384, both tidally overfilled, suggests severe mass loss.
  
Taking the complete sample of investigated OCs (single or in interacting pair candidates), most of them present $f_h\gtrsim0.1$ and, consequently, they are more strongly influenced by the tidal field than those with smaller $f_h$ \citep{Baumgardt:2010}. In addition, all OCs present $f_h\gtrsim0.05$, therefore they are in the ``tidal regime", as stated by \cite{Gieles:2008}; from their equations 8 and 12, these OCs dissolution times are expected to scale with the number of stars and orbital angular frequency, independent of $r_h$.

Considering the whole sample in panel (d) of Fig.~\ref{Fig:plots_radius_age_mass1}, there is no clear correlation between $r_h/r_t$ ratio and the cluster evolutionary stage (as determined by the age/$t_{rh}$ ratio). Despite this, considering each of the gravitationally bound pair candidates (namely, NGC\,5617$-$Trumpler\,22, Collinder\,394$-$NGC\,6716, Ruprecht\,100$-$Ruprecht\,101 and NGC\,659$-$NGC\,663, these four aggregates containing OCs with compatible ages), the more evolved OC is also more compact (that is, smaller $r_h/r_t$ ratios) than its companion, thus suggesting that the internal relaxation also plays a role in the mass distribution across the tidal radius of each OC in a physical binary. Analogous comments can be drawn for the potentially interacting (but gravitationally unbound) pair King\,16$-$Berkeley\,4, the former presenting slightly smaller $r_h/r_t$ and larger log\,(age/$t_{rh}$) compared to its companion.



Except for NGC\,2384, the investigated aggregates follow the general dispersion of literature OCs data in the $r_c\times$\,age/$t_{rh}$ plot (panel (a) of Fig.~\ref{Fig:plots_radius_age_mass2}). Apparently, their central structures are not severely biased due to the presence of a companion cluster since, for each of the interacting pair candidates, whether they are gravitationally bound or not, there are OCs at compatible dynamical stage and with similar $r_c$. The apparent anticorrelation between $r_c$ and age/$t_{rh}$ (as evidenced by the black filled stars, which show the mean and dispersion of the $r_c$ values in three age/$t_{rh}$ bins) seems more importantly determined by the internal dynamics (\citeauthor{Ferraro:2019}\,\,\citeyear{Ferraro:2019}; \citeauthor{Heggie:2003}\,\,\citeyear{Heggie:2003}). 

Panel (b) of Fig.~\ref{Fig:plots_radius_age_mass2} is analogous to panel (c) of Fig.~\ref{Fig:plots_radius_age_mass1}, but this time employing the ratios $f_t=r_t/R_{roc}$ and $f_t=r_t/R_{J}$ for, respectively, interacting pairs and single OCs. Since $f_t$ would be close to 1 for clusters filling their Roche volume \citep{Ernst:2013}, we can see that almost all OCs in our sample are tidally underfilled (that is, $f_t\lesssim1$), with the exceptions of Ruprecht\,100, NGC\,2383 and NGC\,2384. It is expected that Roche lobe underfilling clusters survive for a much larger number of relaxation times than Roche lobe filling ones \citep{Gieles:2008}. 


\begin{figure}
\begin{center}

\parbox[c]{0.37\textwidth}
  {
   \begin{center}
    \includegraphics[width=0.37\textwidth]{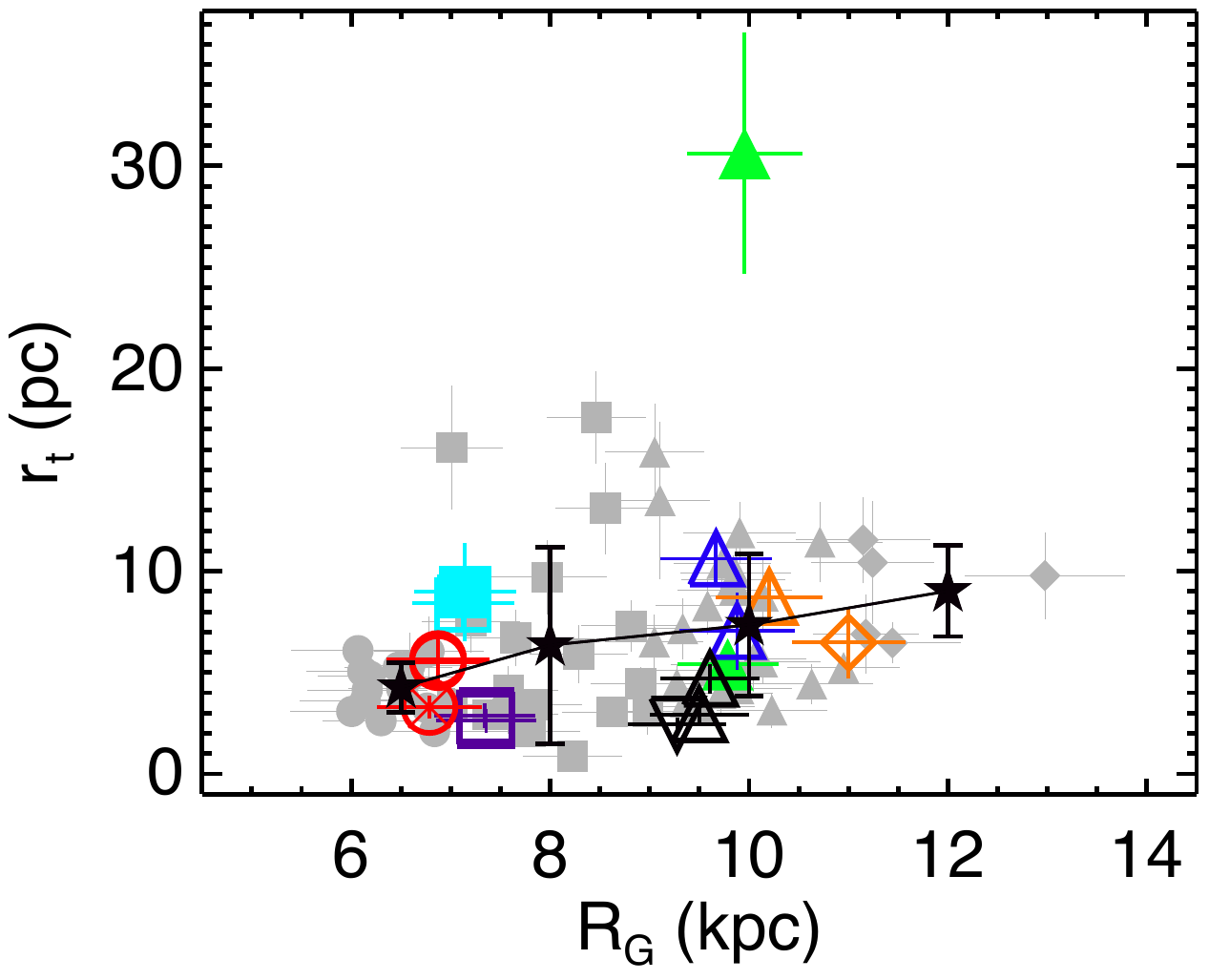}  
    \end{center}    
  }
\caption{ $r_t\times R_G$ plot. The symbol convention of Table~\ref{tab:symbols_convention} has been employed. Mean $r_t$ values and their associated dispersions have been determined from the \textit{literature} OCs divided in 4 $R_G$ bins: below 7.0\,kpc, between 7.0$-$9.0\,kpc, between 9.0$-$11.0\,kpc and greater than 11.0\,kpc. }

\label{Fig:plots_radius_age_mass3}
\end{center}
\end{figure}

Fig.~\ref{Fig:plots_radius_age_mass3} allows to identify the impact of variations in the external tidal field on the clusters' external structure. As stated by \cite{Madrid:2012}, the external tidal field experienced by a cluster strongly affects its dissolution timescale, since the rate at which stars escape from the system depends on the cluster's galactocentric distance. Considering the whole sample, OCs located at smaller Galactocentric distances ($R_G\lesssim7\,$kpc), and thus subject to a more intense Galactic gravitational pull, tend to present smaller and less-dispersed $r_t$ values. The location of the binaries NGC\,5617$-$Trumpler22 ($R_G\sim6.9\,$kpc) and Collinder\,394$-$NGC\,6716 ($R_G\sim7.4\,$kpc) is consistent with this locus, which favors their survival against tidal disruption. The OCs in the binary candidate Ruprecht\,100$-$Ruprecht\,101 ($R_G\sim7.1\,$kpc), in turn, present higher $r_t$ compared to this group. Despite this, as stated previously, Ruprecht\,101 is tidally underfilled and Ruprecht 100, although tidally filled, presents a compact internal structure (as evidenced by its small $r_h/r_t$ ratio), which contributes to its stability against intense mass loss due to external destructive effects. NGC\,2384 presents the largest $r_t$ in our whole sample, as a consequence of its dissolution process. The other investigated OCs present $r_t$ values that are well contained within the data range defined by isolated OCs, that is, $r_t$ between $\sim3-12\,$pc for $R_G$ between $\sim9.5-11\,$kpc.

\subsection{Comparison with other cluster pair candidates in the literature}
\label{sec:comp_with_other_pairs_lit}

\begin{figure*}
\begin{center}

\parbox[c]{0.65\textwidth}
  {
   \begin{center}
    \includegraphics[width=0.65\textwidth]{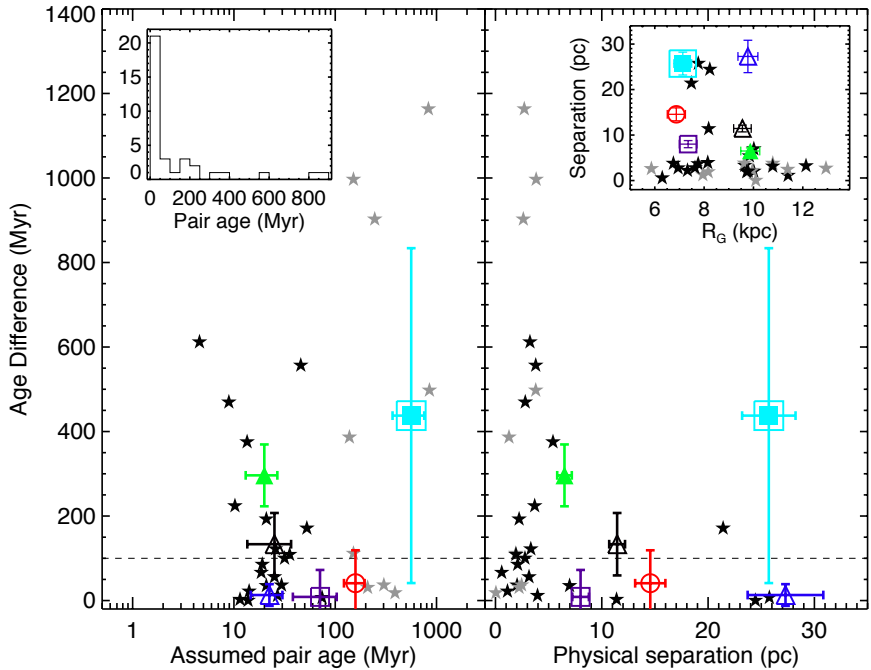}  
    \end{center}    
  }
\caption{ Left panel: Age difference between OCs in each cluster pair as function of the assumed pair age. The inset shows the distribution of pair ages. Filled stars are cluster pairs (black symbols for $t\leq100\,$Myr and grey ones for $t>100\,$Myr) taken from the list of \protect\cite{Piecka:2021}. Other symbols according to Table~\ref{tab:symbols_convention}. The pair Ruprecht\,100$-$Ruprecht\,101 is represented by a double square, since it contains a tidally filled (Ruprecht\,100) and a tidally underfilled (Ruprecht\,101) cluster. The horizontal line delimits a rough superior limit ($\sim100\,$Myr) for the age difference between OCs presenting a correlated origin (MM09). Right panel: same as left panel, but showing age difference of OC pairs as function of their physical separation. The inset shows their separation as function of the Galactocentric distance. }

\label{Fig:pair_age_separation}
\end{center}
\end{figure*}

In order to deepen our analysis, in this section we compared the results obtained for our sample with other binary cluster candidates taken from the list of PP21 (their table A.1), which has been previously matched with the mean cluster parameters as catalogued by CG20. We have restricted PP21' sample to those pairs that contain available age information. For each OC, we summed in quadrature a systematic uncertainty of $\sim$0.1\,mas (appropriate for \textit{Gaia} DR2 data; section 2.3 of \citeauthor{Luri:2018}\,\,\citeyear{Luri:2018}) to the catalogued parallax dispersion ($\Delta\varpi$) of member stars and picked up those pairs satisfying the condition $\vert\langle\varpi_{clu1}\rangle-\langle\varpi_{clu2}\rangle\vert\leq\sqrt{(\Delta\varpi_{clu1})^2+(\Delta\varpi_{clu2})^2}$, where $\langle\varpi\rangle$ is the mean cluster parallax as informed in CG20. The results are shown in Fig.~\ref{Fig:pair_age_separation}; each plot contains 35 cluster pairs: 29 taken from PP21's list, identified as filled stars, and 6 pairs investigated in this paper, shown as coloured symbols (Table~\ref{tab:symbols_convention}).


The left panel of Fig.~\ref{Fig:pair_age_separation} shows the age difference of the OC pairs as function of the pair age, assumed here as the age of its youngest member. We note that $\sim$70\% of these pairs (i.e., 24 systems) present ages smaller than 100\,Myr (black filled symbols together with Collinder\,394$-$NGC\,6716, NGC\,659$-$NGC\,663, King\,16$-$Berkeley\,4 and NGC\,2383$-$NGC\,2384), which is also clear from the analysis of the inset histogram. The few amount of pairs formed by coeval OCs older than this limit may imply that primordial systems do not survive for long (MM09; \citeauthor{Subramaniam:1995}\,\,\citeyear{Subramaniam:1995}). According to DAK21, the combination of mass loss and angular momentum conservation leads several binary clusters to evolve into separated systems. Despite this, longevities of over 100\,Myr are suggested (as in the case of NGC\,5617$-$Trumpler\,22), which agrees with previous works (e.g., \citeauthor{Priyatikanto:2016}\,\,\citeyear{Priyatikanto:2016}).     

Besides, Fig.~\ref{Fig:pair_age_separation} (left panel) suggests that close encounters between OCs with very large age differences ($\Delta\,t\gtrsim500\,$Myr), and therefore with distinct origins, are less common. A probable explanation is that the whole set of pairs are located at small Galactic latitudes ($b\lesssim10^{\circ}$), where clusters are prone to dynamical heating due to destructive external effects, such as close encounters with molecular clouds and/or passages through the disc (\citeauthor{Spitzer:1958}\,\,\citeyear{Spitzer:1958}; \citeauthor{Theuns:1991}\,\,\citeyear{Theuns:1991}; \citeauthor{Gieles:2007}\,\,\citeyear{Gieles:2007}). A typical disruption timescale for OCs in the Galactic disc is $\sim$100\,Myr (e.g., \citeauthor{Bica:2001}\,\,\citeyear{Bica:2001}; \citeauthor{Ahumada:2000}\,\,\citeyear{Ahumada:2000}), which unfavours the formation of stable pairs containing considerably older OCs.



The right panel of Fig.~\ref{Fig:pair_age_separation} shows that most pairs in the complete sample present separations smaller than $\sim$10\,pc, regardless their Galactocentric distance (see also the inset plot in the same panel). These small separations can be attributed to OCs either presenting a common origin (pairs with small age difference) or undergoing close encounters that favor the formation of pairs due to, e.g., tidal capture (which could explain the large spread in age differences), possibly resulting in merger processes (MM10). Particularly, all the older pairs from PP21' sample ($t\gtrsim100\,$Myr; grey filled stars) present separations smaller than $\sim6\,$pc, which favors their survival against disruptive interactions. Pairs with separations larger than $\sim10\,$pc, in turn, are much less numerous and tend to present age differences $\lesssim$100\,Myr (that is, possible correlated origins).  

As shown by MM10, pairs with wider separations tend to be separated under the influence of the Galactic tidal field. Considering the complete timespan of their simulations ($t<$ 200\,Myr; their figures 1, 5, 6 and 7), almost all realizations that show separations above $\sim10\,$pc at a given age result in separated clusters (that is, ``ionized"\,final state). This is the likely evolutionary course of the pairs NGC\,5617$-$Trumpler\,22, NGC\,659$-$NGC\,663, King\,16$-$Berkeley\,4 and Ruprecht\,100$-$Ruprech\,101. Particularly, the old binary Ruprecht\,100$-$Ruprecht\,101 (assumed pair age $\sim$\,560\,Myr) is more likely a transient one, possibly formed by tidal capture (which may operate within resonant trapping; MM09), since long-term stability of primordial binary open clusters appears not to be possible. Only equal-mass pairs formed with originally small separations and nearly circular orbits are likely to be observed as genuine primordial binaries for such long lifetimes (MM10). 

On the other hand, almost all simulated binaries presenting separations smaller than $\sim10\,$pc at a given time evolve into a merger final state, which seems the probable fate of the binary Collinder\,394$-$NGC\,6716 and the pair NGC\,2383$-$NGC\,2384. In the case of the latter, its assumed age ($t\sim20\,$Myr) and separation ($\sim6.5\,$pc) is compatible with other strong candidates to be undergoing merging from MM09 study (see figure 10 of MM10).    


It is worth mentioning that the statements outlined in this section should be considered with some care, since the sample investigated here (Table~\ref{tab:investig_sample}) contains few objects. Besides, the pairs taken from PP21's list need a more detailed characterization, ideally following homogeneous techniques analogous to those performed in the present work and with the fundamental parameters for some of them being possibly reviewed.

\section{Summary and concluding remarks}
\label{sec:conclusions}

The present paper was devoted to a detailed investigation of 16 Galactic OCs distributed as seven stellar aggregates. We determined structural ($r_c$, $r_t$ and $r_{h}$ radii) and time-related ($t_{rh}$ and age/$t_{rh}$ ratio) parameters and searched for possible evolutionary connections among them. We also built unambiguous lists of member stars and derived fundamental astrophysical parameters (age, distance, interstellar reddening) from decontaminated CMDs. This allowed us to establish the physical connection of each binary candidate and, for those potentially interacting ones, we explored their dynamical state. 

The 16 investigated OCs span Galactocentric distances and ages in the ranges $7\lesssim\,R_G(\textrm{kpc})\lesssim11$ and $7.3\leq\textrm{log}\,t\leq9.2$. From the outcomes of our analysis, 4 of the investigated pairs (namely, NGC\,5617 $-$ Trumpler\,22, NGC\,659 $-$ NGC\,663, Collinder\,394 $-$ NGC\,6716 and Ruprecht\,100 $-$ Ruprecht\,101) were considered genuine binary candidates, that is, gravitationally bound structures. The pair King\,16 $-$ Berkeley\,4 seems a physically interacting one, but gravitationally unbound. The OCs NGC\,2383 and NGC\,2384 seem to constitute a dissolving pair. In turn, Pismis\,19, Dias\,1, Czernik\,20 and NGC\,1857 seem not associated to any stellar aggregates.

The analysis of the structural parameters suggests that clusters within bound and dynamically evolved pairs tend to present ratios of half-light to tidal radius slightly larger than single clusters located at similar $R_G$, which suggests that mutual tidal interactions could possibly affect their structural mass distributions. They also tend to be more tidally filled than most of their single OCs counterparts. Although less compact and subject to a stronger external tidal field ($R_G\lesssim7.5\,$kpc), these OCs are tidally underfilled (except for Ruprecht\,100, which presents a compact internal structure with $r_h/r_t\simeq0.25$), that is, their stellar content are contained well within the allowed Roche volume, making them less prone to disruption due to tidal stripping.  

On the other hand, unbound or dynamically unevolved systems apparently present less noticeable signature of tidal forces on their structure. Moreover, the core radius seems more importantly correlated with the clusters' internal dynamical relaxation process. For all pairs containing coeval OCs, the more evolved cluster is also more compact (smaller $r_h/r_t$ ratio), thus suggesting that, in fact, the internal relaxation also plays a role and differences in their dynamical state may be traced back to their formation conditions. 

In turn, the investigated OCs' external structure (as measured by their $r_t$) follow the general trend established by single clusters, which is, on average, larger $r_t$ for clusters located at larger $R_G$. Particularly, those submitted to a more intense external tidal field ($R_G\lesssim7\,$kpc) present smaller and considerably less dispersed $r_t$ values, which favor their survival against tidal disruption.    

Combining parameters from the set of investigated aggregates together with binary candidates taken from the literature revealed that few pairs containing coeval OCs are older that $\sim100\,$Myr. This result is consistent with the outcomes from $N-$body simulations, which show that primordial binaries do not survive for timespans considerably larger than this limit. Besides, close encounters between OCs with very large age differences ($\Delta t\sim500\,$Myr) are rare, probably due to faster dissolution timescales for clusters located at low Galactic latitudes. Most pairs in the complete sample present separations $\Delta r\lesssim10\,$pc between their component OCs. $N-$body simuations show that these systems will more probably evolve into mergers; those with larger separations may evolve into separated clusters. 

There is previous observational evidence that the fraction of OCs located in gravitationally interacting pairs is not negligible and therefore the characterization of an increasingly larger number of candidates is a fundamental step towards a proper understanding of cluster pair formation in the Galaxy. In this context, the availability of spectroscopic data (e.g., $V_r$ and metallicities) in the next releases of the \textit{Gaia} catalogue and also from ground-based spectroscopic surveys will provide additional observational constraints, thus allowing even more precise determinations of their parameters and enlightening our comprehension of the physical processes that rule their dynamical evolution.


\section{Acknowledgments}

The authors thank the anonymous referee for useful suggestions, which helped improving the clarity of the paper. This research has made use of the VizieR catalogue access tool, CDS, Strasbourg, France. This research has made use of the SIMBAD database, operated at CDS, Strasbourg, France. This work has made use of data from the European Space Agency (ESA) mission \textit{Gaia} (https://www.cosmos.esa.int/gaia), processed by the \textit{Gaia} Data Processing and Analysis Consortium (DPAC, https://www.cosmos.esa.int/web/gaia/dpac/consortium). Funding for the DPAC has been provided by national institutions, in particular the institutions participating in the \textit{Gaia} Multilateral Agreement. This research has made use of \textit{Aladin sky atlas} developed at CDS, Strasbourg Observatory, France. The authors thank the Brazilian financial agencies CNPq and FAPEMIG. This study was financed in part by the Coordena\c{c}\~ao de Aperfei\c{c}oamento de Pessoal de N\'ivel Superior $-$ Brazil (CAPES) $-$ Finance Code 001.

\subsection*{Data availability}
\textit{The data underlying this article are available in the article and in its online supplementary material.}

\bibliographystyle{mn2e}
\bibliography{referencias}

\newpage
\newpage
\newpage

\appendix

\section{A simple approach to estimate Roche limit for clusters in a physical binary}
\label{sec:simple_approach_Roche_limit}

\begin{figure}
\begin{center}

\parbox[c]{0.47\textwidth}
  {
   \begin{center}
    \includegraphics[width=0.47\textwidth]{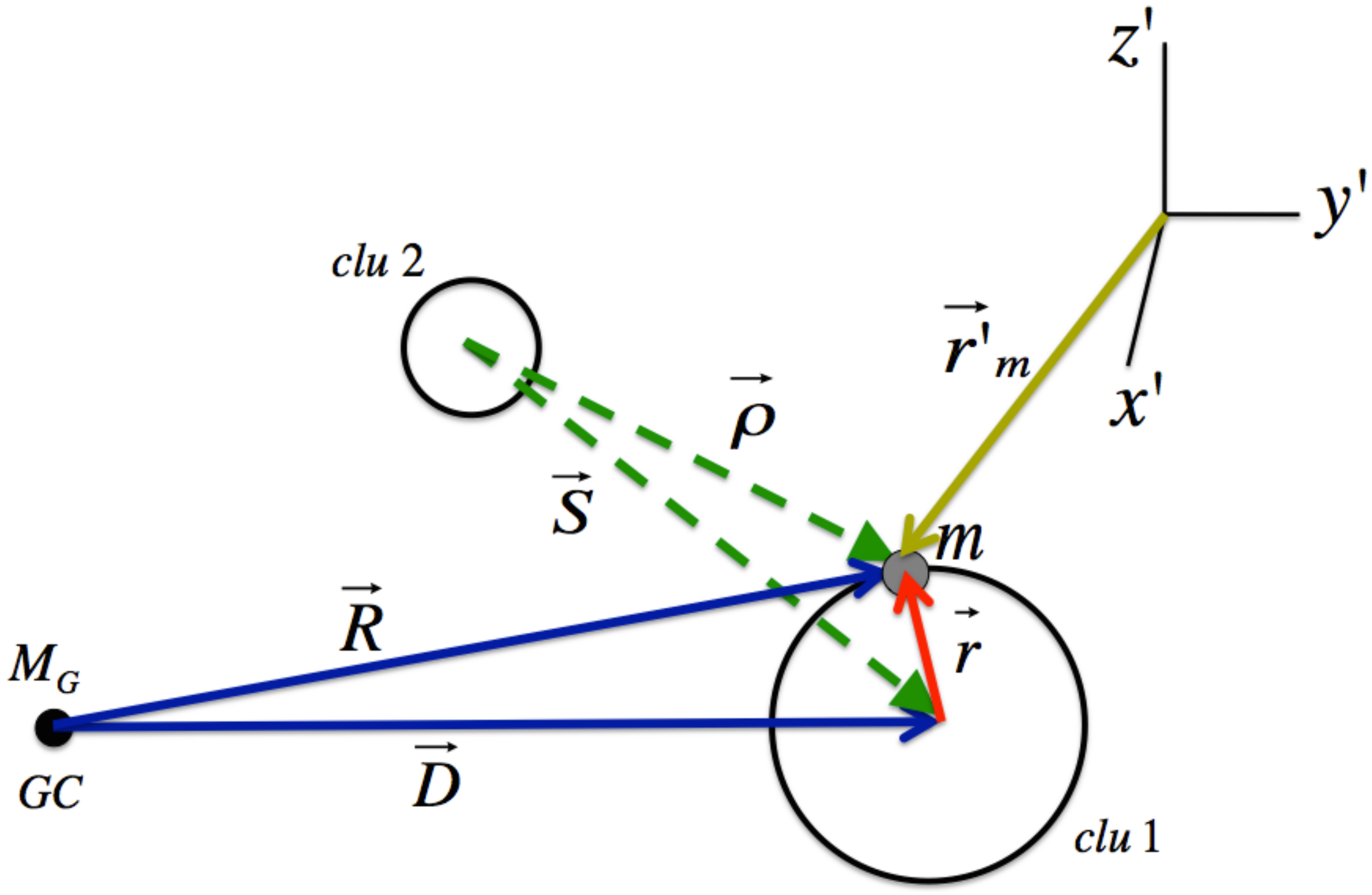}  
    \end{center}    
  }
\caption{ Geometry to analyse tidal interactions between cluster\,1 (identified as $clu1$), cluster\,2 (identified as $clu2$) and the host galaxy (point-mass $M_G$). An inertial reference system ($x'y'z'$) is also represented. See text for details regarding each of the represented vectors in the figure.  }

\label{Fig:three_body}
\end{center}
\end{figure}

In this section, we provide a simplified approach to estimate the limit of the gravitational influence of an OC in a binary system, taking into account the presence of the Galactic tidal field. Fig.~\ref{Fig:three_body} represents schematically the interaction between two stellar clusters ($clu1$ and $clu2$, with masses $M_1$ and $M_2$, respectively) subject to the gravitational pull of the host galaxy (treated here as a point-mass $M_{G}$). In this representation, $\mathbf{D}$ is a vector connecting the galaxy centre and the $clu1$ centre, $\mathbf{R}$ connects $M_G$ and a small test mass $m$ (located close to $clu1$), $\mathbf{S}$ is the position of $clu1$ from the point of view of $clu2$ and \text{\boldmath$\rho$} is the position of the test mass $m$ relatively to $clu2$. Vector $\mathbf{r}$ is the position of the test mass $m$ as measured from $clu1$. An inertial reference system ($x'\,y'\,z'$ in the figure) is also represented and $\mathbf{r'_m}$ is the position vector of mass $m$ in this frame.

The resulting force acting on mass $m$ is the result of gravitational the pull as exerted by $clu1$, $clu2$ and $M_G$, therefore 

\begin{equation}
   m\,\mathbf{\ddot{r'}_m}=-\frac{GM_{1}m}{r^2}\mathbf{\hat{r}}-\frac{GM_{2}m}{\rho^2}\text{\boldmath$\hat{\rho}$}-\frac{GM_{G}m}{R^2}\mathbf{\hat{R}},
   \end{equation}

\noindent
since $\mathbf{\ddot{r'}_m}$ is measured from an inertial reference frame. In order to derive an expression for $\mathbf{\ddot{r}}$, as measured in the \textit{noninertial system} placed at the centre of $clu1$, it is useful to relate $\mathbf{r'_m}$, $\mathbf{r}$ and $\mathbf{r'_{clu1}}$, where the latter is the vector position (not represented in Fig.~\ref{Fig:three_body}, for better visualization) of $clu1$'s centre, as measured from the origin of the inertial reference system:

\begin{equation}
      \mathbf{r'_{clu1}}+\mathbf{r}=\mathbf{r'_m}\,\Rightarrow\,\mathbf{\ddot{r}}=\mathbf{\ddot{r'}_m}-\mathbf{\ddot{r'}_{clu1}}
      \label{eq:posicoes_clu1_m}
\end{equation}

The acceleration of $clu1$ is due to the gravitational pull of $M_G$ and $clu2$, that is:

 \begin{equation}
      \mathbf{\ddot{r'}_{clu1}}=-\frac{GM_{G}}{D^2}\mathbf{\hat{D}}-\frac{GM_2}{s^2}\mathbf{\hat{s}}.
\end{equation}

\noindent
This expression can be inserted in eq.~\ref{eq:posicoes_clu1_m}, thus providing an expression for $\mathbf{\ddot{r}}$:

\begin{equation}
   \mathbf{\ddot{r}}=-\frac{GM_{1}}{r^2}\mathbf{\hat{r}}-GM_G\left[\frac{\mathbf{\hat{R}}}{R^2}-\frac{\mathbf{\hat{D}}}{D^2}\right]-GM_2\left[\frac{\text{\boldmath$\hat{\rho}$}}{\rho^2}-\frac{\mathbf{\hat{s}}}{s^2}\right]
   \label{eq:aceleracao_r_dotdot}
\end{equation}

\noindent
The first term is the acceleration of test mass $m$ due to $clu1$'s gravitational pull and the other two terms are the acceleration from the tidal forces.

\begin{figure}
\begin{center}

\parbox[c]{0.47\textwidth}
  {
   \begin{center}
    \includegraphics[width=0.47\textwidth]{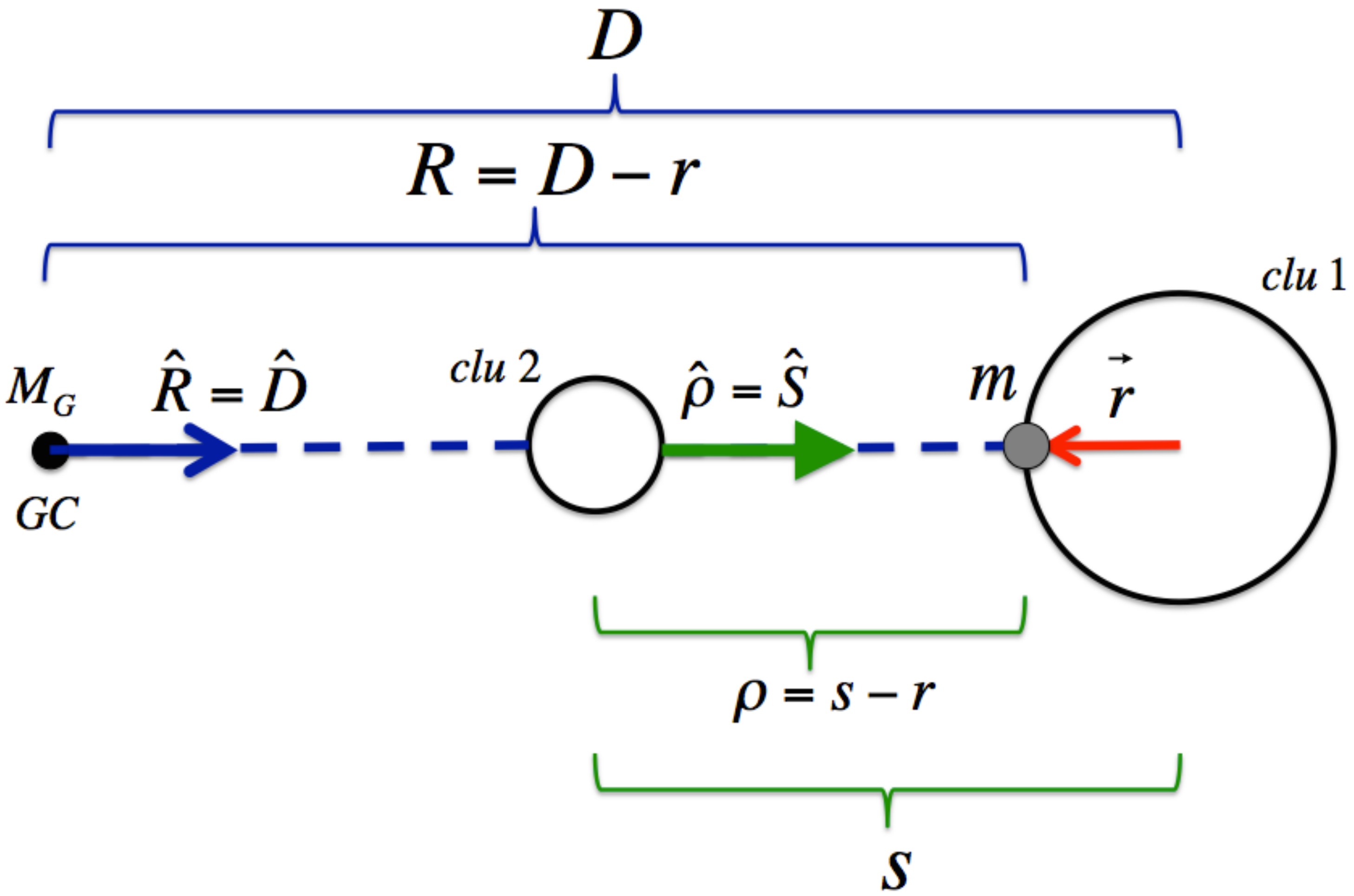}  
    \end{center}    
  }
\caption{  A simplified version of Fig.~\ref{Fig:three_body}, with the four masses ($M_G$, $M_1$, $M_2$ and $m$) lying along the same line. See text for details. }

\label{Fig:three_body_2parte}
\end{center}
\end{figure}

Eq.~\ref{eq:aceleracao_r_dotdot} can be considerably simplified by employing a construction in which the four masses ($M_G$, $M_1$, $M_2$ and $m$) lie along the same line. In this case, both tidal accelerations in eq.~\ref{eq:aceleracao_r_dotdot} point out to the left (since $\mathbf{\hat{R}}=\mathbf{\hat{D}}$, with $R\,<\,D$, and \text{\boldmath$\hat{\rho}$}\,=\,$\mathbf{\hat{S}}$, with $\rho\,<\,s$) in Fig.~\ref{Fig:three_body_2parte} and the contribution of $clu2$ (second therm in brackets in eq.~\ref{eq:aceleracao_r_dotdot}) is the largest possible, since $\rho$ assumes its smallest value when it is equal to $s-r$.

Since in Fig.~\ref{Fig:three_body_2parte} we have $\mathbf{\hat{R}}=\mathbf{\hat{D}}=\mathbf{-\hat{r}}$,  \text{\boldmath$\hat{\rho}$}\,=\,$\mathbf{\hat{S}}=\mathbf{-\hat{r}}$, $\rho=s-r$ and $R=D-r$, we can write

\begin{equation}
   \mathbf{\ddot{r}}=-\frac{GM_{1}}{r^2}\mathbf{\hat{r}}\,+\,GM_G\left[\frac{1}{(D-r)^2}-\frac{1}{D^2}\right]\mathbf{\hat{r}}\,+\,GM_2\left[\frac{1}{(s-r)^2}-\frac{1}{s^2}\right]\mathbf{\hat{r}}   
   \label{eq:rdotdot_completa}
\end{equation}

\noindent
The first term in brackets in eq.~\ref{eq:rdotdot_completa} can be simplified due to the condition $D\gg r$:

\begin{equation}
   \frac{1}{(D-r)^2}=\frac{1}{D^2\left(1-\frac{r}{D}\right)^2}\,\simeq\,\frac{1}{D^2}\left(1+\frac{2r}{D}\right).  
\end{equation}

\noindent
Therefore, eq.~\ref{eq:rdotdot_completa} can be rewritten

\begin{equation}
   \mathbf{\ddot{r}}=-\frac{GM_{1}}{r^2}\mathbf{\hat{r}}\,+\,\frac{2GM_G\,r}{D^3}\mathbf{\hat{r}}\,+\,GM_2\left[\frac{1}{(s-r)^2}-\frac{1}{s^2}\right]\mathbf{\hat{r}}   
   \label{eq:rdotdot_modified}
\end{equation}

In the noninertial frame placed at the centre of $clu1$, the test mass $m$ is in equilibrium when $ \mathbf{\ddot{r}}=\mathbf{0}$, that is, when the gravitational pull due to $clu1$ is balanced by the tidal forces due to $M_G$ and $clu2$. Here we assume as the Roche limit (namely $R_{roc}$) the distance $r$ that results in null acceleration of $m$. This value can be obtained by solving the equation below:

\begin{equation}
  \left[\frac{M_2}{(s-r)^2}-\frac{M_1}{r^2}\right]_{r=R_{roc}}\,=\,\left[\frac{M_2}{s^2}-\frac{2M_G\,r}{D^3}\right]_{r=R_{roc}}.
   \label{eq:numerical_solution_R_roche}
\end{equation}

It is interesting to note that, for a single cluster subject to the tidal field of the host galaxy (that is, employing $M_2=0$ in eq.~\ref{eq:rdotdot_modified}), we have (for $\mathbf{\ddot{r}}=\mathbf{0}$)

\begin{equation}
   \frac{2M_G\,r}{D^3}=\frac{M_{clu}}{r^2}\,\Rightarrow\,r=D\left(\frac{M_{clu}}{2M_G}\right)^{1/3},
\end{equation}

\noindent
which is an analytical expression for the tidal radius of a cluster of mass $M_{clu}$ located at a distance $D$ from the centre of the host galaxy \citep{von-Hoerner:1957}.

\section{NGC\,5617, Trumpler\,22 and Pismis\,19}
\label{sec:N5617_Tr22_Pis19}
\begin{figure*}
\begin{center}

\parbox[c]{1.00\textwidth}
  {
   \begin{center}
    \includegraphics[width=0.33\textwidth]{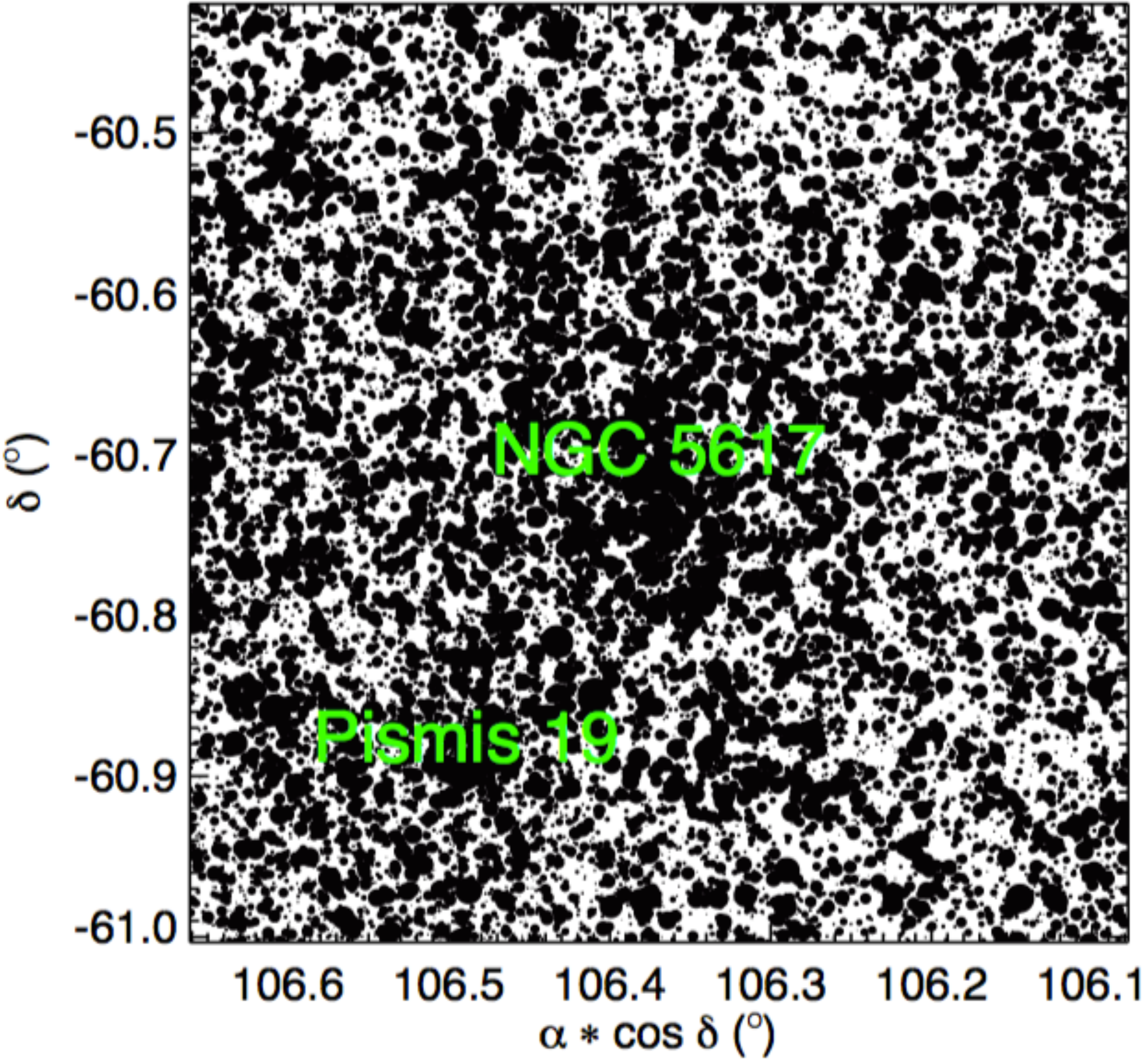}  
    \includegraphics[width=0.33\textwidth]{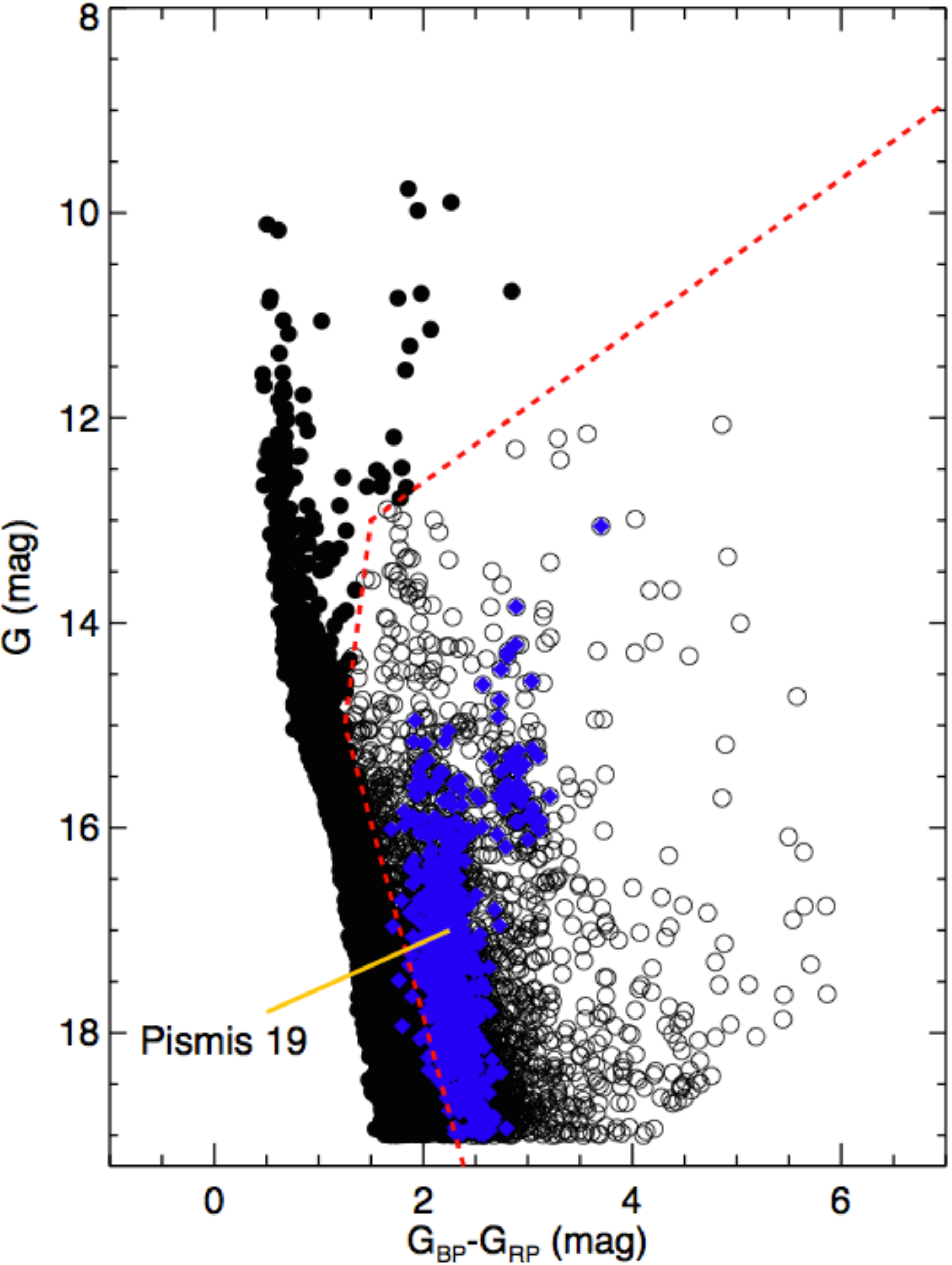}  
    \includegraphics[width=0.33\textwidth]{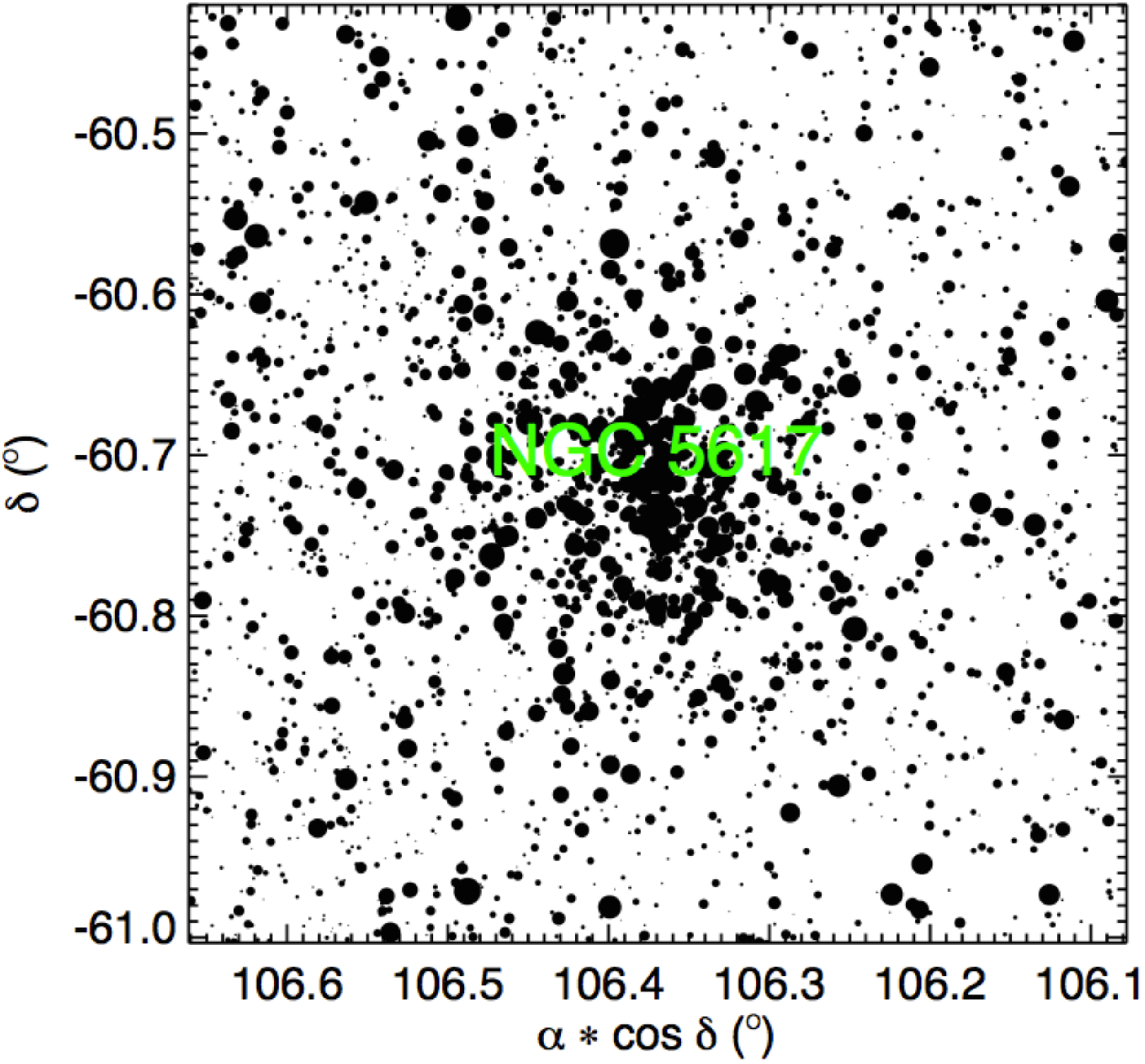}  
    \end{center}    
  }
\caption{ Preliminary analysis of NGC\,5617. Left panel: skymap for stars in a squared area of $35\arcmin\times35\arcmin$ centered in NGC\,5617. Note the proximity with the OC Pismis\,19. After applying the VPD box filter (Section~\ref{sec:struct_params}), which does not eliminate the contamination by Pismis\,19 stars (due to their close proximity with NGC\,5617 even in the VPD), we build the resulting CMD (middle panel). The blue symbols are Pismis\,19 member stars. Open (filled) circles represent stars (not) removed by the colour filter (red dashed lines). The skymap in the rightmost panel corresponds to the sample of stars shown in the middle panel, but after the use of the colour filter. This `filtered' skymap is then used in the structural analysis of NGC\,5617. }

\label{fig:pre_analysis_NGC5617}
\end{center}
\end{figure*}

\begin{figure*}
\begin{center}

\parbox[c]{1.00\textwidth}
  {
   \begin{center}
    \includegraphics[width=0.90\textwidth]{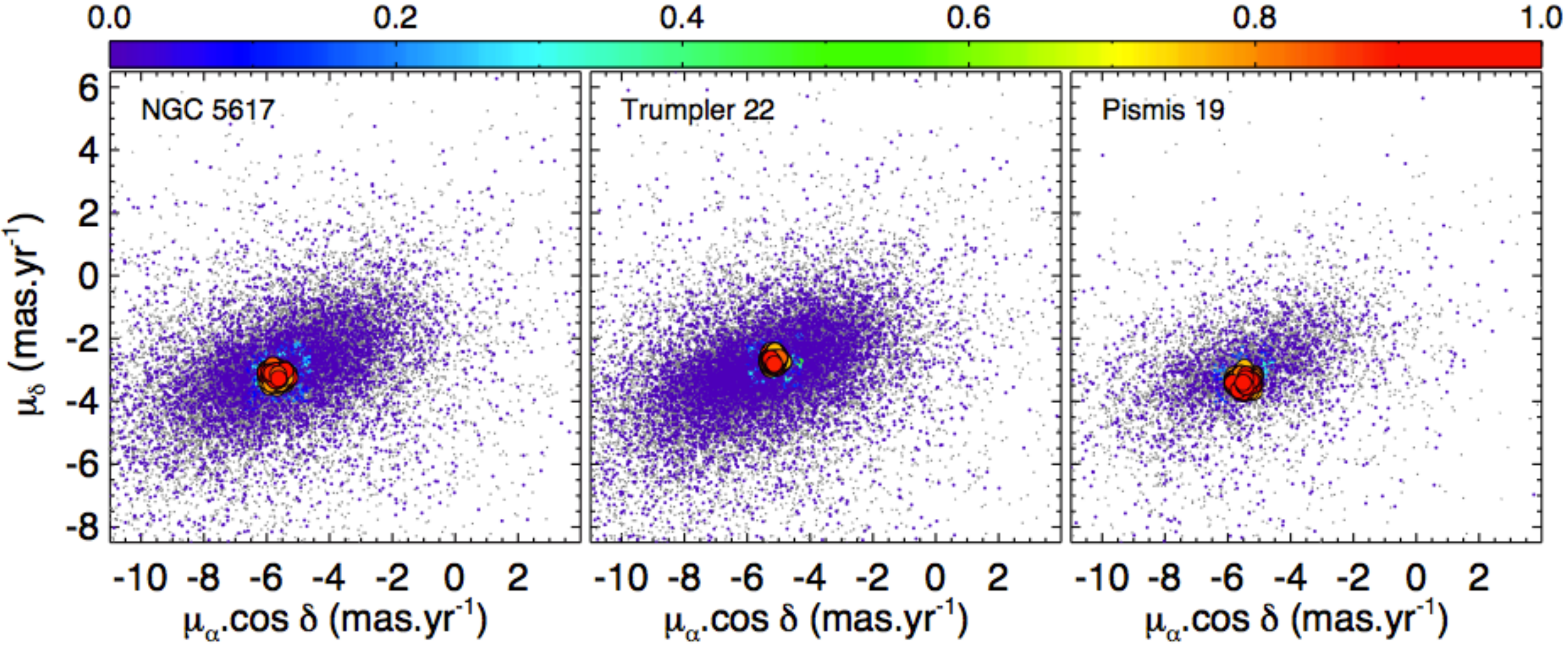}  
    \end{center}    
  }
\caption{ VPDs for stars in the clusters' areas ($r\leq r_t$; coloured symbols) and in the control field (grey dots) for the OCs NGC\,5617, Trumpler\,22 and Pismis\,19. Symbol convention is the same as that of Fig.~\ref{fig:decontam_CMDs_part1}. }

\label{Fig:VPDs_binaries_part3}
\end{center}
\end{figure*}

\begin{figure*}
\begin{center}

\parbox[c]{1.00\textwidth}
  {
   \begin{center}
    \includegraphics[width=0.90\textwidth]{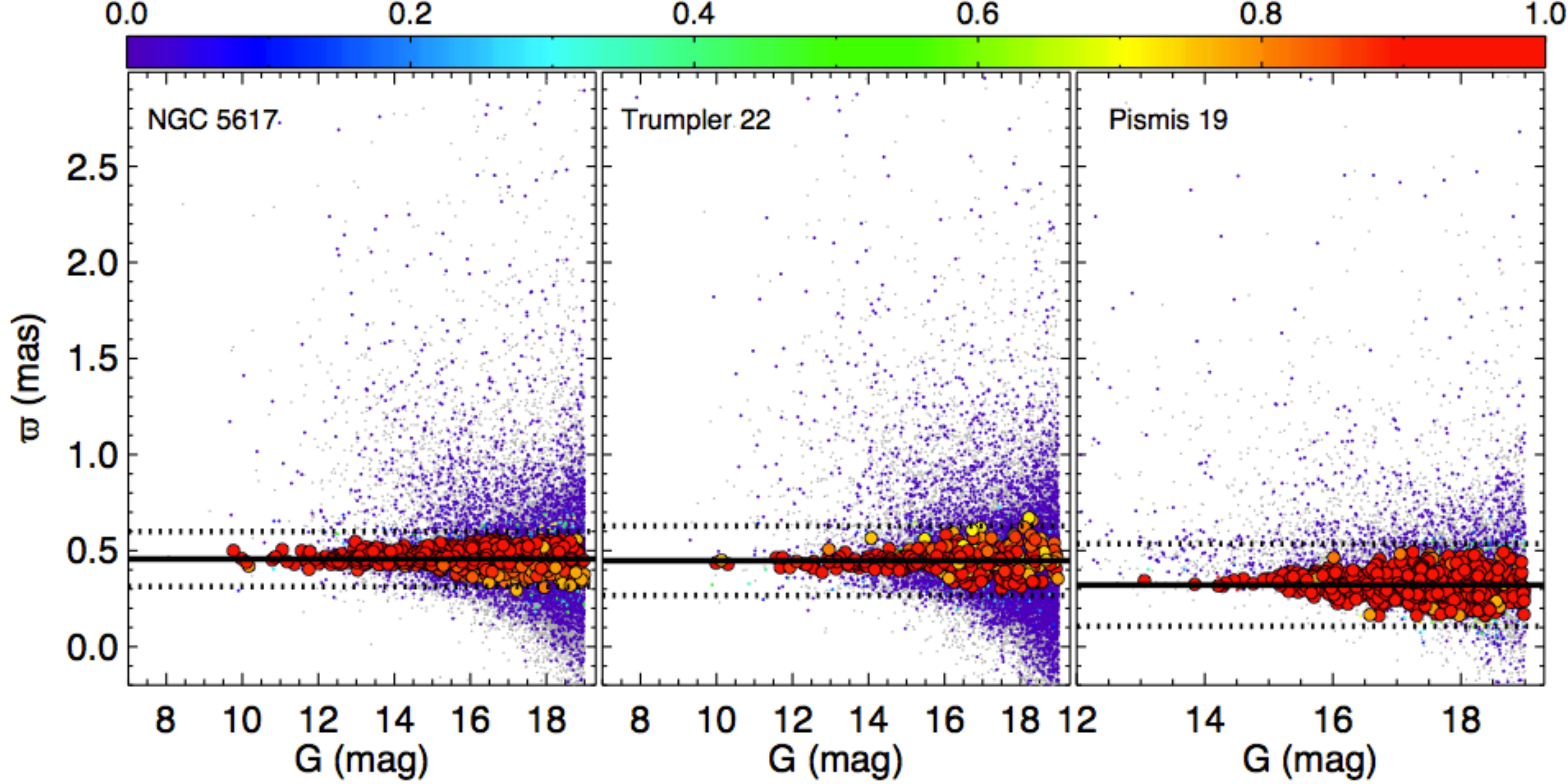}  
    \end{center}    
  }
\caption{ $\varpi$ versus $G$ magnitude for the OCs NGC\,5617, Trumpler\,22 and Pismis\,19. Symbol convention the same as that of Fig.~\ref{fig:decontam_CMDs_part1}. The horizontal continuous line represents the mean $\varpi$ value for member stars and the dotted ones represent the 3$\sigma$ limits. }

\label{Fig:plxvsGmag_binaries_part3}
\end{center}
\end{figure*}

These three OCs are projected in the same region, as can be seen in Fig.~\ref{Fig:skymap_members_binaries_part1}. The analysis of the OCs NGC\,5617 and Pismis\,19 is hampered by the fact that both clusters are nearly indistinguishable in the VPD (Fig.~\ref{Fig:VPDs_binaries_part3}), since the centroids defined by their member stars are located at similar positions. Consequently, the use of VPD filters, as explained in Section~\ref{sec:struct_params}, does not allow to promptly distinguish both clusters. Besides, we can see an overlap in their outermost structures (Fig.~\ref{Fig:skymap_members_binaries_part1}), which affects star counts and therefore the determination of structural parameters.

To overcome these difficulties, we firstly searched for member stars of Pismis\,19, following those procedures explained in Section~\ref{sec:method}. We chose to analyze this cluster first because it is the most compact one compared to NGC\,5617 and Trumpler\,22. This way, as can be inferred from Pismis\,19 RDP (Fig.~\ref{fig:RDPs_part1}), star counts in its outermost parts are not critically contaminated by stars in the neighboring OCs. 

Then we proceeded to the preliminary analysis of NGC\,5617. The skymap in the left panel of Fig.~\ref{fig:pre_analysis_NGC5617} shows stars in an area of $35\arcmin\times35\arcmin$ centered in NGC\,5617. We applied the VPD box filter (Section~\ref{sec:struct_params}), which also includes stars in Pismis\,19 area, to this sample and built the resulting CMD (middle panel of Fig.~\ref{fig:pre_analysis_NGC5617}). After that, a colour filter was applied in order to remove the contamination by Pismis\,19's member stars. This is illustrated in Fig.~\ref{fig:pre_analysis_NGC5617} (middle panel), where we can see, to the left and above the colour filter (red dashed lines), an extended main sequence in the bluer portion of this CMD together with some evolved sequences. The complete set of member stars of Pismis\,19 are overplotted as filled blue diamonds. The use of this colour filter was possible since Pismis\,19 present more reddened evolutionary sequences. Open circles in this CMD represent stars removed by the colour filter. Filled circles represent those stars that have been kept.

Applying the colour filter resulted in the skymap on the rightmost panel of Fig.~\ref{fig:pre_analysis_NGC5617}, where the severe contamination by Pismis\,19 stars (and also other reddened stars, which are part of the Galactic disc) is eliminated. With this filtering procedure, we were able to build the RDP of NGC\,5617 (Fig.~\ref{fig:RDPs_part1}). The rest of the analysis procedure of NGC\,5617 was executed as described in Section~\ref{sec:method}. Analogous procedures were employed in the case of Trumpler\,22. The results are shown in Figs.~\ref{Fig:VPDs_binaries_part3} and \ref{Fig:plxvsGmag_binaries_part3}.



\section{Supplementary figures}
\label{sec:supplementary_figures}

This section contains the whole sets of plots not shown in the main text and neither in the previous appendices.


\begin{figure*}
\begin{center}

\parbox[c]{1.00\textwidth}
  {
   \begin{center}
        \includegraphics[width=0.500\textwidth]{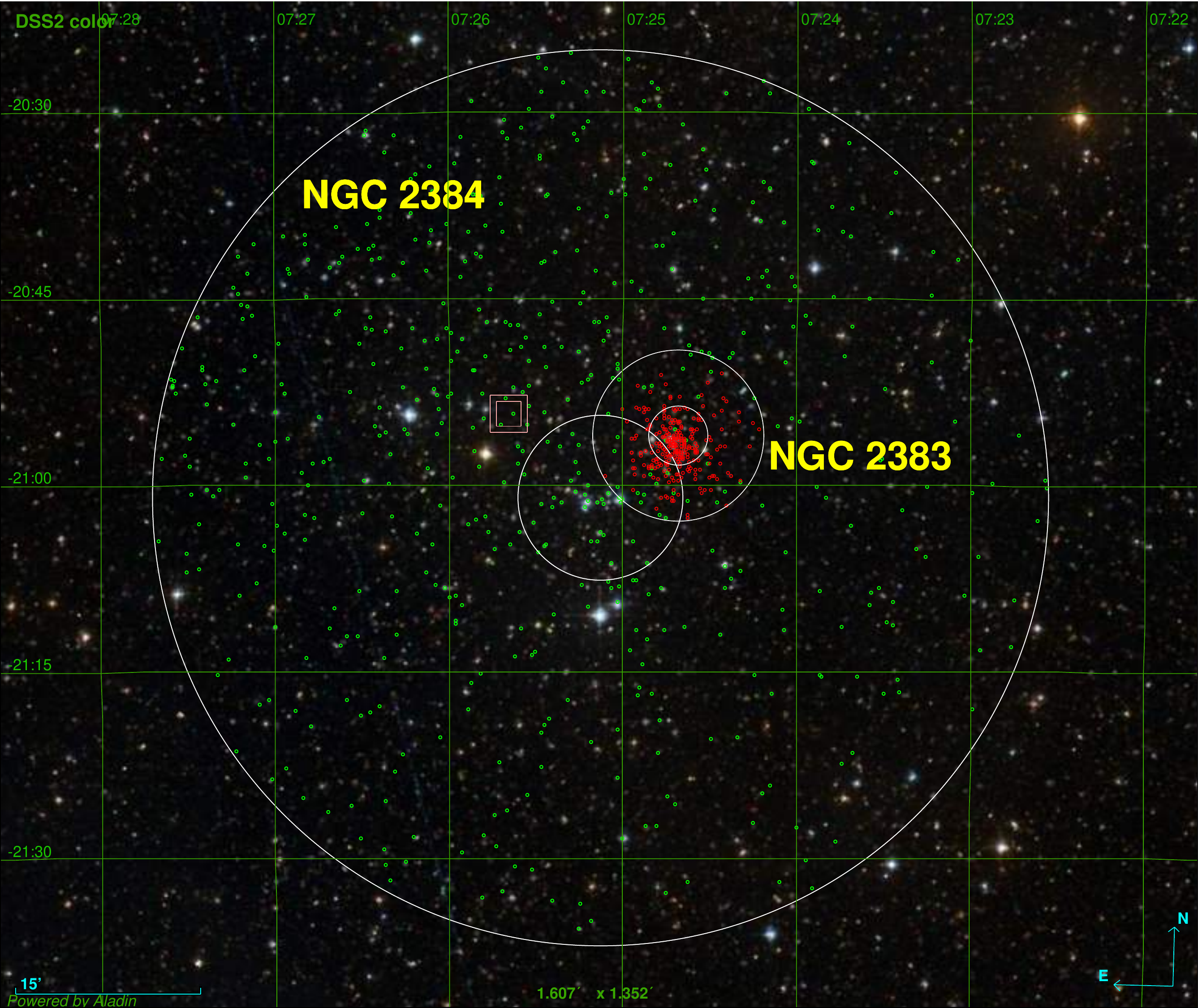}  
        \includegraphics[width=0.460\textwidth]{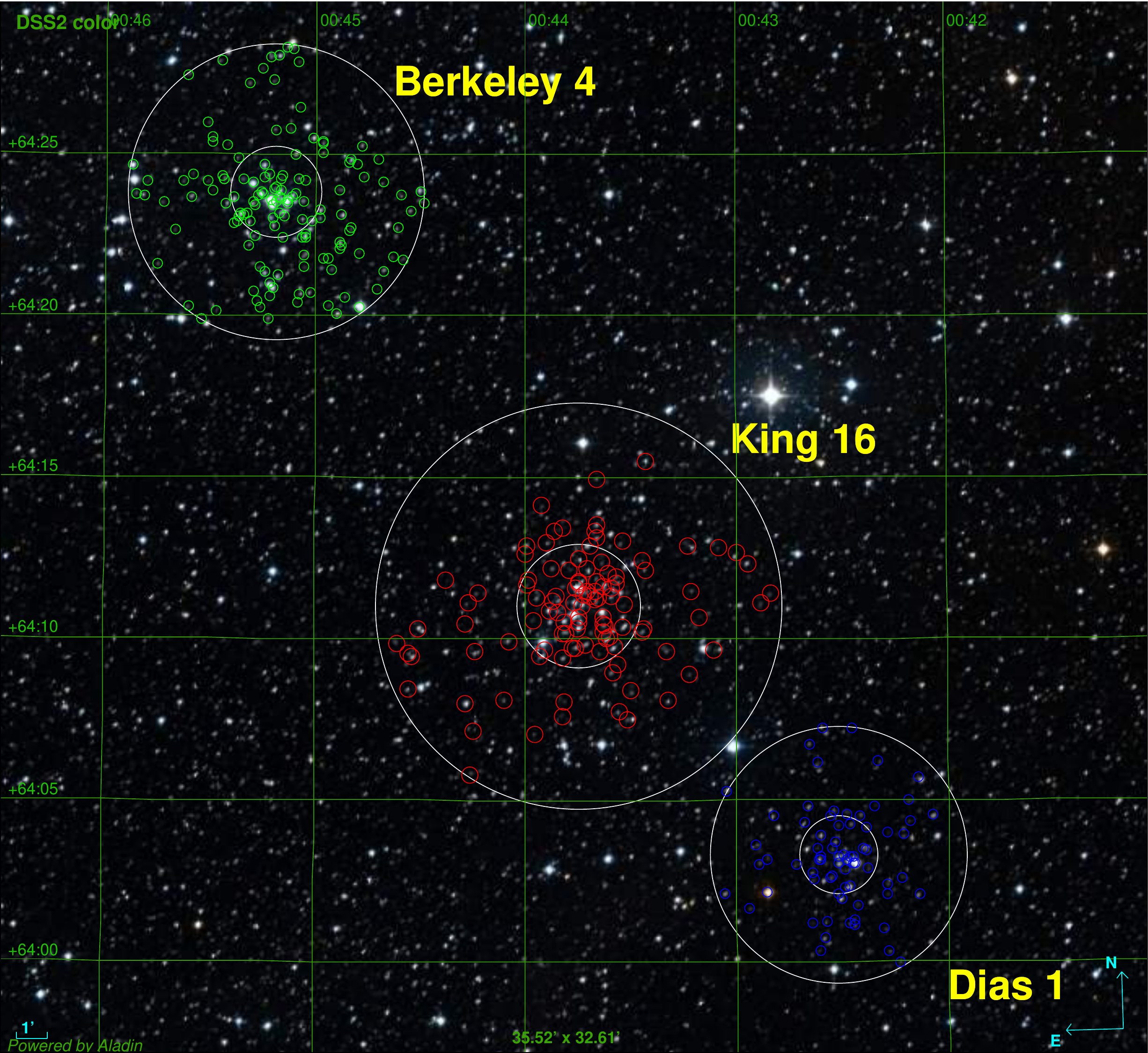} 
        
           \begin{center}
                \includegraphics[width=0.480\textwidth]{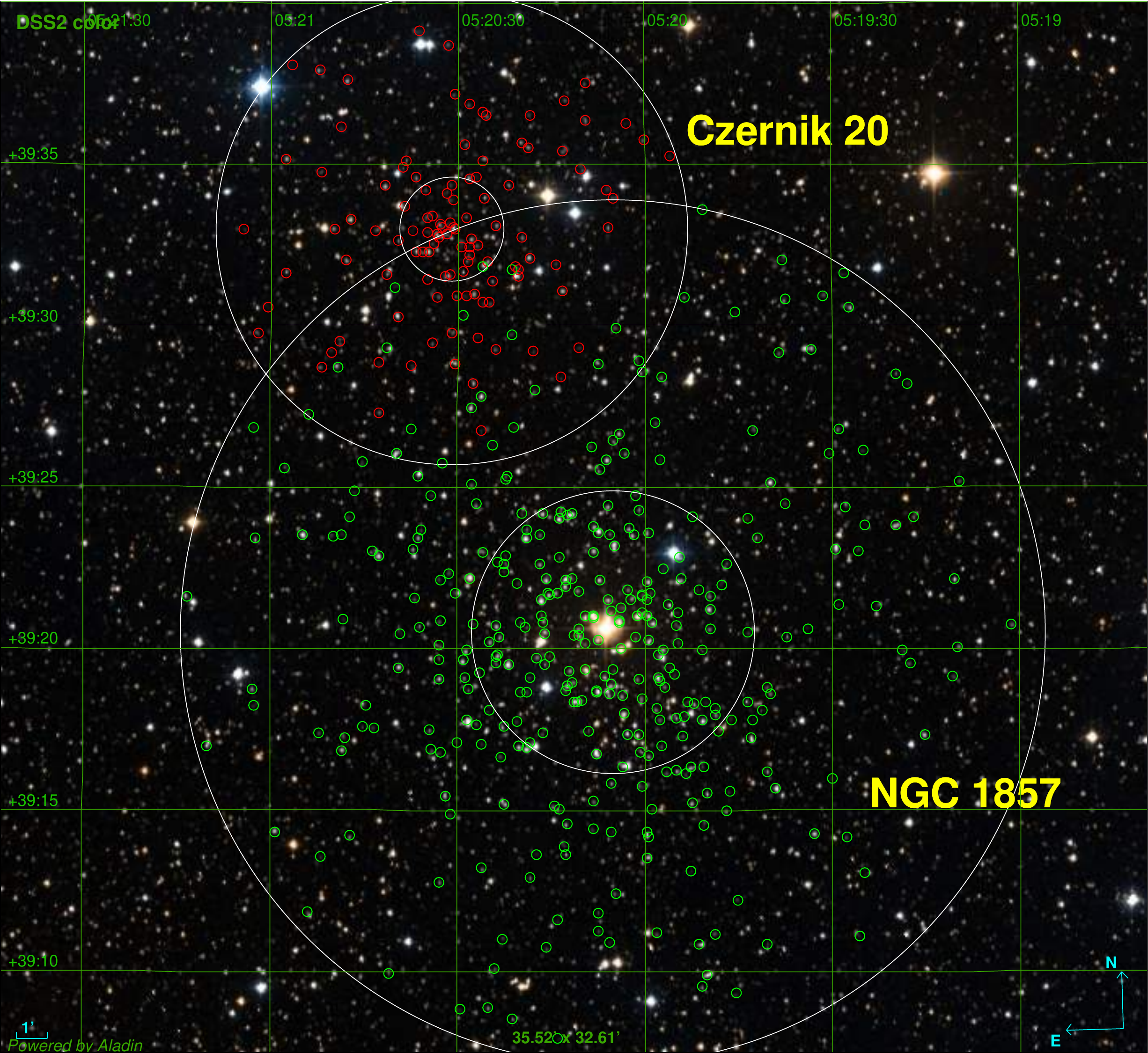}
           \end{center}    

   \end{center}    
  }
\caption{ Same as Fig.~\ref{Fig:skymap_members_binaries_part1}, but showing DSS2 images for other 3 studied cluster pairs with member stars encircled. From top left to bottom right panel: NGC\,2383 (red)$-$NGC\,2384 (green; the smaller circle shows its central region, as inferred from the cluster's RDP in Fig.~\ref{fig:RDPs_part2}; the pink double square shows the central coordinates of UBC\,224, as listed in CG20; see also Section~\ref{sec:particular_procedures_N2383_N2384}; image size: 1.6$^{\circ}\times1.4^{\circ}$), King\,16 (red)$-$Berkeley\,4 (green; image size: 36$\arcmin\times33\arcmin$); the blue circles identify Dias\,1's member stars, projected in the same area; Czernik\,20 (red)$-$NGC\,1857 (green; image size: 36$\arcmin\times33\arcmin$). }

\label{Fig:skymap_members_binaries_part2}
\end{center}
\end{figure*}

\begin{figure*}
\begin{center}

\parbox[c]{1.0\textwidth}
  {
   \begin{center}
    \includegraphics[width=0.70\textwidth]{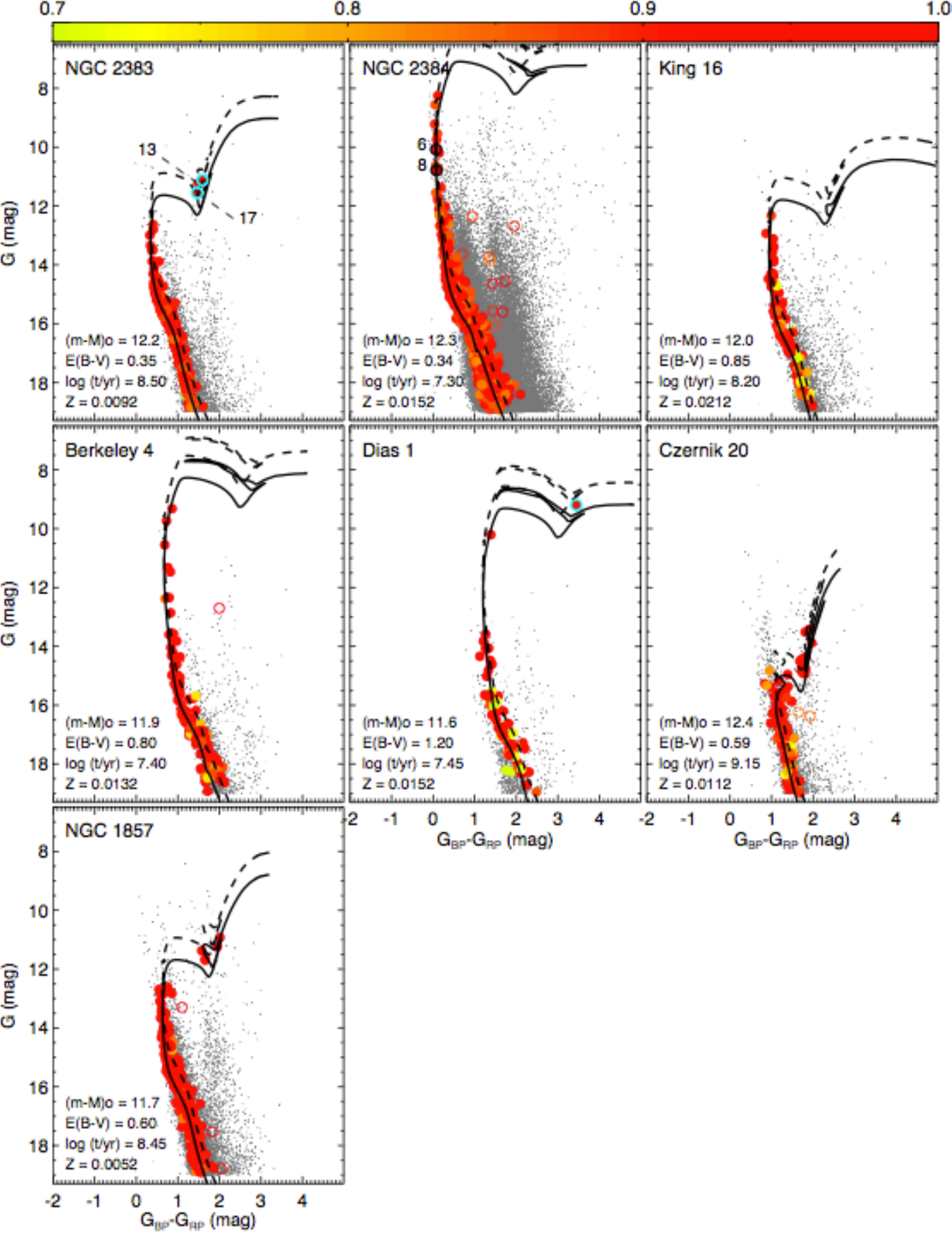} 
    \end{center}
    
  }
\caption{ Same as Fig.~\ref{fig:decontam_CMDs_part1}, but showing the decontaminated CMDs for the OCs: NGC\,2383, NGC\,2384, King\,16, Berkeley\,4, Dias\,1, Czernik\,20 and NGC\,1857. Stars 6, 8, 13 and 17  present spectral types (respectively, B1.5V, B1.5V, K3III and K2III, obtained by \citeauthor{Subramaniam:1999}\,\,\citeyear{Subramaniam:1999}) informed in table 3 of \protect\cite{Vazquez:2010}. }

\label{fig:decontam_CMDs_part2}
\end{center}
\end{figure*}

\begin{figure*}
\begin{center}

\parbox[c]{1.0\textwidth}
  {
   \begin{center}
    \includegraphics[width=0.70\textwidth]{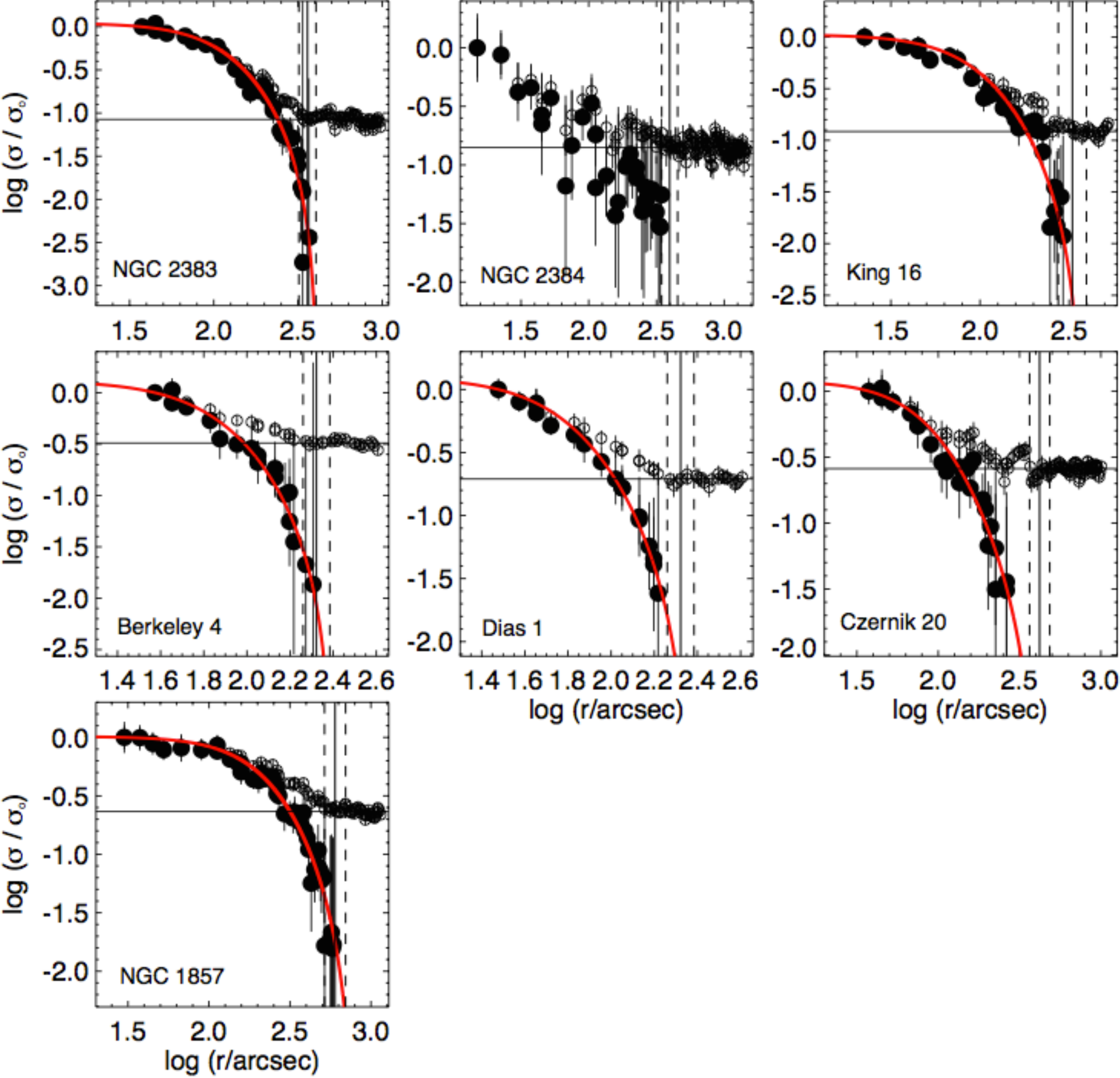} 
    \end{center}
    
  }
\caption{ Same as Fig.~\ref{fig:RDPs_part1}, but showing the RDPs for the OCs: NGC\,2383, NGC\,2384, King\,16, Berkeley\,4, Dias\,1, Czernik\,20 and NGC\,1857. }

\label{fig:RDPs_part2}
\end{center}
\end{figure*}

\begin{figure*}
\begin{center}

\parbox[c]{1.00\textwidth}
  {
   \begin{center}
    \includegraphics[width=0.70\textwidth]{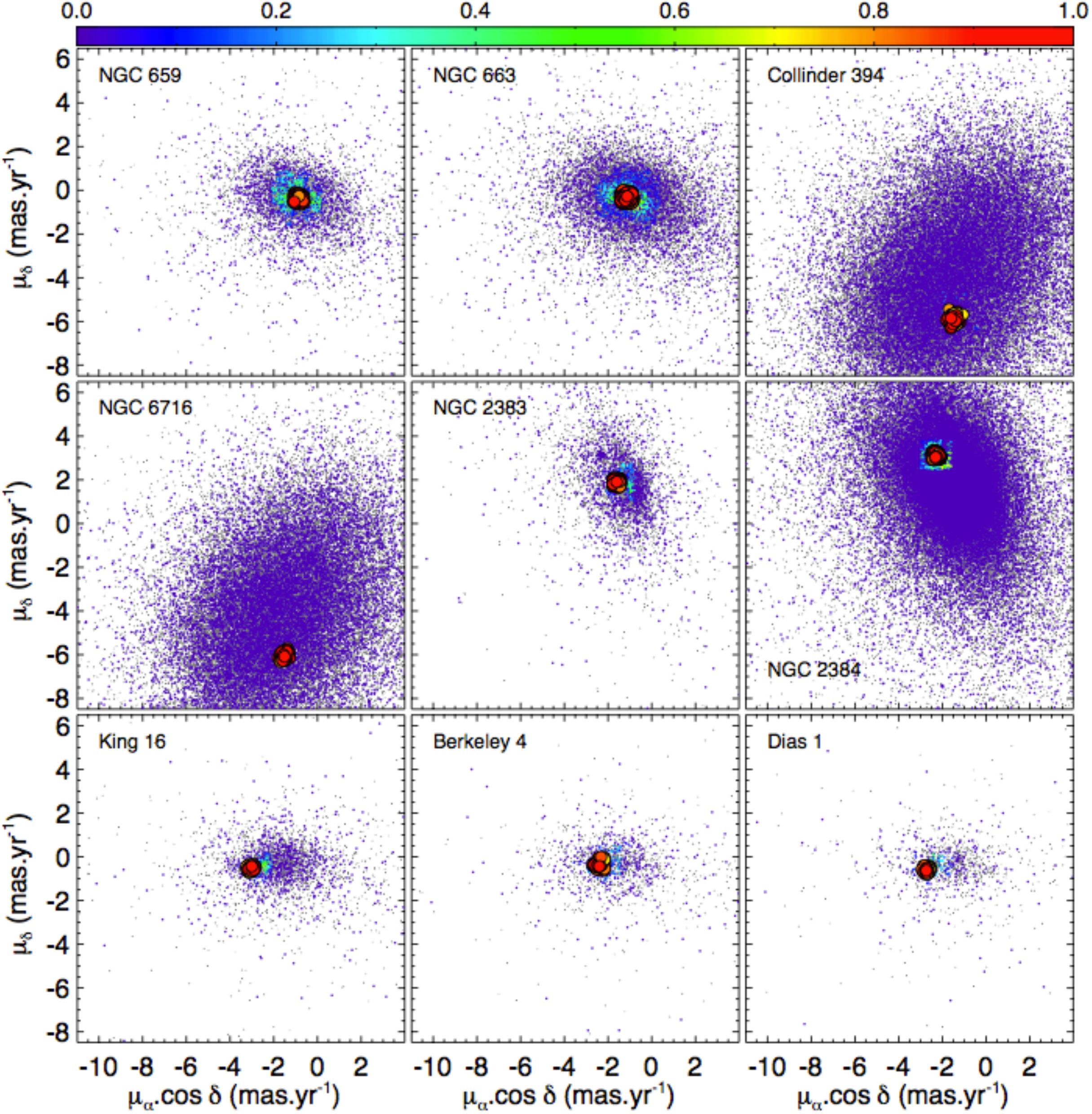}  
    \end{center}    
  }
\caption{ VPDs for stars in the clusters' areas ($r\leq r_t$; coloured symbols) and in the control field (grey dots) for 9 investigated OCs. Symbol convention is the same as that of Fig.~\ref{fig:decontam_CMDs_part1}. }

\label{Fig:VPDs_binaries_part1}
\end{center}
\end{figure*}

\begin{figure*}
\begin{center}

\parbox[c]{0.75\textwidth}
  {
   \begin{center}
    \includegraphics[width=0.75\textwidth]{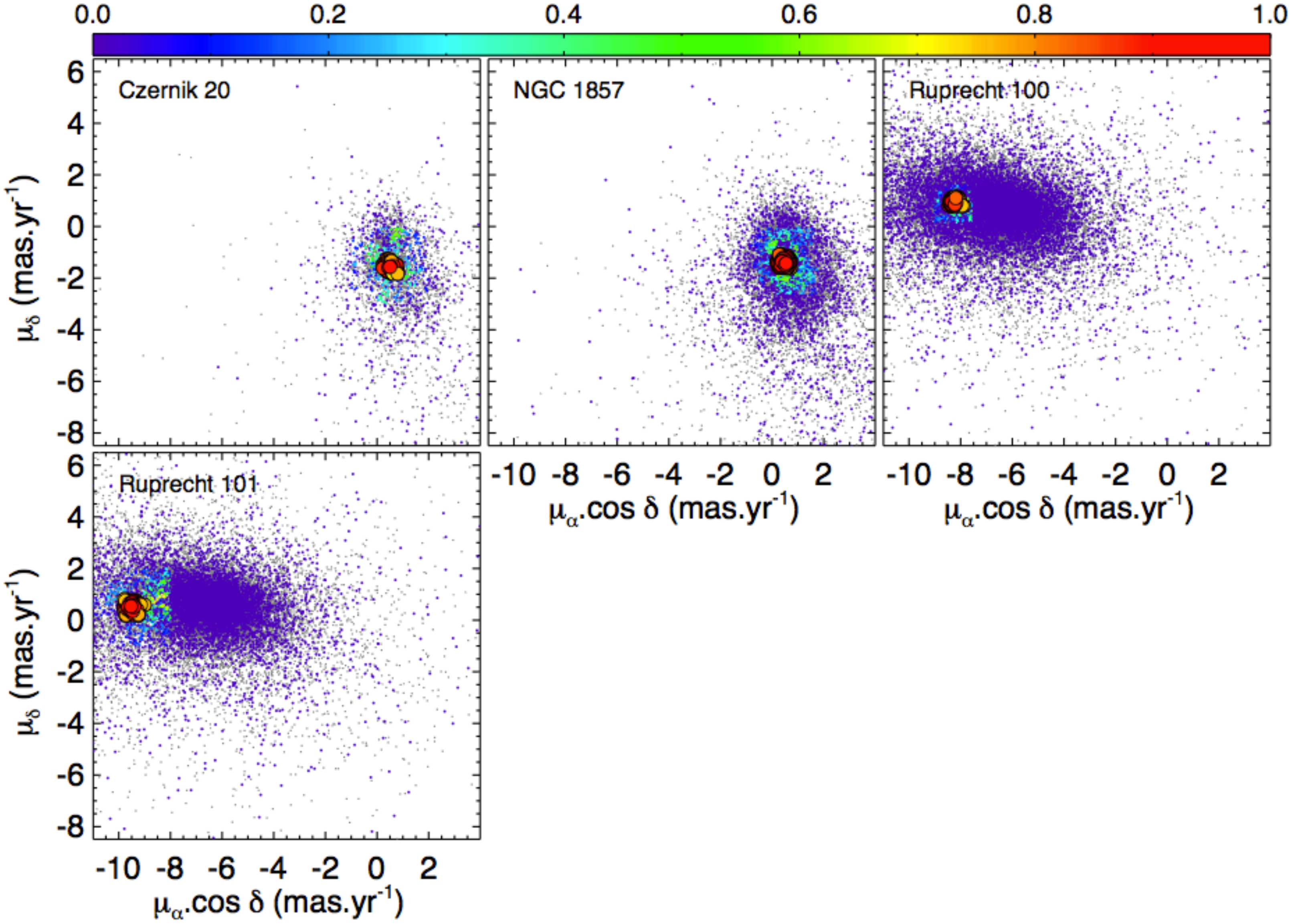}  
    \end{center}    
  }
\caption{ Same as Fig.~\ref{Fig:VPDs_binaries_part1}, but showing the VPDs for the OCs Czernik\,20, NGC\,1857, Ruprecht\,100 and Ruprecht\,101. }

\label{Fig:VPDs_binaries_part2}
\end{center}
\end{figure*}

\begin{figure*}
\begin{center}

\parbox[c]{0.65\textwidth}
  {
   \begin{center}
    \includegraphics[width=0.65\textwidth]{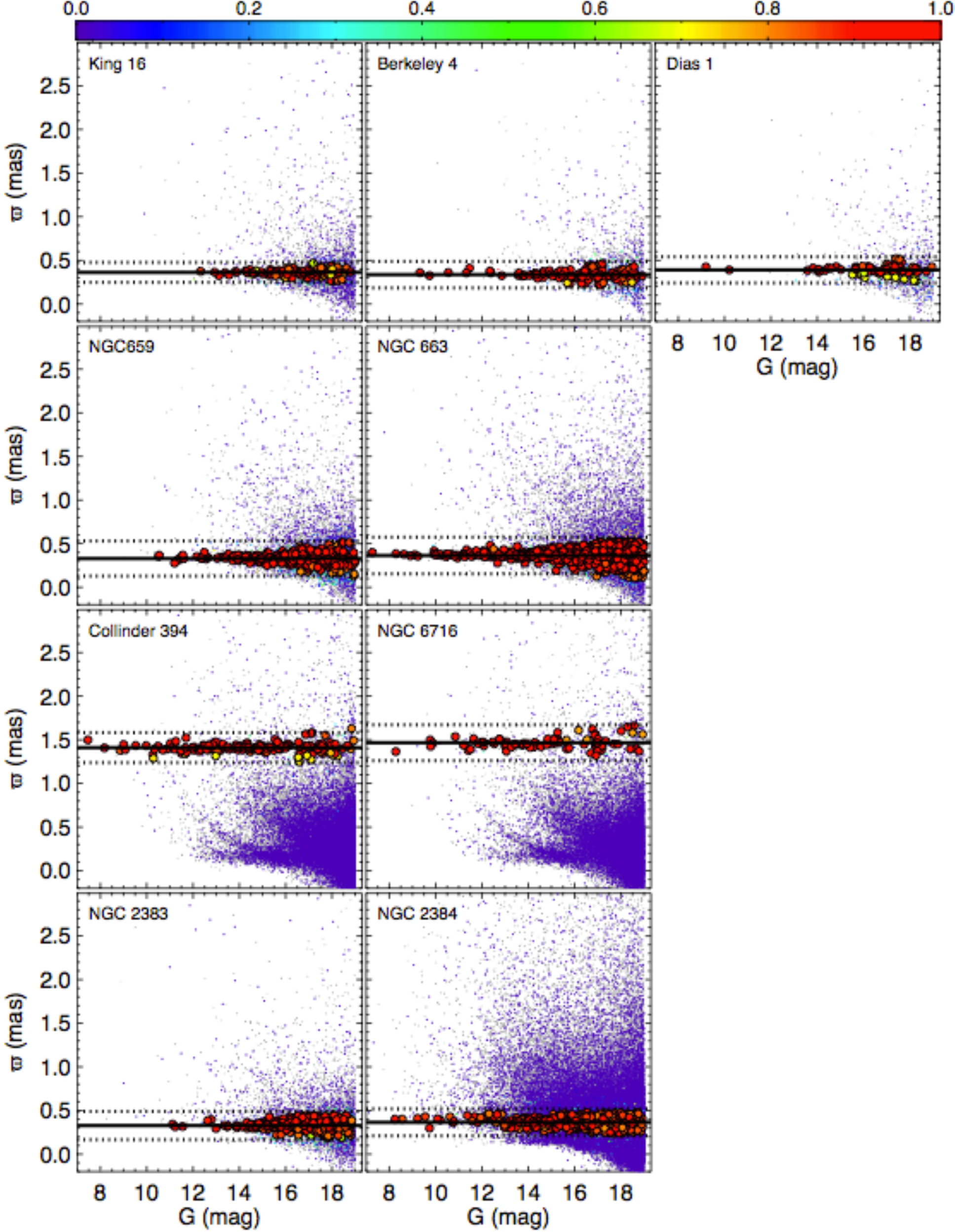}  
    \end{center}    
  }
\caption{ $\varpi$ versus $G$ magnitude for 4 investigated stellar cluster aggregates. Symbol convention the same as that of Fig.~\ref{fig:decontam_CMDs_part1}. The horizontal continuous line represents the mean $\varpi$ value for member stars and the dotted ones represent the 3$\sigma$ limits. }

\label{Fig:plxvsGmag_binaries_part1}
\end{center}
\end{figure*}

\begin{figure*}
\begin{center}

\parbox[c]{0.55\textwidth}
  {
   \begin{center}
    \includegraphics[width=0.55\textwidth]{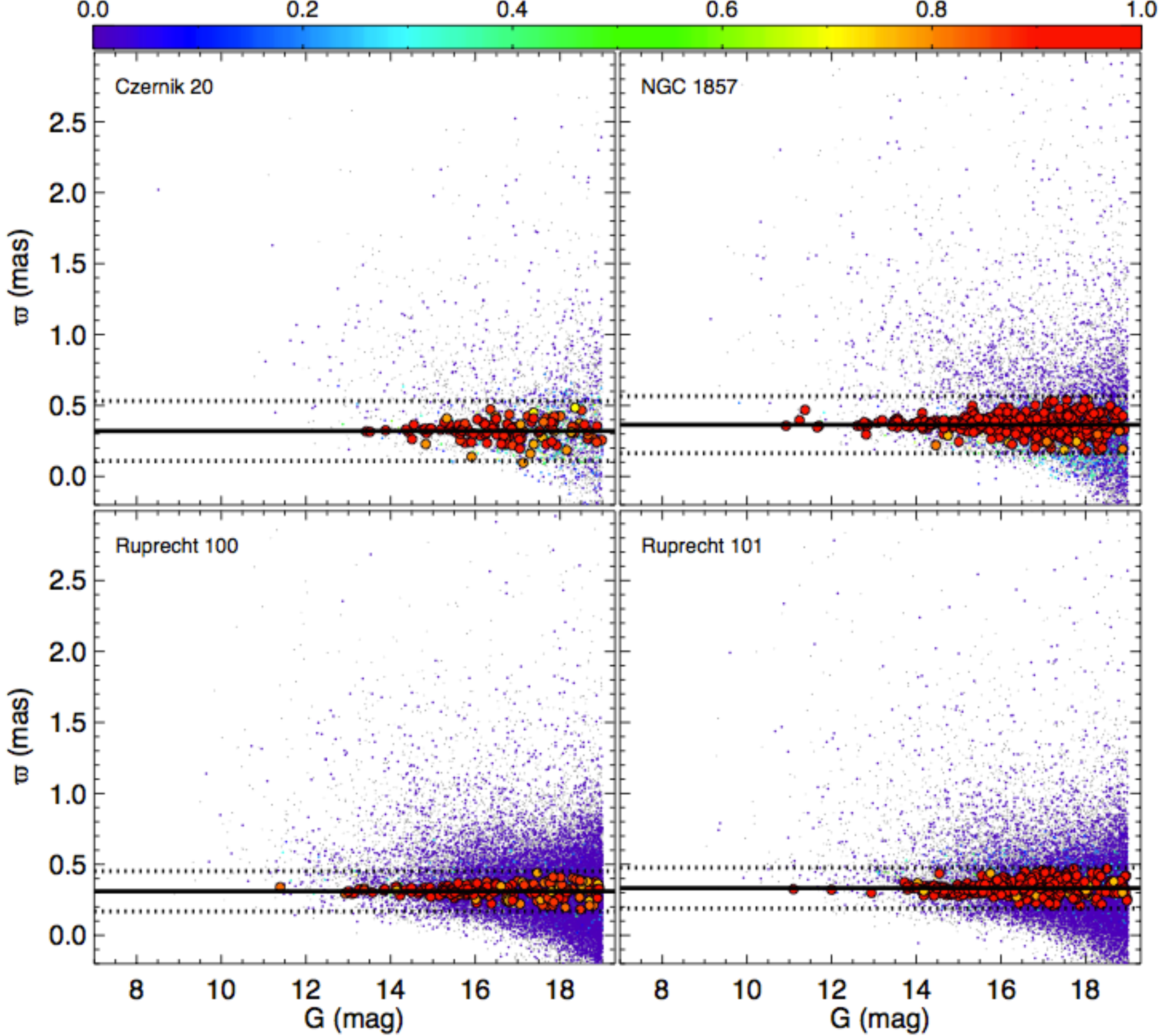}  
    \end{center}    
  }
\caption{ Same as Fig.~\ref{Fig:plxvsGmag_binaries_part1}, but showing the $\varpi$ versus $G$ for the OCs Czernik\,20, NGC\,1857, Ruprecht\,100 and Ruprecht\,101. }

\label{Fig:plxvsGmag_binaries_part2}
\end{center}
\end{figure*}

\begin{figure*}
\begin{center}

\parbox[c]{0.55\textwidth}
  {
   \begin{center}
    \includegraphics[width=0.45\textwidth]{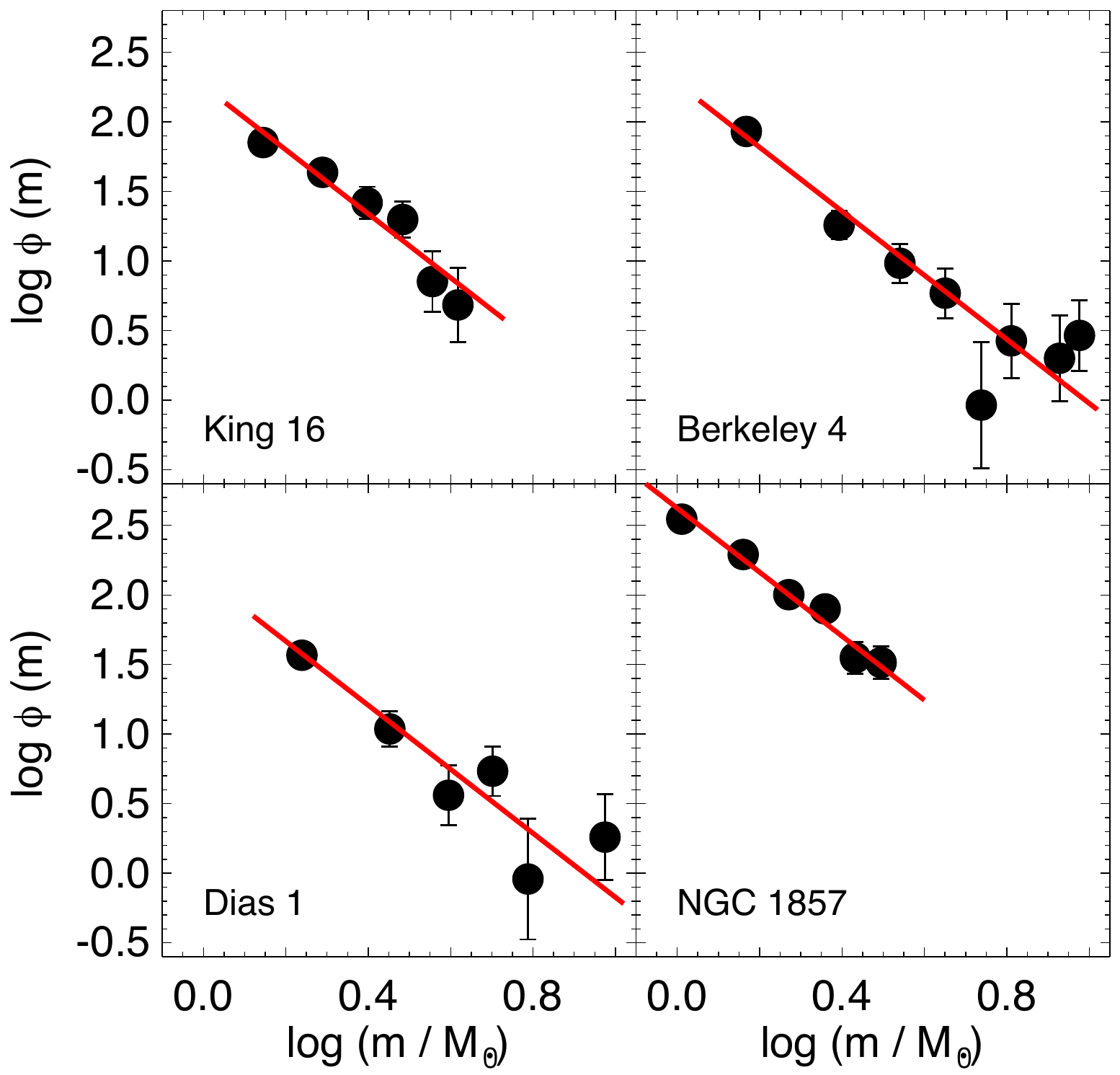}  
    \end{center}    
  }
\caption{ Same as Fig.~\ref{Fig:MF_binaries_part1}, but showing the mass functions for the OCs King\,16, Berkeley\,4, Dias\,1 and NGC\,1857. }

\label{Fig:MF_binaries_part2}
\end{center}
\end{figure*}

\begin{figure*}
\begin{center}

\parbox[c]{0.85\textwidth}
  {
   \begin{center}
     \includegraphics[width=0.85\textwidth]{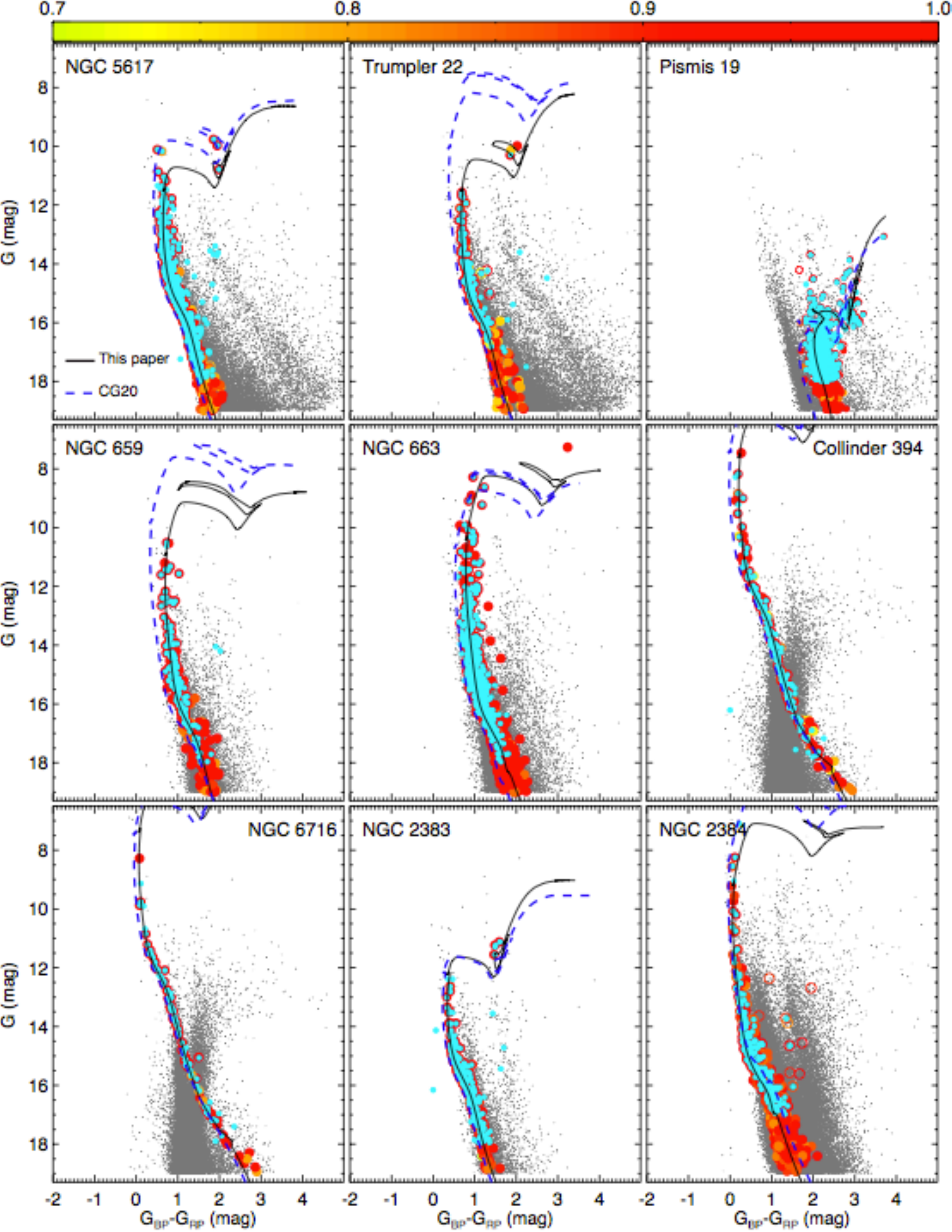}    
    \end{center}
    
  }
\caption{ Same as Figs.~\ref{fig:decontam_CMDs_part1} and \ref{fig:decontam_CMDs_part2}, but overplotting the list of members of CG20 (turquoise circles; $G, G_{BP}$ and $G_{RP}$ magnitudes in the \textit{Gaia} DR2 photometric system) and their isochrone \citep{Bressan:2012} fit solution (blue dashed lines). }

\label{fig:decontam_CMDs_compCGaudin_part1}
\end{center}
\end{figure*}

\begin{figure*}
\begin{center}

\parbox[c]{0.85\textwidth}
  {
   \begin{center}
     \includegraphics[width=0.85\textwidth]{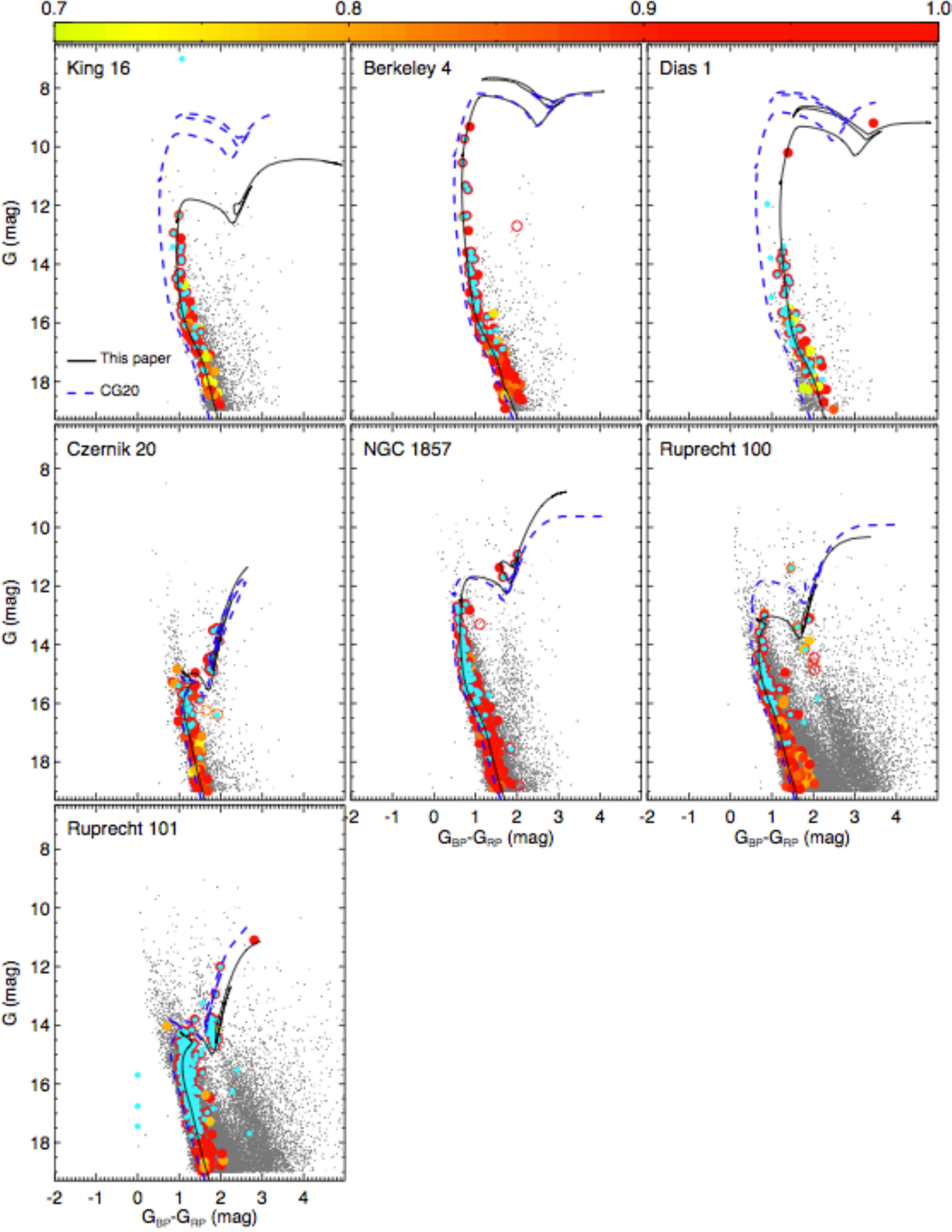}    
    \end{center}
    
  }
\caption{ Same as Fig.~\ref{fig:decontam_CMDs_compCGaudin_part1}, but showing the OCs King\,20, Berkeley\,4, Dias\,1, Czernik\,20, NGC\,1857, Ruprecht\,100 and Ruprecht\,101. }

\label{fig:decontam_CMDs_compCGaudin_part2}
\end{center}
\end{figure*}




\section{Radial velocity data}

Table~\ref{tab:Vrads} presents a compilation of radial velocities from the \textit{Gaia}\,EDR3 catalogue and other sources.

 \begin{table*}
 \small
  \caption{ Radial velocities ($V_r$) for the investigated clusters. }
  \label{tab:Vrads}
 \begin{tabular}{lcccl}

\hline

Cluster         & Source ID (\textit{Gaia} EDR3)     &  $V_{r}$          & $\Delta\,V_{r}$         & Reference                         \\
                &                                    & (km.s$^{-1}$)     & (km.s$^{-1}$)           &                                   \\ 
\hline                                                                                                                                 
NGC\,5617       & 5878621213798169728                & -36.21            & 0.31                    & \textit{Gaia} DR2/EDR3            \\
                & 5878644745900633088                & -34.32            & 0.17                    & \textit{Gaia} DR2/EDR3            \\
                & 5878622141511126400                & -35.52            & 0.29                    & \textit{Gaia} DR2/EDR3            \\
                &                                    &                   &                         &                                   \\
Trumpler22      & 5878407461881147264                & -42.22            & 0.230                   & \textit{Gaia} DR2/EDR3            \\ 
                & 5878406396729283584                & -42.94            & 0.760                   & \textit{Gaia} DR2/EDR3            \\
                & 5878420621661898112                & -45.19            & 3.190                   & \textit{Gaia} DR2/EDR3            \\
                &                                    &                   &                         &                                   \\
NGC\,659        & 509677829088466048                 & 77.81             & 17.61                   & \textit{Gaia} DR2/EDR3            \\
                &                                    &                   &                         &                                   \\
NGC\,663        & 511189863732885248                 & -37.04            & 0.37                    & \textit{Gaia} DR2/EDR3            \\   
                & 511189863732885248                 & [-37.28,-28.35]   & $-$                     & \cite{Mermilliod:2008}            \\
                & 511193471505349504                 & -45.6             & $-$                     & \cite{Liu:1991}                   \\
                & 511193334066395904                 & [-28.3,-26.2]     & $-$                     & \cite{Liu:1991}                   \\
                & 511216904848319104                 & [-43.9,-35.2]     & $-$                     & \cite{Liu:1991}                   \\
                & 511264424368755584                 & [-32.6,-29.0]     & $-$                     & \cite{Liu:1989};\,\cite{Liu:1991}   \\
                & 511264699246655488                 & [-34.9,-30.6]     & $-$                     & \cite{Liu:1989};\,\cite{Liu:1991}   \\
                & 511217828255340672                 & [-36.5,-30.0]     & $-$                     & \cite{Liu:1989};\,\cite{Liu:1991}   \\
                & 511265145924693760                 &  -38.6            & $-$                     & \cite{Liu:1991}                   \\
                & 511213262716085120                 & [-37.1,-33.2 ]    & $-$                     & \cite{Liu:1989};\,\cite{Liu:1991}   \\
                &                                    &                   &                         &                                   \\
Collinder\,394  & 4085939587124734848                &  31.12            & 3.50                    & \textit{Gaia} DR2/EDR3            \\
                & 4085931203348582016                &  21.89            & 8.14                    & \textit{Gaia} DR2/EDR3            \\
                & 4085946802669788416                &  45.53            & 13.84                   & \textit{Gaia} DR2/EDR3            \\
                &                                    &                   &                         &                                   \\
NGC\,6716       & 4086041536793495808                &  2.24             & 3.00                    & \textit{Gaia} DR2/EDR3            \\   
                & 4085290948306525312                &  21.33            & 7.24                    & \textit{Gaia} DR2/EDR3            \\
                &                                    &                   &                         &                                   \\
NGC\,2383       & 5619923307641438848                &  72.11            & 0.35                    & \textit{Gaia} DR2/EDR3            \\   
                & 5619924063555291008                &  72.60            & 0.35                    & \textit{Gaia} DR2/EDR3            \\   
                &                                    &                   &                         &                                   \\
NGC2384         & 5619907055484721024                & [2.7,5.4]         & $-$                     & \cite{Huang:2006}                 \\
                & 5619906810662019840                & [47.8,55.4]       & $-$                     & \cite{Huang:2006}                 \\
                & 5619909769904079360                & [43.2,43.3]       & $-$                     & \cite{Huang:2006}                 \\
                & 5619910010422265216                & [51.5,51.6]       & $-$                     & \cite{Huang:2006}                 \\
                & 5619909903038471296                & [12.9,37.9]       & $-$                     & \cite{Huang:2006}                 \\
                & 5619907570880970752                & [53.2,67.2]       & $-$                     & \cite{Huang:2006}                 \\
                & 5619909907343072896                & [53.8,71.8]       & $-$                     & \cite{Hron:1985}                  \\  
                & 5619910147861188864 	             &  49.4             & $-$                     & \cite{Liu:1991}                   \\
                &                                    &                   &                         &                                   \\
Dias\,1         & 523960107175429632                 & -47.14            & 0.89                    & \textit{Gaia} DR2/EDR3            \\
                &                                    &                   &                         &                                   \\
Ruprecht\,100   & 6057450148662497152                & -10.42            & 1.05                    & \textit{Gaia} DR2/EDR3            \\  
                & 6057401082950880640                & -12.06            & 1.69                    & \textit{Gaia} DR2/EDR3            \\
                &                                    &                   &                         &                                   \\
Ruprecht\,101   & 6054363750798812928                & -12.71            & 5.98                    & \textit{Gaia} DR2/EDR3            \\  
                & 6054367083693515520                & -16.03            & 0.99                    & \textit{Gaia} DR2/EDR3            \\  
                & 6054363643368186368                & -13.96            & 0.29                    & \textit{Gaia} DR2/EDR3            \\  
                & 6054367083693514112                & -14.87            & 1.13                    & \textit{Gaia} DR2/EDR3            \\  
\hline
\multicolumn{5}{l}{\textit{Note:}}																																																     \\
\multicolumn{5}{l}{Numbers between brackets indicate a $V_r$ range instead of a single value.}	  												             \\

\end{tabular}

\end{table*}

\bsp

\label{lastpage}

\end{document}